\documentclass[final,3p,times]{elsarticle}

\usepackage{amsmath,amsthm,amsfonts,amssymb,amscd}
\usepackage{subfig}
\usepackage{graphicx,bm,color}
\usepackage{algorithm}
\usepackage{algpseudocode}
\usepackage{stmaryrd}
\usepackage{booktabs,ctable,multirow,longtable} 

\usepackage{amsthm}
\usepackage{graphicx}
\usepackage{textcomp}
\usepackage{bm}
\usepackage{booktabs}
\usepackage{mathtools}

\usepackage{commath} 

\definecolor{gray}{gray}{0.6}

\def\R{\mathbb R}
\newcommand{\calB}{\mathcal{B}}
\newcommand{\calT}{\mathcal{T}}

\newcommand{\calR}{\mathcal{R}}
\newcommand{\calC}{\mathcal{C}}

\newcommand{\calH}{\mathcal{H}}

\newcommand{\bs}[1]{{\mathbf{#1}}} 
\newcommand{\Bu}{\bs{u}}
\newcommand{\Bn}{\bs{n}}
\newcommand{\Bg}{\bs{g}}
\newcommand{\Bx}{\bs{x}}
\newcommand{\Ba}{\bs{a}}
\newcommand{\BM}{\bs{M}}

\newcommand{\BI}{\bs{I}}
\newcommand{\BG}{\bs{G}}

\newcommand{\Bve}{\bs{\varepsilon}}
\newcommand{\Bvarphi}{{\boldsymbol{\varphi}}}
\newcommand{\BX}{\bs{X}}
\newcommand{\BF}{\bs{F}}
\newcommand{\BC}{\bs{C}}
\newcommand{\BL}{\bs{L}}
\newcommand{\BN}{\bs{N}}
\newcommand{\BP}{\bs{P}}
\newcommand{\BK}{\bs{K}}

\newcommand{\Bb}{\bs{b}}
\newcommand{\BB}{\bs{B}}
\newcommand{\BV}{\bs{V}}
\newcommand{\BR}{\bs{R}}
\newcommand{\xx}{\bs{x}}
\newcommand{\yy}{\bs{y}}
\newcommand{\BA}{\bs{A}}
\newcommand{\AND}{\quad\text{and}\quad}                    
\newcommand{\WITH}{\quad\text{with}\quad}
\newcommand{\Btau}{{\boldsymbol{\tau}}}
\newcommand{\Div}{\mbox{Div}}
\newcommand{\Bzero}{{\boldsymbol{\mathit 0}}}
\newcommand{\BcalF}{\bm{\mathcal{F}}}
\newcommand{\req }[1]{(\ref{#1})}
\DeclareMathAlphabet{\Bgothic}{U}{euf}{b}{n}
\DeclareRobustCommand{\BfrakU}{{\Bgothic U}}
\DeclareRobustCommand{\BfrakC}{{\Bgothic C}}

\DeclareRobustCommand{\Bsigma     }{{\boldsymbol{\sigma}}}
\DeclareRobustCommand{\Bzero }{{\boldsymbol{\mathit 0}}}
\DeclareRobustCommand{\Bn}{{\boldsymbol{\mathnormal n}}}
\DeclareRobustCommand{\Btau       }{{\boldsymbol{\tau}}}

\newcommand{\var}{\texttt}

\newcommand{\PP}{\mathcal{P}}
\newcommand{\RNum}[1]{\uppercase\expandafter{\romannumeral #1\relax}}

\newcommand{\fterm}[1]{\fbox{$\displaystyle#1$}}
\renewcommand{\div}{\text{div}}

\begin{document}

\begin{frontmatter}

\title{Bayesian Inversion for Anisotropic Hydraulic Phase-Field Fracture
}

\author[a]{Nima Noii}
\ead{noii@ifam.uni-hannover.de}

\author[a]{Amirreza Khodadadian\corref{cor1}}
\ead{khodadadian@ifam.uni-hannover.de}

\author[a]{Thomas Wick}
\ead{thomas.wick@ifam.uni-hannover.de}

\cortext[cor1]{Corresponding author}

\address[a]{Leibniz Universit\"at Hannover, Institute of Applied Mathematics,\\
	Welfengarten 1,
	30167 Hannover, Germany}

\begin{abstract}
In this work, a Bayesian inversion framework for hydraulic phase-field transversely isotropic and orthotropy anisotropic fracture is proposed. Therein, three primary fields are pressure, displacements, and phase-field while direction-dependent responses are enforced (via penalty-like parameters). A new crack driving state function is introduced by avoiding the compressible part of anisotropic energy to be degraded. 
For the Bayesian inversion, we employ the delayed rejection adaptive Metropolis (DRAM) algorithm to identify the parameters. 
We adjust the algorithm to estimate parameters according to a hydraulic fracture observation, i.e., the maximum pressure. The focus is on uncertainties arising from different variables, including elasticity modulus, Biot's coefficient, Biot's modulus, 
dynamic fluid viscosity, and Griffith's energy release rate in the case of the isotropic hydraulic fracture while in the anisotropic setting, 
we identify additional penalty-like parameters. Several numerical examples are employed to substantiate our algorithmic developments.\\[1em]


\end{abstract}


\end{frontmatter}

\section{Introduction} \label{sec:Intro}
Hydraulic fracturing is a widely used technique to intentionally create fracture networks in rock materials by fluid injections. This process commonly used in low permeability rocks, for instance, \textit{shale structure} and frequently used in the oil and gas industry \cite{kiparsky2013regulation}. Through fracking process water is injected with very high pressure in the well, to create intentionally a fracture network induced by pressure flow.  The fracture network expands from natural cracks found in the vicinity of the well.  Finally, the fracturing fluid is drained off \cite{moosavi2018initiation}.  

Shale is one of the most abundant sedimentary rocks in the Earth's crust and constitutes a large proportion of the clastic fill in sedimentary basins \cite{kuila2011stress}. These types of rocks are known to be characterized by low porosity within the sedimentary rocks and also very low permeability \cite{goral2020confinement}. 
Shale structures behave naturally in an anisotropic fashion \cite{kuila2011stress} 
and therefore fracture responses highly depend on the interaction 
between structural properties of the material constituent \cite{he2018experimental}. 
Thus, the goal of the fracking process in 
the shale gas reservoir is to activate and open natural fractures and create an anisotropic fracture network \cite{hu2017prediction}. 

In this work, fracture modeling is achieved with a phase-field method.
The variational-based model to fracture by \cite{FraMar98} and  the  related  regularized  formulation,  commonly  referred  to  a variational  phase-field  formulation \cite{miehe+welschinger+hofacker10a,BourFraMar08,BourFraMar00}  
of  brittle  fracture is a widely accepted framework for fracture modeling. 
For strongly anisotropic fracture, e.g. shale structures, a higher-order phase-field framework with smooth local maximum entropy is proposed in  \cite{li2015phase} for crack propagation in brittle materials. In \cite{teichtmeister2017phase},
a phase-field fracture setting for modeling anisotropic brittle material behavior under small and finite deformations was developed.  
The crack phase-field model was extended in \cite{gultekin2018numerical} to include anisotropic fracture employing an anisotropic volume-specific fracture surface function. In \cite{zhang2017modification}, a modified phase-field model that can recognize between the critical energy release rates for mode I and mode II cracks was proposed for simulating mix-mode crack propagation in rock-like materials. A robust and efficient multiscale treatment (according to the Global-Local approach) was developed by the authors to model phase-field fracture in anisotropic 
heterogeneous materials \cite{NoiiAldakheelWickWriggers2019}.

Pressurized and fluid-filled fractures using phase-field modeling was subject in numerous papers in recent years. 
These studies range from 
mathematical modeling 
\cite{BourChuYo12,MiWheWi19,WickLagrange2014,singh2018finite,NoiiWick2019,CHUKWUDOZIE2019957,LeeWheWi16,LeeMiWheWi16}, 
mathematical analysis \cite{MiWheWi15b,MiWheWi14,MiWheWi15c,WiSiWhe15,Wilson2016264}, 
numerical modeling and simulations 
\cite{Miehe2015186,MieheMauthe2015,ehlers17,heider2019phase,HEIDER2018116,LeeMinWhe2018,wang2017unified,Cajuhi2017,LeeWheWiSri17,zhou2018phase,ZHOU2019169,Wick15Adapt}, and 
up to (adaptive) global-local formulations \cite{AlNoWiWr20} 
(see here in particular also \cite{GePlTuDo20} and \cite{NoiiAldakheelWickWriggers2019}
for non-pressurized studies)
and high performance parallel computations 
\cite{HeiWi18_pamm,JoLaWi20}. Extensions 
towards multiphysics phase-field fracture in porous media were proposed in which 
various phenomena couple as for instance proppant \cite{LeeMiWheWi16},
two-phase flow formulations \cite{LeeMiWheWi18} or given temperature variations \cite{NoiiWick2019}.
Recent reviews include \cite{WheWiLee20} and \cite{Wi20_book}.

In the previously described situations, 
uncertainties in parameters arise from several sources, e.g., heterogeneity of rock mass, variability in
geologic formations, such as oil and gas reservoirs,
formation and fluid properties, damage parameters, dynamic viscosity, or critical dissipation.
Furthermore. material parameters fluctuate randomly in space. The mechanical material parameters (e.g., elasticity) are spatially variable and, hence, the uncertainty related to spatially varying properties can be represented by random fields. For instance, the material stiffness property has spatial variability. In order to provide a robust model, their uncertainty should be taken into consideration. 
Thus, the necessity of using an inverse approach (here we used the DRAM algorithm \cite{green2001delayed}) to estimate/identify the influential parameters, simultaneously, seems to be particularly demanding.

Bayesian inversion (as an inverse method) is a statistical technique to identify the various unknown parameters based on prior information (primary knowledge). The great advantage of the method is identifying the various parameters (at the same time) that they can no be measured straightly or only with significant experimental endeavors. 
We use the observations (i.e., experiment values
or synthetic measurements \cite{noii2019characterization}) to update the prior information, then obtain the posterior knowledge. Markov chain
Monte Carlo (MCMC) is a common computational method to extract information according to an inverse problem. Metropolis-Hastings method is the most popular MCMC technique used by the authors for the first time in phase-field fracture \cite{khodadadian2019bayesian}. Here, employing a reference value, i.e., load-displacement curve (estimated by a sufficiently fine mesh), the mechanical parameters (including Lam\'e constants and Griffith's critical elastic energy release rate) identified precisely.

The main objectives of the underlying work are two-fold. First, a modular framework for a variational phase-field formulation of a hydraulic fracture toward anisotropic setting is formulated. We mainly extend the hydraulic phase-field fracture for the transversely isotropic poroelastic material and the layered orthotropic poroelastic materials. Here, direction-dependent responses due to the preferred fiber orientation in the poroelasticity material are enforced via additional anisotropic energy density function for both mechanical and phase-field equations. We derived a new consistent additive split for the bulk anisotropic energy density function to take into account only the tensile part of the energy. Thus, a modified crack driving state function is proposed such that it is only affected by the tensile part of both isotropic and also an anisotropic contribution, as well. Accordingly, a fully coupled monolithic approach for solving pressure and displacement is used while the computed results alternately fixed by solving the weak formulation corresponds to the crack phase-field. A detailed consistent linearization procedure with finite element discretization is further elaborated.

 Second, in this work, a parameter estimation framework using Bayesian inversion for a hydraulic phase-field fracture of the isotropic/anisotropic setting is provided. Here, to enhance the performance of the MCMC method, we employ an adaptive Metropolis \cite{haario1999adaptive}, delayed rejection \cite{green2001delayed} named DRAM. The method has been employed by the authors to estimate the effective physical and biological parameters in silicon nanowire sensors  \cite{khodadadian2020bayesian,mirsian2019new}. Here, the main aim is to determine several effective parameters in the phase-field hydraulic fracture. For this, the maximum pressure during the fluid injection is chosen, and we strive to estimate the peak point and predict the crack behavior. The interested reader can refer to  \cite{elsheikh2014efficient,blaheta2020bayesian} for the application of Bayesian inversion in porous media.

The outline of the paper is as follows. In Section \ref{Section2}, we present the mathematical framework and the variational phase-field model. The explain the model for isotropic materials and then derive a new setting to model hydraulic fractures in anisotropic materials. At the last step, as variational formulation will be derived for the coupled multi-field problem. 
We described how we used the finite element method to discretize the weak formulation and obtain the solutions, detailed in Appendix A. In Section \ref{Section3}, we present the DRAM algorithm and explain how it will be adjusted to identify the parameters in hydraulic fractures. In Section 4 four specific test experiments are given to verify the efficiency of the developed model, where the first two examples cover the isotropic materials. In the next numerical experiments, we consider the hydraulic fracture approach for transversely isotropic and orthotropy anisotropic fracture. Finally, the last section concludes the paper with some remarks.

\section{Phase-field formulation of anisotropic hydraulic fracture}\label{Section2}
In this part, we model variational anisotropic phase-field fracture model toward poroelastic media, considering small deformations.
Three governing equations are employed to characterize the constitutive formulations for the mechanical deformation, fluid pressure as well as the fracture phase-field. Afterward, we describe strong and variational formulations of the coupled multi-physics system. To formulate direction-dependent responses due to the preferred fiber orientation in the poroelasticity material, mechanical and phase-field equations are enforced via additional anisotropic energy density function.  
\subsection{Governing equations of poroelasticity}\label{Section21}

%
Let us consider $\calB\in{\calR}^{\delta}$ a solid in the Lagrangian (reference) configuration with dimension $\delta = 2,3$ in the spacial direction, time $t\in \calT = [0,T]$ and $\partial\calB$ its surface boundary. Regarding the boundary condition, we  assume Neumann conditions on $\partial_N \calB := \Gamma_N \cup \mathcal{C}$, where $\Gamma_N$  indicates the outer domain boundary and additionally Dirichlet boundary conditions on $\partial_D\calB $. 

The given boundary-value-problem (BVP) is a coupled multi-field system for the fluid-saturated porous media of the fracturing material. Fluid-saturated porous media can be formulated based on a coupled three-field system. 
At material points $\Bx\in\mathcal{B}$ and time $t\in\calT$, the BVP solution indicates  the displacement field $\Bu(\Bx,t)$ of the solid, the fluid pressure field $p(\Bx,t)$ as well as the crack which can be represented by
\begin{equation}
\Bu: 
\left\{
\begin{array}{ll}
\calB \times \calT \rightarrow \calR^{\delta} \\
(\Bx, t)  \mapsto \Bu(\Bx,t)
\end{array}
\right.
,\quad \
p:
\left\{
\begin{array}{ll}
\calB \times \calT \rightarrow \calR \\
(\Bx, t)  \mapsto p(\Bx,t)
\end{array}
\right.
,\quad \
d: 
\left\{
\begin{array}{ll}
\calB \times \calT \rightarrow [0,1] \\
(\Bx, t)  \mapsto d(\Bx,t)
\end{array}.
\right.
\label{phi-p-d-fields}
\end{equation}

Considering $d(\Bx,t)=1$ in addition to $d(\Bx,t)=0$ are referred to the unfractured as well as completely fractured part of the material, respectively. Following Figure \ref{Figure1}, the regularized fracture surface $\mathcal{C}_l$ is estimated in $\calB_L\subset\calB$  named \textit{fractured area}. The unbroken area without fracture is defined as:
\begin{align}
\calB_C:=\calB \backslash \calB_L\subset\calB\quad\text{such~that}\quad {\calB}_C\cup{\calB}_L=:\calB~\text{and}~{\calB_C}\cap{\calB_L}=\varnothing.
\end{align}
For stating the variational formulations, we 
now introduce:
\begin{equation}\label{space1}
\begin{aligned}
\bm{V}_{\Bu}&:= \{ {\bf H}^1(\calB)^\delta:\bm u=\bar{\bm u}\; \mathrm{on} \; \partial_D\calB  \}, \quad\\ 
W&:= \text{H}^1(\calB) ,\quad
W_{in} := \{ d \in \text{H}^1(\calB)^{\delta-1} | \; 0 \leq
d\leq d^{old} \},\\
{{V}_{p}}&:=\{ {H}^1(\calB):  p=\bar{p} \; \mathrm{on} \; \partial_D\calB \}.
\end{aligned}
\end{equation}

As typical in problems with inequality constraints (see e.g.,
\cite{KiOd88,KiStam00}), $W_{in}$ is a nonempty, closed, convex,
subset of the linear function space $W$, which is no longer a linear space.

\begin{figure}[!t]
	\centering
	{\includegraphics[clip,trim=0cm 7cm 0cm 1cm, width=15cm]{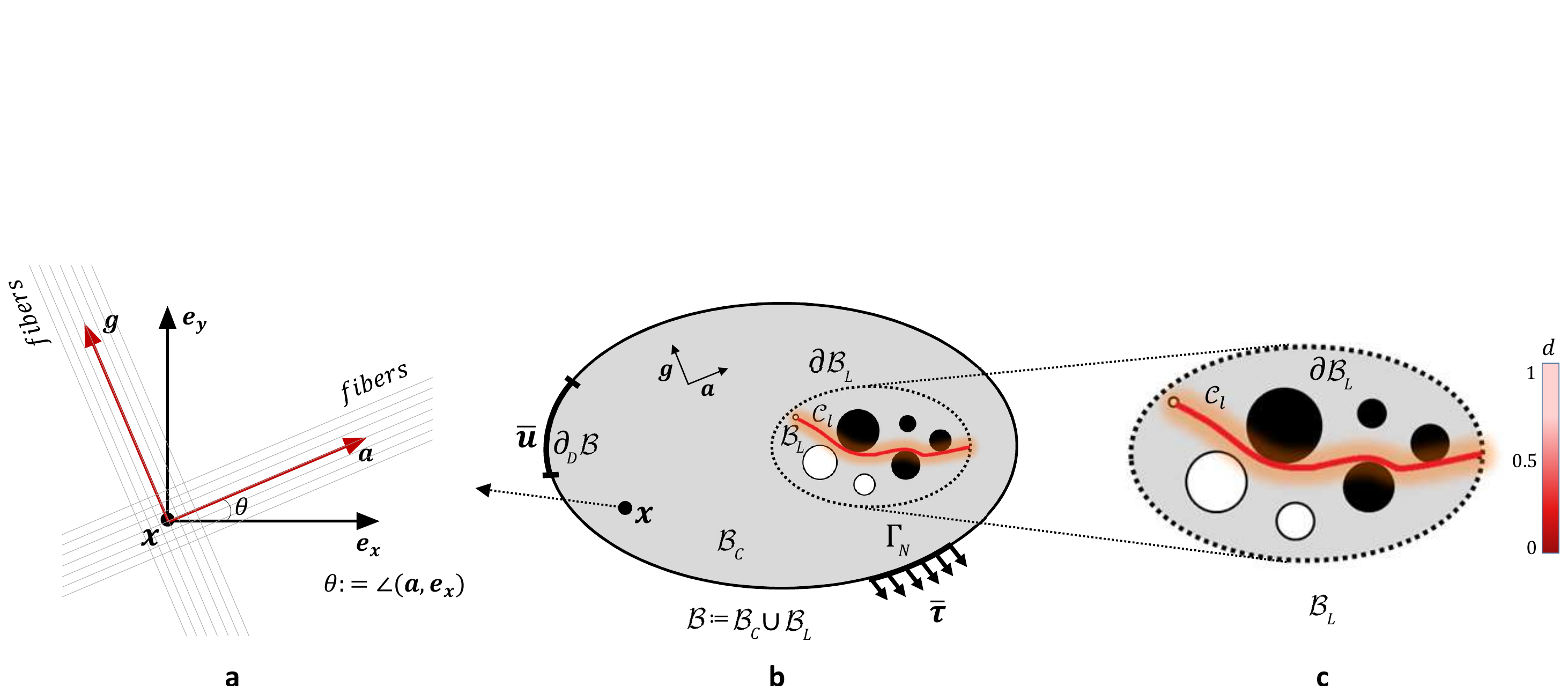}}  
	\caption{ (a) Geometry and loading setup in a domain denoted by $\calB$ whereas ${\calB}_C$
		stands for the unbroken domain, (b) zoom into the fractured region ${\calB}_L\subset\calB$ such that regularized ${\mathcal{C}}_l$ (the red curved crack surface) is approximated in this closed area, and (c) global Cartesian coordinate system with unit vectors
		$(\bm{e}_{\bm{x}},\bm{e}_{\bm{y}})$ and local orthogonal principal
		material coordinates corresponds to the first and second families of
		fibers $(\bm{a},\bm{g})$.}
	\label{Figure1}
\end{figure}

%
\subsubsection{Mechanical contribution}

Here, we represent the governing equations for brittle fracture in elastic solids at small strains. For isotropic materials, we can specify the energy stored in a bulk strain density
the following constants
\begin{equation}\label{eq1}
I_1(\bm{\varepsilon})=\text{tr}(\bm{\varepsilon}), \quad \quad I_2(\bm{\varepsilon})=\text{tr}(\bm{\varepsilon}^2).
\end{equation}
Let the solid material is strengthened by two groups of fibers denoted as a orthotropic solid materials. Thus, an anisotropic material is reinforced by two fibers
namely $\bm{a}$ and $\Bg$ with $\| \bm{a} \| = 1$ and $\| \Bg \| = 1$; see Figure \ref{Figure1}. These materials have the highest strength in the fiber direction (i.e.,  either in $\bm{a}$ or $\bm{g}$). Therefore, at the material point $\bm {x}$, the stress state relates to the deformation in addition to the given direction which leads to a \textit{deformation-direction-dependent} framework.
For this, we impose a penalty-like parameter $\chi_a>0$ and $\chi_g>0$ corresponding to  $\bm{a}$ or $\bm{g}$ which restrict a deformation on the normal plane to $\Ba$ or $\Bg$. Therefore, we can define three specific second-order tensorial quantities, i.e., the strain $\bm{\varepsilon}$ in addition to $\BM$ and $\Bg$ tensors to specify the effective bulk free energy
\begin{equation}\label{eq2}
\Bve = \nabla^{sym} \Bu = \text{sym}[ \nabla \Bu ]
,\quad
\BM:=\Ba \otimes \Ba,
\AND
\BG:=\Bg \otimes \Bg.
\end{equation}
We employ two deformation-direction-dependent constants to clarify them as
they can be represented by additional two 
\begin{equation}\label{eq3}
I_4=tr(\bm{\varepsilon\cdot M}), \quad 
I_6=tr(\bm{\varepsilon \cdot G}).
\end{equation}
Considering the symmetric strain tensor $\bm{\varepsilon}$ and the structural tensors $\BM$ and $\BG$, we have the scalar-valued function ${\Psi}(\bm \varepsilon;\BM,\BG)$. Therefore, the scalar-valued effective strain density function denotes an invariant in spatial and temporal directions between two sets of points in a specific domain under rotation. As a result, ${\Psi}(\bm \varepsilon;\BM,\BG)$ is expressed by the seven principal invariants as
\begin{equation}\label{eq5}
\begin{aligned}
{\Psi}(\bm\varepsilon;\BM,\BG)={\Psi}\big(I_1,I_2,I_4,I_6\big)=\widetilde{\Psi}^{iso}\big(I_1,I_2\big)+\widetilde{\Psi}^{aniso}\big(I_4,I_6\big).
\end{aligned}
\end{equation}
In this case, the isotropic  free-energy function relates to
\begin{equation}\label{eq6}
\widetilde{\Psi}^{iso}\big(I_1,I_2\big)
:=(\frac{K}{2})I^2_1
-\mu~\Big(\frac{I^2_1}{\delta}- I_2\Big)
\quad\text{with}\quad K>0\quad\text{and}\quad\mu>0,
\end{equation}
where $K= \lambda+\frac{2}{3}\mu>0$ is the bulk modulus and including shear modulus $\mu$ with dimension  in the spacial direction. Note, in our formulation, instead of using elastic Lam\'e's first constant denoted by $\lambda >-\frac{2}{\mu}$ which has a lower bound, we used a shear bulk modulus as a positive quantity to avoid unnecessary condition. This has an advantage for our next goal; that is Bayesian estimation for the material parameters.

The anisotropic free-energy function can be specified by 
the anisotropic free-energy function for orthotropic materials reads 
\begin{equation}\label{eq7}
\widetilde{\Psi}^{aniso}\big(I_4,I_5,I_6,I_7\big):=\frac{1}{2}\chi_a I^2_4+\frac{1}{2}\chi_gI^2_6.
\end{equation}
Again, $\chi_a$ and $\chi_g$ point out the anisotropic penalty-like material parameters. 

To construct the mechanical BVP, let the geometry to be enforced by prescribed deformations $\overline{\Bu}$ and additionally the traction vector $\overline{\bm{\tau}}$ at the surface $\partial\calB$ of the reference configuration, which are denoted by the time-dependent Dirichlet- and Neumann type boundary conditions as
\begin{equation}
\Bu = \bar\Bu(\Bx,t) \ \textrm{on}\ \partial_D\calB
\AND
\Bsigma\Bn = \bar\Btau(\Bx,t)\ \textrm{on}\ \partial_N\calB.
\label{mech-bcs}
\end{equation}
Here, $\Bn$ points out the unit normal vector in the reference setting such that the Cauchy stress tensor $\Bsigma(\Bve, p, d)$ denotes the thermodynamic dual to $\Bve$. The global mechanical form of the equilibrium equation for the solid body can be represented through {\it first-order} PDE for the multi-field system as
\begin{equation}
\fterm{
	\Div\,\Bsigma(\Bu, p, d) + \overline\Bb = \Bzero \, ,
	\label{equil:defo}
}
\end{equation}
such that dynamic motion is neglected (i.e., quasi-static response), and we denote $\overline\Bb$ as a prescribed body force.
\subsubsection{Fluid contribution}\label{Section212}
To formulate the constitutive equation for the poromechanics, let us move forward with a {\it biphasic} fully saturated porous material, which includes of pore fluid and a solid matrix within the bulk material. A local volume element denoted by $dV$ in the reference configuration is additively decomposed into a fluid portion $dV_{F}$ in addition to a solid portion $dV_{S}$. Thus, the volume fraction is introduced via $n_c:=dV_{c}/dV$, where $c=\{S,F\}$. Concerning the fully saturated porous medium the \textit{saturation condition} reads
\begin{equation}
\sum_{c}n_{c} = n_{F}+n_{S} = 1, 
\end{equation}
where $n_{F}(\Bx,t)$ indicates the porosity, which point out the volume occupied by the fluid is {\it same as} the pore volume. In the {\it fracture zone} we have
\begin{equation}
d = 0 \quad \mbox{leading to} \quad n_S =0\quad \mbox{also} \quad n_F = 1.
\end{equation}
The volume fraction in the porous medium, i.e., $n_{c}$, depends on the physical density (i.e., material, effective, intrinsic) $\rho_{c R}$ to the partial density $\rho_{c}$ through
\begin{equation}
\rho_{c}=n_{c}\;\rho_{c R} \quad \text{with} \quad \rho_{c R}:=dm_{c}/dV_{c} \quad \text{and} \quad \rho_{c}:=dm_{c}/dV,
\end{equation}
where $dm_{c}$ denotes the mass of the phase $c$. 
Denoting the initial porosity $n_{F,0}$, for a constant fluid material density, the porosity (i.e., fluid volume fraction) $n_F$ is related to the fluid volume ratio $\theta$  (fluid content) per unit volume of the reference configuration $\calB$ via
\begin{equation}
n_F = n_{F,0} + \theta,
\label{porosity-ratio}
\end{equation}
where $\theta$ prescribes the {\it first local internal} variable (history field); see \cite{Bio72,coussy95,Markert2007}. Also, the evolution equation for the fluid volume ratio $\dot{\theta} = \dot{n}_F$ can be obtained by means of the fluid pressure field $p$. Prescribed Dirichlet boundary condition  and Neumann boundary condition for the pressure can be described by
\begin{equation}
p = \bar{p}(\Bx,t) \ \textrm{on}\ \partial_D\calB~~~\text{and}~~~
\BcalF\cdot\Bn = \bar{f}(\Bx,t)\ \textrm{on}\ \partial_N\calB,
\label{flu-bcs}
\end{equation}
through the fluid volume flux vector $\BcalF$, the imposed fluid pressure $\bar{p}$ on the boundary surface, fluid transport $\bar{f}$ on the Neumann boundary surface. 
Because the fluid-filled equation denotes a time-dependent problem, the initial condition needs to be set for the fluid volume rate and hence by $\theta(\Bx,t_0)=0$ yielding $n_F = n_{F,0}$ in $\calB$. Moreover, the fluid flux vector in \req{flu-bcs} can be described through the negative direction of the material gradient of the fluid pressure $\nabla p$ through the permeability, based on Darcy-type fluid's:
\begin{equation}
\BcalF := - \BK(\Bve,d)\; \nabla p.
\end{equation}
Here, the second-order permeability tensor is given by anisotropic second-order tensor $\BK$ that described based on the strain tensor $\Bve$ as well as the crack phase-field $d$. To denote the effect of the fracture on the fluid contribution, we decompose the permeability tensor  into a {\it Darcy-type flow for the unfractured porous medium} $\boldsymbol{K}_{Darcy}$ in addition to a {\it Poiseuille-type flow in a completely fractured material} $\boldsymbol{K}_{crack}$ which is explained as follows
\begin{equation}\label{permability}
\begin{array}{ll}
\boldsymbol{K}(\Bve,d) &=
{\boldsymbol{K}_{Darcy}}(\Bve)+(1-d)^{\zeta}
{\boldsymbol{K}_{frac}}(\Bve,d), \\ [3mm]
{\boldsymbol{K}_{Darcy}}(\Bve) &= K_D \BI, \\ [3mm]
{\boldsymbol{K}_{frac}}(\Bve,d) &= \big(\frac{\omega_d^2}{{12\eta_F}}-{{K}_D}\big)\; \; \big[
\BI - \Bn \otimes  \Bn 
\big],
\end{array}
\end{equation}
with $\omega_d$ the so-called crack aperture (or the crack opening displacement) \cite{nguyen2016phase} defined as
\begin{equation}\label{COD}
\omega_d :=\llbracket\Bu(\Bx) \rrbracket.\Bn =(\Bn\cdot\Bve\Bn) h_{e},
\end{equation}
denoting the outward unit normal to the fracture surface $\Bn=\nabla d/|\nabla d|$ , ${K}_D$ in $\boldsymbol{K}_{Darcy}$ and  $\boldsymbol{K}_{frac}$ indicates the isotropic intrinsic permeability of the pore space, $\eta_F$ represents the dynamic fluid viscosity, and $\zeta \ge 1$ denotes a
permeability transition exponent. The characteristic element length $h_{e}$ in \req{COD} typically set as a minimum discretized element size, i.e., diameter of an element in the fractured region; see \cite{MieheMauthe2015}. Notably, the second-order permeability tensor in \req{permability} in the intact region,  i.e. $d=1$, recover $\BK(\Bve, d)\approx{\BK_{Darcy}}(\Bve)$. Following \cite{MieheMauthe2015}, 
The conservation of the fluid mass which reflects the {\it second} PDE within hydraulic fracturing setting reads
\begin{equation}
\dot{n}_F - \bar{r}_F + \Div[\BcalF] = 0 \, ,
\label{2nd-be}
\end{equation}
by a given/imposed fluid source $\bar{r}_F$ per unit volume of the initial setting $\calB$ describing the fluid injection process in the hydraulic fracturing.
\subsubsection{Phase-field contribution}\label{Section2110}
Within regularized fracture framework, a sharp-crack surface topology denoted by $\calC$ to guarantee the continuity of the fracture field is further specified by the smeared fracture surface functional shown by $\calC_l$ thus $\calC \rightarrow \calC_l$. Hence we have
{\begin{equation}
	\calC_l(d) = \int_{\calB} \gamma_l(d, \nabla d;\BM,\BG) \, dV
	\WITH \gamma_l(d, \nabla d;\BM,\BG):=\gamma^{iso}_l(d, \nabla d)+\gamma^{aniso}_l(\nabla d;\BM,\BG),
	\label{s2-gamma_l}
	\end{equation}}
where the isotropic part is 
{\begin{equation}\label{eq15}
	\gamma^{iso}_l(d, \nabla d) :=  
	\frac{1}{2l} {(1-d)^2} + \frac{l}{2}  \nabla d\cdot\nabla d,
	\end{equation}}
and  anisotropic part of phase-field density function reads
\begin{equation}\label{eq151}
\gamma^{aniso}_{l}(\nabla d;\BM,\BG):=  \beta_a\frac{l}{2}  \nabla d\cdot\BM\cdot\nabla d+\beta_g\frac{l}{2}  \nabla d\cdot\BG\cdot\nabla d,
\end{equation}
considering the anisotropic penalty-like material parameters $\beta_a$ and $\beta_g$. 
Here, $\gamma_l$ denotes the regularized crack surface density function per unit volume of the solid, the regularization item $l$ indicates the length scale (also named regularization parameter), which captures the fracture diffusivity.
%
%
%
%
Therefore, we can derive $d$ (the crack phase-field) by minimizing  diffusive crack surface $\calC_l(d)$, as follows
{\begin{equation}
	d = \mbox{argmin} \{ \inf_{d}
	\calC_l(d) \}\quad \text{where}\quad
	d=0 \; \mbox{on} \; \calC\subset \calB\;.
	\label{min-d-geo}
	\end{equation}}
The outcome Euler-Lagrange differential system is
\begin{equation}
d-1 - l^2 \Delta d- \beta_a l^2 \div(\nabla d\cdot \BM)- \beta_g l^2 \div(\nabla d\cdot \BG) = 0\quad in~ \calB,
\end{equation}
augmented by the homogeneous NBC that is $\nabla d \cdot \Bn = 0$ on $\partial\calB$. %
We then consider the smeared crack phase-field functional given in \req{s2-gamma_l} to ensure the fracture Kuhn-Tucker conditions \cite{NoiiWick2019,NoiiAldakheelWickWriggers2019}. To that end, the constitutive
functions response by means of a global evolution system of regularized crack fracture gives rise to the global crack dissipation functional
{\begin{equation}
	\fterm{
		\vphantom{\int_{\calB}}
		\frac{d}{dt} \calC_l(d)  := 
		\frac{1}{l} \int_{\calB} [\; -g^{\prime}(d_+) \calH + \eta  \dot{d}\; ]\;
		\dot{d} \, dV=\int_{\calB}\;\delta\gamma_l(d,\nabla d;\BM,\BG)\dot{d} \, dV \ge 0\;,}
	\label{gamma-evol}
	\end{equation}}
{results to the two inequity conditions for the crack phase-field through
	\begin{equation}\label{eq29_H5-2}
	\delta\gamma_l(d,\nabla d;\BM,\BG)\le0\quad\text{with}\quad\dot{d}\le0,
	\end{equation}}
{with the functional derivative of $\gamma_l$ with respect to $d$ by,
	\begin{equation}\label{eq29_H5-1}
	\int_\calB \delta_{d} \gamma_{l}(d,\nabla d;\BM,\BG) \mathrm{d}{V}=\int_\calB \frac{1}{l}[(d-1)-l^2 \Delta d-\beta_a l^2 \div(\nabla d\cdot\BM)-\beta_g l^2 \div(\nabla d\cdot\BG)] \mathrm{d}{V}.
	\end{equation}}
Additionally, in \req{gamma-evol},  $\calH$ indicates the crack driving force and represent
\begin{equation}
\calH = \max_{s\in [0,t]} D(\Bx,s) \ge 0,
\label{driving-force}
\end{equation}
where $D$ indicates the fracture driving state function. Furthermore, $\calH$ considers the irreversibility of the crack phase-field evolution by filtering out a maximum value of $D$. This is referred to the {\it local history variable}.
Also, an artificial/numerical material parameter (denoted by $\eta  \ge 0$) is employed to specify
\textcolor{black}{the viscosity term of crack growth}. 

The local evolution of the crack phase-field equation in the given domain $\calB$ resulting from 
\req{gamma-evol} augmented with its homogeneous NBC, i.e. $\nabla d \cdot \Bn = 0$ on $\partial\calB$ yields
{\begin{equation}
	\fterm{
		[ \, d-1 - l^2 \Delta d- \beta_a l^2 \div(\nabla d\cdot\BM)- \beta_g l^2 \div(\nabla d\cdot\BG) \, ] - \eta \dot{d} + 2(1-\kappa)d {\calH} = 0 \, ,
	}
	\label{euler-eq-d}
	\end{equation}}
which states the {\it third} equation in the coupled system.

\subsection{Constitutive functions}\label{Section22}

The coupled BVP is formulated through three specific fields (i.e., unknown solution fields) to illustrate the hydro-poro-elasticity of fluid-saturated porous media in the fracturing material by
\begin{equation}
\mbox{Global Primary Fields}: \BfrakU := \{ \Bu, p, d \}.
\label{global-fields}
\end{equation}
Here, $\Bu$ is the displacement (mechanical deformation), $p$ denotes the pressure, and $d$ is the crack phase-field ($0\le d\le1$). For the numerical implementation standpoint, to guarantee $0\le d\le1$ holds, we project $d>1$ to 1 and $d<0$ to 0 to avoid unphysical crack phase-field solution \cite{NoiiWick2019}. The constitutive formulations for the hydraulic phase-field fracture are written in terms of the following set
\begin{equation}
\mbox{Constitutive State Variables}: 
\BfrakC := \{ \Bve, \theta, d, \nabla d \}
\ ,
\label{state}
\end{equation}
which shows the response of the poroelasticity material modeling with a first-order gradient damage model. A pseudo-energy density function denoted by ${W}(\BfrakC)$ for the poroelastic media per unit volume reads
\begin{equation}
{W}(\BfrakC) = 
{W}_{elas}(\Bve, d;\BM,\BG) + {W}_{fluid}(\Bve,\theta) + 
{W}_{frac}(d, \nabla d;\BM,\BG).
\label{pseudo-energy}
\end{equation}
%

\subsubsection{Fluid contribution}
Following \cite{miehe2016phase}, the fluid density function takes the following form
\begin{equation}
{W}_{fluid}(\Bve, \theta) = \frac{M}{2} \Bigg[ B^2tr^2[\Bve] - 2\, \theta\,tr[\Bve] +  \theta^2
\Bigg]=\frac{M}{2}\big(B\;tr[\Bve]-\theta\big)^2,
\label{fluid-part}
\end{equation}
based on the given fluid coefficient including which includes Biot's coefficient $B$ and Biot's modulus $M$. By employing the Coleman-Noll inequality condition in thermodynamics, the fluid pressure $p$ is derived from the first-order derivative of the pseudo-energy density function ${W}$ given in \req{pseudo-energy} by
\begin{equation}\label{p-piola-stresses}
p(\Bve, \theta) := \frac{\partial{W}}{\partial \theta } =\frac{\partial{{W}_{fluid}}}{\partial \theta }= \theta M  - MB  tr(\Bve),
\end{equation}
for the isotropic solid material. Employing the above-mentioned pressure in addition to the second equation in \req{2nd-be} as well as \req{porosity-ratio}, the conservation of mass takes the following form
\begin{equation}
\fterm{
	\frac{\dot{p}}{M} + B~ \partial_t{tr(\Bve)} - \bar{r}_F+ \Div[\BcalF] = 0,
}
\label{pres-pde}
\end{equation}
which now depends on the fluid pressure $p$ and not fluid volume fraction (porosity). 
\subsubsection{Mechanical contribution}
Here, modified elastic density function ${W}_{elas}$ is degraded elastic response resulting from the fractured state, a fluid density function contribution ${W}_{fluid}$, and fracture density function denoted by ${W}_{frac}$ which contain the accumulated dissipative energy are accordingly used. For a compressible isotropic elastic solid,
the elastic density function is formulated through a linear elasticity strain energy function as
\begin{equation}\label{elast_energy}
{W}_{elas}(\Bve, d;\BM,\BG)= g(d)\; {\Psi}(\Bve;\BM,\BG),
\end{equation} 
whereas $\Psi$ given in \req{eq5}. {Here, the standard monotonically decreasing quadrature degradation function, reads as $g(d):=(1-\kappa)d^2 + \kappa$.
	\subsubsection{Strain-energy decomposition for the bulk free energy}
	Since the fracturing materials behave significantly different in \textit{tension}
	and {\textit{compression}}, a consistent additive split for the strain energy
	density function given in \req{elast_energy}
	for the isotropic and anisotropic counterpart of energy are accordingly described. Thus, compared to other studies \cite{gultekin2018numerical, teichtmeister2017phase, NoiiAldakheelWickWriggers2019}, we derived a new crack driving state function, which mainly includes the tensile part of anisotropic energy density function.
	\begin{itemize}	
		\item \textbf{Strain-energy decomposition for the isotropic term}.	
	\end{itemize}
	To derive an additive decomposition of the isotropic strain energy function, i.e. $ {\Psi}^{iso}\big(I_1(\bm{\varepsilon}),$ $I_2(\bm{\varepsilon})\big)$, we carry out an additive split of the strain tensor $\bm\varepsilon(\bm u)$ through 
	\[
	\bm\varepsilon(\bm u)=\bm\varepsilon^{+}(\bm u)+\bm\varepsilon^{-}(\bm
	u)
	\quad\text{where}\quad
	\bm\varepsilon^{\pm}(\bm
	u):=\sum_{i=1}^{\delta} \langle\varepsilon_i\rangle^{\pm} {\textbf{N}_i} \otimes {\textbf{N}_i},	\]
	in the term of the tension strain $\bm\varepsilon^{+}$ 
	and compression strain $\bm\varepsilon^{-}$. Also, $\langle x \rangle_{\pm} := \frac{ x {\pm} |x|}{2}$ indicates a ramp function of $\R_{\pm}$ explained by the Macauley bracket, {$\{\varepsilon_i\}$ point out the principal strains, and $\{\textbf{N}_i\}$ denote the principal strain directions.} The tension/compression fourth-order projection tensor can be expressed by
	\begin{equation}\label{eq16}
	\mathbb{P}^\pm_{\bm {\varepsilon}}:=\frac{\partial \bm {\varepsilon}^\pm}{\partial \bm {\varepsilon}}=\frac{\partial \big(\sum_{i=1}^{\delta} \langle\varepsilon_i\rangle^{\pm}  {\textbf{N}_i} \otimes {\textbf{N}_i}\big)}{\partial \bm {\varepsilon}},
	\end{equation}
	here $\mathbb{P}^\pm_{\bm {\varepsilon}}$ projects the total strain to the positive and negative features, i.e., $\bm {\varepsilon}^{\pm}=\mathbb{P}^\pm_{\bm {\varepsilon}}:\bm {\varepsilon}$. Therefore, a decoupled explanation of the isotropic strain-energy function into a named tension and compression contribution reads
	\begin{equation}\label{eq17aa}
	{\Psi}^{iso}\big(I_1,I_2\big):=
	\underbrace{\widetilde{\Psi}^{iso,+}\big(I^{+}_1,I^{+}_2\big)}_{\text{tension term}}+\underbrace{\widetilde{\Psi}^{iso,-}\big(I^{-}_1,I^{-}_2\big)}_{\text{compression term}},
	\end{equation}
	with the positive and negative principal invariants take
	\begin{equation}\label{eq18}
	I_1^{\pm}:=\langle{I_1(\bm{\varepsilon})}\rangle_{\pm}, \quad I^{\pm}_2(\bm{\varepsilon}):=I_2(\bm{\varepsilon}^{\pm}).
	\end{equation}
	
	\begin{itemize}	
		\item \textbf{Strain-energy decomposition for the anisotropic term}.	
	\end{itemize}
Now, a decoupled explanation of the anisotropic strain-energy function of a namely tension and compression contribution is introduced. Here, we mainly aim to derive the new crack driving state function, which mainly includes the tensile part of the anisotropic energy density function. Thus, the anisotropic strain-energy function can be additively decomposed as
	\begin{equation}\label{eq17}
	{\Psi}^{aniso}\big(I_4,I_6\big):=\underbrace{\widetilde{\Psi}^{aniso,+}\big(I^{+}_4,I^{+}_6\big)}_{\text{tension term}}+\underbrace{\widetilde{\Psi}^{aniso,-}\big(I^{-}_4,I^{-}_6\big)}_{\text{compression term}},
	\end{equation}
	where, the positive and negative principal invariants are
	\begin{equation}
	I_4^{\pm}:=\langle{I_4(\bm{\varepsilon}{;\BM})}\rangle_{\pm}, \quad I_6^{\pm}:=\langle{I_6(\bm{\varepsilon}{;\BG})}\rangle_{\pm}~.
	\end{equation}
	Now, using \req{eq17aa} and \req{eq17}, the bulk work density function for the orthotropic materials with two families of fibers used in \req{elast_energy} modified through 
	\begin{equation}\label{elastc_enery_modified}
	\begin{aligned}
	{W}_{elas}(\Bve, d;\BM,\BG)=g(d_+)\Big[\widetilde{\Psi}^{iso,+}(I^{+}_1,I^{+}_2)
	&+ \widetilde{\Psi}^{aniso,+} (I^{+}_4,I^{+}_6) \Big]\\
	&+\widetilde{\Psi}^{iso,-}(I^{-}_1,I^{-}_2)+\widetilde{\Psi}^{aniso,-} (I^{-}_4,I^{-}_6).
	\end{aligned}
	\end{equation}
	The constitutive stresses corresponding to \req{elastc_enery_modified} read:
	\begin{equation}
	\begin{aligned}
	{\bm \sigma}(\bm{\varepsilon},p,d ;\BM,{\BG}):=\frac{\partial{W}}{\partial\Bve } = \bm {\sigma}_{eff}-Bp\BI.
	\label{eq21}
	\end{aligned}
	\end{equation}
	Here, the second-order Cauchy stress tensor $\Bsigma$ is further decomposed in an additive manner into the effective stress tensor $\BP_{eff}$ and additionally a pressure part. This additive decomposition is written based on the classical Terzaghi split, as outlined in \cite{terzaghi1943theoretical,de1990development}
	\begin{equation}
	\begin{aligned}
	&\bm {\sigma}_{eff}=\bm {\sigma}^{iso} + \bm {\sigma}^{aniso},
	\\[0.5mm]
	&\bm {\sigma}^{iso}
	=g(d_+){\bm {\widetilde{\sigma}}^{iso,+}}+{\bm {\widetilde{\sigma}}^{iso,-}},
	\\[0.5mm]
	&\bm {\sigma}^{aniso}
	= g(d_+){\bm {\widetilde{\sigma}}^{aniso,+}}+{\bm {\widetilde{\sigma}}^{aniso,-}},
	\label{eq211}
	\end{aligned}
	\end{equation}
	where
	\begin{equation}
	\begin{aligned}
	&\bm {\widetilde{\sigma}}^{iso,\pm}:=
	K~I_1^{\pm}(\bm{\varepsilon}) -\mu~\Big(\frac{2}{\delta}I_1^{\pm}(\bm{\varepsilon})  {\boldsymbol{I}} 
	- 2 \bm\varepsilon_\pm\Big)
	\quad\text{with}\quad K>0\quad\text{and}\quad\mu>0
	\\[0.5mm]
	&{\bm {\widetilde{\sigma}}^{aniso,\pm}}
	:=\frac{\partial \widetilde{\Psi}^{aniso,\pm}}{\partial \bm \varepsilon}=
	\chi_a I_4^{\pm} \BM +\chi_g I_6^{\pm} {\BG}.
	\label{eq24}
	\end{aligned}
	\end{equation}
	Note, the identities $\partial_\Bve I_4=\BM$ and $\partial_\Bve I_6=\BG$ are used. 
	\subsubsection{Fracture contribution}
	The fracture contribution of pseudo-energy density given in \req{pseudo-energy} takes the following explicit form
	\begin{equation}
	{W}_{frac}(d, \nabla d;\BM,\BG) =  G_c{\gamma}_l(d, \nabla d;\BM,\BG),
	\label{frac-part}
	\end{equation}
	where ${G}_c > 0$ is so-called a Griffith's energy release rate where ${\gamma}_l$ is given in \req{s2-gamma_l}. Following \cite{miehe2016phase,NoiiAldakheelWickWriggers2019}, by taking the first variational derivative $\delta_d W$ of \req{pseudo-energy}, the positive crack driving state function $D$ reads
	\begin{equation}
	D:= \frac{l}{G_c}\big[\widetilde{\Psi}^{iso,+}+
	\widetilde{\Psi}^{aniso,+}
	\big]{\ge0}.
	\label{h-history-field}
	\end{equation}
	\subsection{Variational formulations derived for the coupled multi-field problem}\label{Section23}
	The primary fields $\BfrakU$ given in \req{global-fields} for the coupled poroelastic media of the fracturing material is obtained by equations in \req{equil:defo}, \req{pres-pde} as well as \req{euler-eq-d} in a strong form framework. Here, the PDE models are governed in a temporal domain $[t_n, t_{n+1}]$ such that time step $\Delta t=t_{n+1}-t_{n} >0$ holds. Next, three test functions with respect to the deformation $\delta \Bu(\Bx) \in  {\bm{V}_{\Bu}}$, fluid pressure $\delta p(\Bx) \in{{V}_{p}}$ and crack phase-field $\delta d(\Bx)\in W_{in}$ are defined, see \req{space1}. The variational formulations with respect to the three PDEs for the coupled poroelastic media of the fracturing material are derived by
	\begin{equation}
	\begin{array}{ll}
	\mathcal{E}_\Bve(\BfrakU, \delta \Bu) &= \displaystyle\int_\calB \Big[ \Bsigma:\delta \Bve - \bar{\Bb} \cdot \delta \Bu \Big] dV - \int_{\Gamma_N} \bar{\Btau} \cdot \delta \Bu  \; dA
	= 0, \\ [4mm]
	\mathcal{E}_p(\BfrakU, \delta p) &= \displaystyle\int_\calB \Big[\Big(\frac{1}{M}(p-p_n) + B \big(tr(\Bve)-tr(\Bve_n)\big) -\Delta t \;\bar{r}_F
	\Big)\delta p + (\Delta t \;\BK \;\nabla p) \cdot \nabla \delta p \Big] dV \\ [3mm]
	& + \displaystyle\int_{\partial_N\calB} \bar{f} \;\delta p\; dA = 0, \\ [4mm]
	\mathcal{E}_d(\BfrakU, \delta d) &= (1-\kappa)\Delta t\displaystyle\int_{\calB}\Big[ 2d \calH\cdot \delta dV\Big]+
	\displaystyle\int_{\calB}\Big[   \Delta t(d-1)\cdot\delta d-\eta(d-d_n)\cdot\delta d\Big]\;dV\\ [4mm]
	&+\displaystyle\int_{\calB}\Big[  l^2 \Delta t\nabla d\big(1+\beta_a\cdot\BM+\beta_g\cdot\BG\big)\cdot\nabla(\delta d)\Big]\,
	dV = 0.\\
	\end{array}
	\label{weakForm}
	\end{equation}
	Here, the Cauchy stress tensor $\Bsigma$,  the second-order permeability tensor $\BK$ and the crack driving force $\calH$ are given in given \req{eq21}, \req{permability} and \req{driving-force}, respectively. The fully coupled variational multi-field problem to describe hydraulic fractures in porous media is formulated in \req {weakForm}. Following \req {weakForm}, the compact variational form for the hydraulic phase-field brittle fractures in porous media reads
	\begin{equation}\label{compat_argmin}
	\fterm{ 
		\mathcal{E}_{\BfrakU}(\BfrakU, \delta \BfrakU)
		=\mathcal{E}_\Bve(\BfrakU, \delta \Bu)+\mathcal{E}_p(\BfrakU, \delta p)+\mathcal{E}_d(\BfrakU, \delta d)=0 \quad\forall\;\; (\delta \Bu,\delta p,\delta d)\in ({\bm{V}_{\Bu}},{{V}_{p}},{{V}_{d}}).
	}
	\end{equation}
	
	In order to solve the phase-field hydraulic fracture system \eqref{compat_argmin}, we first solve the first two equations monolithically (simultaneously obtain $(\bm u, p)$). Then, a staggered approach is used to obtain the phase-field fracture $d$. To that end, we fix alternately  $(\bm u, p)$ and estimate $d$ and vice versa. The procedure is continued until its convergence (using given $\texttt{TOL}_\mathrm{Stag}$). We provide a summary of the algorithm steps in Algorithm 1. Accordingly, a detailed consistent linearization formulation for \eqref{compat_argmin} which is frequently used in the Newton-Raphson iterative solver including a finite element discretization is illustrated in Appendix A.

	
	%
	\begin{algorithm}
		{\bf Input:} loading data $(\bar{f},\bar{\bm t}_{n})$ on $\mathcal{C}$ and $\Gamma_{N}$, respectively; \\[2mm] 
		\hspace{1.4cm}solution $(\bm u_{n-1},p_{n-1},d_{n-1})$ from step $n-1$. \\[2mm]
		\quad\quad Initialization, $k=1$:\\ 
		
		\quad\quad \textbullet\; set $(\bm u^0,p^0,d^0):=(\bm u_{n-1},p_{n-1},d_{n-1})$.\\ 
		
		Staggered iteration between $(\bm u,p)$ and $d$:\\
		
		\quad\quad \textbullet~ solve following system of equations (in \eqref{weakForm}) in a monolithic manner given $d^{k-1}$, 
		\begin{align*}
		\begin{cases}
		\mathcal{E}_{\Bve}(\bm u,p,d^{k-1};\delta \Bu)=0,\\\\
		\mathcal{E}_{p}(\bm u,p,d^{k-1};\delta p)=0,
		\end{cases} 
		\end{align*}
		
		\quad\quad \quad for $(\bm u,p)$, set $(\bm u,p)=:(\bm u^k,p^k)$,\\
		
		\quad\quad \textbullet\; given $(\bm u^k, p^k)$, solve $\mathcal{E}_d(\bm u^k, p^k,d;\delta d)=0$ for $d$, set $d=:d^k$,\\

		\quad\quad \textbullet\; for the obtained pair $(\bm u^k,p^k,d^k)$, check {staggered residual} by 
		\begin{align*}
		\quad \quad \;\; \mathrm{Res}_\mathrm{Stag}^k:=|\mathcal{E}_{\Bve}(\bm u^k,p^k,d^k;\delta \Bu)|
		+|\mathcal{E}_{p}(\bm u^k,p^k,d^k;\delta p)|
		\leq\texttt{TOL}_\mathrm{Stag}, \; \forall \; (\delta \Bu,\delta p)\in({\bf V}_\Bu,V_p),
		\end{align*}
		\quad\quad \textbullet\; if fulfilled, set $(\bm u^k,p^k,d^k)=:(\bm u_n,p_n,d_n)$ then stop; \\
		
		\quad\quad\quad {\color{white}\textbullet}\; else $k+1\rightarrow k$. \\[2mm]
		{\bf Output:} solution $(\bm u_n,p_n,d_n)$ at $n^{\text{th}}$ time-step. \\[2mm]
		\caption{\em The staggered iterative solution process for \eqref{weakForm} at a fixed time-step $n$.}
		\label{alg_upd}
	\end{algorithm}
	
	\begin{figure}[!ht]
		\centering
		{\includegraphics[clip,trim=5.5cm 2.5cm 4cm 7cm, width=18cm]{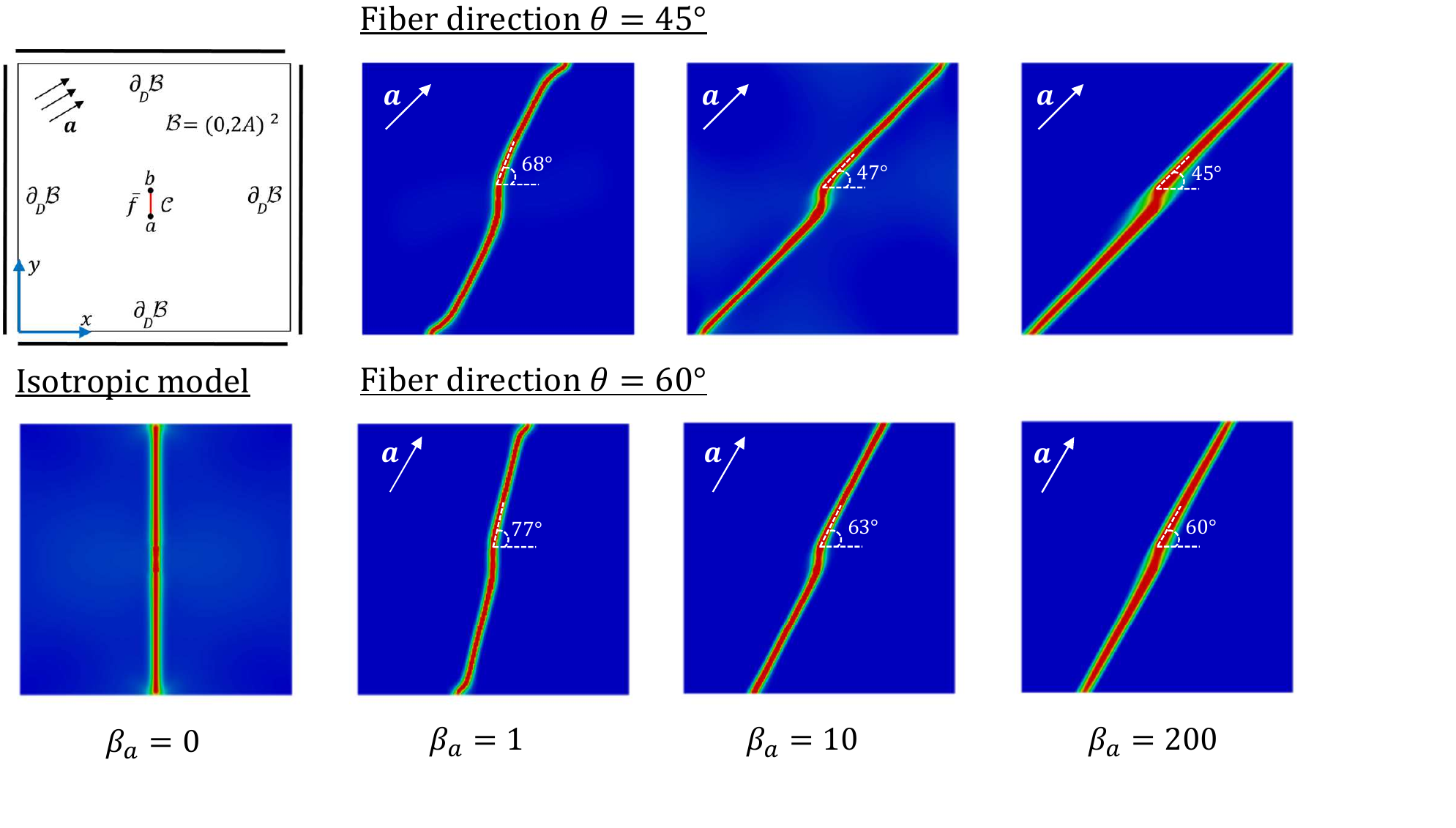}}  
		\caption{Influence of the different values of penalty-like parameters (here $\beta_a$ while $\beta_g=0$) in the hydraulically induced crack driven by fluid volume injection. Here, the computation domain is shown in the first plot and two different fiber direction are studied.
		}
		\label{example__modelP}
	\end{figure}
	
	\subsection{Numerical illustration}\label{Section26} 
	
	We now investigate the influence of the penalty-like parameters for the computed solution from \req{weakForm} to the crack phase-field solution state. Here, we consider only transversely isotropic poroelastic material responses induced by the fluid volume injection. We vary the parameters for the anisotropic modeling as $\beta_a=1, 10, 200$ to represent the transverse isotropy characterized by the normal vector defined as $\Ba:=\big(cos(\phi), sin(\phi)\big)^{T}$. Also, by letting $\beta_g=\chi_g=0$ we recover the isotropic hydraulic fracture simulation. Furthermore, we fixed and set $\chi_a=10$, because of its  negligible effect on the fracture model. Consider the boundary value problem given in Figure \req{example__modelP} (depicted in the first row from left). The material properties same as \cite{xia2017phase} and shown in Table \ref{material-parameters}. 
	
	To model transversely isotropic poroelastic material, structural stiffness is continuously augmented with a unidirectional fiber. We consider two cases. First, the fiber is inclined under an angle $\phi=+45^\circ$ while in the second one fiber is inclined under $\phi=+60^\circ$ with respect to the $x$-axis of a fixed Cartesian coordinate system. The two cases are shown in Figure \ref{example__modelP} in the first row and second row, respectively. 
	
	For both cases, the first important observation is that, increasing the penalty-like parameter $\beta_a$ in \req{weakForm} results in the fracture direction to be aligned with the highest strength direction of the poroelastic material which is $\Ba$. Another impacting factor that should be noted the less value for $\beta_a$ behave as an isotropic model (inclined vertically) that is shown in Figure \req{example__modelP}, depicted in the second row. Note, there is very slight changes in the crack profile between $\beta_a=10$ and $\beta_a=200$.
	
	Next,  we  use  the anisotropic hydraulic fracture model used in Algorithm 1 and validated in this section as a departure point for the Bayesian inversion framework for the phase-field hydraulic fracture. That is described in a detail in Section \ref{Section3}.
\section{Bayesian inversion for anisotropic phase-field hydraulic fracture  }\label{Section3}

Uncertainties in the description of reservoir lithofacies, porosity, and permeability are major contributors to the uncertainties in reservoir performance forecasting. Moreover, the uncertainties in the characterization of
formation properties, fracture properties, temperature effects, identification of elastic parameters, and flow mechanisms affect the productivity of wells. In this section, we introduce a computational technique based on MCMC to model the uncertainty in hydraulic fractures.

There is usually a lack of information about problem parameters and they undergo
many uncertainties coming e.g. from the heterogeneity of rock formations and complicated realization of experiments for
parameter identification.  The Bayesian approach provides a principal framework for combining the
prior knowledge with dynamic data in order to make predictions on quantities of interest. In most cases, direct measurements of the quantities is not feasible; therefore, using the forward model, an inverse approach enables us to predict these parameters with a low computational cost. As a result in addition to provide a comprehensive model that describes the fracture (crack propagation), the influential parameters are identified.

Here, we introduce a parameter estimation framework to determine different effective hydraulic fracture phase-field parameters. We consider 
the unknown values, i.e., 
elasticity modulus of the rock formation, Biot's coefficients, fluid velocity, and the energy release rate will be
components of a random vector.
We first employ the statistical model
\begin{align}
	\mathcal{R}=\mathcal{M}(\Theta)+\epsilon.
\end{align} 
\begin{itemize}
	\item$\mathcal{R}$ is a random variable that indicates the reference value (i.e., reference observation or measured data).
		\item $\mathcal{M}$ points out a PDE-based model (here the hydraulic phase-field model i.e., \eqref{compat_argmin}). The variable $\Theta$ is a random field and denotes the realizations of random variables (here the unknown material parameters, denoting  $\Theta\in\{\lambda,\mu,B,M,\eta_F,G_c,\beta_a,\beta_g\}$, see Subsection \ref{parameter} for the physical interpretation). In total, $\mathcal{M}(\Theta)$ relates to the solution of the model according to a given set of parameters. Here, we focus on the scalar-valued maximum pressure, that is
	\begin{equation}
	\PP:=\max_{\calB}\{p\},
	\end{equation}
	at a fixed fluid injection time. This value indicates an observation with respect to a given
	set of parameters ($\theta\in \Theta$). In other words, by solving the system of equations \eqref{compat_argmin} (using $\theta$), the function $\mathcal{M}$ transfers the material unknowns to an observation (here $\PP$) which shows the fracture behavior. We should note that the function $\mathcal{M}:\mathbb{R}^{n_r}:\rightarrow \mathbb{R}^{n_m}$, where $n_r$ is the dimension of the unknown parameters and $n_m$ is the dimension of the reference observation (here $\PP$).

	\item $\epsilon$ is  the estimation error and arises from uncertainties in experimental situations and denotes a sample of $\mathcal{N}(0,\sigma^2 I_{n_r})$, where $\mathcal{N}$ is the normal distribution and $\sigma^2$ denotes a fidelity parameter.
\end{itemize}
Several Markov chain Monte Carlo techniques, e.g., Metropolis-Hastings (MH) \cite{smith1993bayesian} or more effective methods such as the delayed-rejection adaptive-Metropolis method are employed to determine the posterior density of the parameters.

Generally, for a sample $\theta$ of the random field $\Theta$ indicating a realization of the parameters (given in Subsection \ref{parameter})  related
to a sample $r$ of the observations $\mathcal{R}$, the posterior
density is expressed as 
\begin{align}
	\pi(\theta|r)=\frac{\pi(r|\theta)\pi_0(\theta)}{\pi(r)}=\frac{\pi(r|\theta)\pi_0(\theta)}{\displaystyle\int_{\mathbb{R}^{n_r}}\pi(r|\theta)\pi_0(\theta)\,d\theta},
\end{align}
where $\pi_0(\theta)$ denotes the prior distribution (prior information) and $\mathbb{R}^{n_r}$ indicates the  parameters space (a normalization parameter) and its estimation is not computationally easy. Hence, we estimate the posterior density neglecting the normalization constant leads to 
\begin{align*}
	\pi(\theta|\,r)\propto\pi(r|\,\theta)\pi_0(\theta).
\end{align*} 
To estimate the posterior density in the above relation which is the probability density of a set of unknown parameters, the likelihood function $(\pi(r|\theta))$ should be formulated. 
The likelihood function reads
\begin{align}
	\label{likelihood}
	\pi(r|\theta):=\frac{1}{(2\pi \sigma^2)^{\bar{n}/2}}\exp\left(- \zeta(\theta)/2\sigma \right), \qquad \text{with}\qquad \zeta(\theta)=\sum_{i=1}^{n} (r_i-\mathcal{M}(\theta))^2,
\end{align}
where $n$ is the number of time-steps.
Here, for a given realization $\theta$ (a set of parameters), higher $\zeta(\theta)$ (more difference between the model solution and the observation) gives rise to a lower likelihood estimation (lower probability). On the other hand, a negligible difference leads to maximum likelihood estimation and a probability close to 1. We choose also the reference observation by a fine mesh to perform the test problems.  We observe the dependence of maximum pressure ($\PP$) to different effective parameters (listed in Subsection \ref{parameter}).  Obviously, the observation can be modified (more precise) by using a measured value.  

In MH algorithms, during every sampling, a new candidate $\theta^*$ based on the proposal density (e.g., uniform or Gaussian) $\phi$ is proposed, and its acceptance rate (denoted by $ \lambda_1 $) concerning the previous candidate (here $\theta^{k-1}$) is computed. The ratio is given as follows:

\begin{equation}
\lambda_1 (\theta^*|\,\theta^{k-1}):=\min\left(1, \cfrac{\pi(\theta^*|\,r)~\phi(\theta^{k-1}|~\theta^*)}{\pi(\theta^{k-1}|\,r)~\phi(\theta^*|\,\theta^{k-1}))}\right).
\end{equation}
A high acceptance ratio means the proposed proposal gets simulation results closer to the (reference) observation; therefore, it will be accepted; otherwise, the algorithm rejects the candidate (if the ratio is low).

\subsection{The DRAM algorithm for hydraulic fracture}\label{Section31}

\begin{algorithm*}[!]
	\label{algorithm1}
	$\bullet$  Produce an initial $\theta^0\sim\pi(\theta^0|\,r)$ value ($k=0$)\quad\qquad  \\[3mm]
	\textbf{for} {$k=1,\ldots, N$}\\
	
	\quad$\bullet$ set $\var{FLAG}$=true \qquad \qquad $\bullet$ set $n=0$\\[2mm]
	
	\quad 1. Propose a new proposal $\theta^{*}=\theta^{k-1}+\mathcal{D}_k\mathcal{Z}_k$.  \\[2mm]
	
	\quad\quad \textbf{while} $\var{FLAG}$~ \textbf{do}\\
	
	\qquad \quad  \RNum{1}. Solve the system of equations (i.e., 	$\mathcal{E}_{\BfrakU}(\BfrakU, \delta \BfrakU)=0$, see \eqref{compat_argmin}) using Algorithm 1\\[1mm]
	\vspace{-0.1cm}	\hspace{1.4cm}  and obtain  $(\bm u,p,d)$ according to the realization  $\theta^*  \in \{\mu, K, M, B, G_c, \eta_F,\beta_a,\beta_g\}$.\\

	\qquad \quad \RNum{2}. Estimate the maximum of the pressure in the geometry  $\mathcal{P}=\text{max} (p)$\\
	
	\qquad \quad \RNum{3}. \textbf{if} $d$ reaches the boundary\\
	
	\qquad \quad \RNum{4}. \textbf{else if}  $t_n>T$\quad \textbf{then}
	\qquad\qquad $\bullet$ set $\var{FLAG}$=false\\
	
	\qquad \quad \RNum{5}.~ \textbf{else}  \quad\quad $\bullet$~ set~$n+1\rightarrow n$ \qquad $\bullet$~ set $t_n=n\Delta t$\\

	\quad  2. Calculate the acceptance/rejection probability  
	$$\lambda_1 (\theta^*|\,\theta^{k-1})=\min\left(1, \cfrac{\pi(\theta^*|\,r)~\phi(\theta^{k-1}|~\theta^*)}{\pi(\theta^{k-1}|\,r)~\phi(\theta^*|\,\theta^{k-1}))}\right)$$.

	\quad 3.$~\textbf{if}~$ $\mathcal{RV}<\lambda_1~$ \textbf{then}\quad accept the proposal $\theta^*$ and put $\theta^k=\theta^*$\quad \textbf{else}\\
	
	\qquad\quad\RNum{1}.  Calculate the alternative proposal
	\quad$\theta^{**}=\theta^{k-1}+\sigma^2\mathcal{D}_k\mathcal{Z}_k$ .\\
	
	\qquad\quad\RNum{2}.	~$\bullet$ set $\var{FLAG}$=true \qquad \qquad $\bullet$ set $n=0$\\  
	
	\qquad\quad ~~~~~$\bullet$ obtain new $\mathcal{P}$ (see the while/do loop) according to the realization $\theta^{**}$\\
	
	\qquad\quad\RNum{3}.   Calculate the acceptance/rejection probability of the delayed rejected candidate~  
	$$\qquad\qquad\lambda_2 (\theta^{**}|\,\theta^{k-1},\theta^*)=\min\left(1, \cfrac{\pi(\theta^{**}|\,r)~\phi(\theta^*|~\theta^{**})\left(1-\lambda_1(\theta^*|\theta^{**})\right)}{\pi(\theta^{k-1}|\,r)~\phi(\theta^*|~\theta^{k-1})\left(1-\lambda_1(\theta^*|\theta^{k-1})\right)}\right)$$.
	
	\qquad\quad\RNum{4}.  \textbf{if} ~$\mathcal{RV}<\lambda_2$~~\textbf{then} \quad accept the proposal $\theta^{**}$ and put $\theta^{k}=\theta^{**}$ \\
	
	\qquad\quad\RNum{5}.  \,\,\textbf{else} \quad\quad $\bullet$ reject the proposal $\theta^{**}$ and put $\theta^{k}=\theta^{k-1}$
	\qquad\quad\\
	
	\quad 4. Update the covariance matrix as $\mathcal{V}_k=\text{Cov}(\theta^0,\theta^1,\ldots\theta^{k})$.\\
	
	\quad 5. Update $\mathcal{D}_k$ \\
	
	\textbf{end}
	
	\caption{The DRAM algorithm for anisotropic hydraulic phase-field fracture.}
\end{algorithm*}

The Metropolis-Hastings technique is a robust and efficient MCMC technique to estimate the posterior density. Its efficiency verified by the authors in \cite{khodadadian2019bayesian} to identify mechanical coefficients (mechanical parameters and the critical energy rate). Despite its efficiency, during the iterations, the covariance function of the proposal should be tuned manually, and the method has a high autocorrelation. To overcome these drawbacks, during each sampling,  the covariance based on the existing samples (adaptive Metropolis) is updated; therefore, the posterior density is not sensitive to the proposal density. We can modify the technique additionally by using a delayed rejection. To this end, a replacement of the rejected proposal is obtained (i.e., $\theta^{**}$); then, the new acceptance/rejection probability (denoted by $\lambda_2$) is computed, that is

\begin{equation}
\qquad\qquad\lambda_2 (\theta^{**}|\,\theta^{k-1},\theta^*)=\min\left(1, \cfrac{\pi(\theta^{**}|\,r)~\phi(\theta^*|~\theta^{**})\left(1-\lambda_1(\theta^*|\theta^{**})\right)}{\pi(\theta^{k-1}|\,r)~\phi(\theta^*|~\theta^{k-1})\left(1-\lambda_1(\theta^*|\theta^{k-1})\right)}\right).
\end{equation}
In other words,  a second-stage move will be used to increase the acceptance chance of the rejected proposal. The algorithm is useful, specifically when the samples have a high-dimensional conditional density \cite{zuev2011modified}.

The DRAM algorithm for parameter estimation in hydraulic fracture is summarized in Algorithm 2. Here, $\mathcal{Z}_k\sim \text{Uniform}~(0,I_{n_r})$ where $I_{n_r}$ denotes the $n_r$-dimensional identity matrix, $\mathcal{D}_k$ indicates the Cholesky decomposition of $\mathcal{V}_k$ (the covariance of the realizations), and $\mathcal{RV}\sim \text{Uniform}~(0,1)$. To enhance the acceptance rate, we update the covariance function of the proposal density.
In order to provide a narrower proposal density, we use $\sigma<1$. The covariance function can be estimated by
\begin{align}
	\text{Cov}(\theta^0,\theta^1,\ldots,\theta^{k})=\frac{1}{k}\left(\sum_{i=0}^{k}
	\theta^i\left(\theta^i\right)^T-(k+1)~\hat{\theta}^k\left(\hat{\theta}^k\right)^T\right),
\end{align}  
where $\hat{\theta}^k=\frac{1}{k+1}\displaystyle\sum_{i=0}^{k}\theta^i$. 

\subsection{Physical interpretation of the parameters}\label{Section32}
\label{parameter}
Here we review the list of important parameters (with their used unit) in hydraulic fracture and explain that are they correlated or not.
\vspace{-0.1cm} 
\begin{itemize}
	\item \textit{Elasticity modulus}\,\text{[GPa]}. 
	Generally, the mechanical material parameters denote the shear modulus $\mu$ and Lam{\'e}'s first parameter $\lambda$. The bound $\lambda>-\frac{2\mu}{3}$ may relate it to the shear modulus. Poisson's ratio $\nu$ also satisfies the condition $-1 < \nu < \frac{1}{2}$.  Hence, these two parameters are not well-suited for the estimation due to their bounds and dependency. Instead, the effective bulk modulus, $K=\lambda+ \frac{2\mu}{3}$ in addition to the shear modulus are chosen as the elasticity parameters.
	Therefore, the only necessary constraint is the positivity of the parameters. Since the mechanical parameters are correlated a joint probability density will be estimated.

Higher shear modulus (due to higher Young's modulus) increases the reservoir hardness; therefore, the fracture initiates faster and the propagation rate is more for the cases with higher Young's modulus. The bulk modulus is the measure of the decrease in volume with an increase in pressure.

	\item \textit{Biot's coefficient $B$} described by Biot \cite{biot1941general} and  represents the change of the bulk volume because of a pore pressure change while the
	stress is constant (the contribution of the pore
	pressure to the stress). 
	The fracture length reduces with a raise in
	the Biot's number. The effect of
	the pore pressure on the fracture propagation can be more pronounced for higher Biot's coefficient \cite{golovin2018influence}.
	\item \textit{Biot's modulus} $M$\,[GPa] considers the combined fluid/solid compressibility. The inverse of $M$ denotes the rate of the volume of fluid
	released from a non-deforming frame to the pore
	pressure drop, therefore determines a storage coefficient \cite{cheng1997material}.
	\item \textit{Dynamic fluid viscosity} $\eta_F$\,[kg/(m.s)]  is the resistance to movement of one layer of a fluid over another.  By raising dynamic viscosity, the breakdown pressure rises noticeably however the fracture initiation pressure rises only slightly.
	\item \textit{Griffith's critical energy release rate} $G_c$\,[GPa] indicates a property of the
	materials that the fracture is propagating in or into. In other words, it  is represented as the decline in total potential energy per increase in fracture surface area.
\end{itemize}

 Due to the nature of subsurface systems, it is hard and time-consuming to provide the
measurements for the inverse problem. Considering the limited resources, it is useful to choose the reference value, via a simulation-based technique. To that end, the quantity of interest in our simulation is the maximum fluid pressure versus the fluid injection time. Findings \cite{LeeWheWi16,yoshioka2020crack,CHUKWUDOZIE2019957} showed that maximum pressure increases within the fractured area before the onset of the crack propagation, which yields into a drop of the fluid pressure, that is well-know observation in the fracking process \cite{LeeWheWi16}.

The typical random distribution of the unknown parameters is log-normal. Hence, in the Bayesian inverse framework,
it is usual to work with their logarithms (i.e., natural logarithm) instead of the original variables and to choose Gaussian distribution as the prior distribution. With that, we can remove the positivity constraint as well.

\section{Numerical experiments }\label{Section5}
In this section, to use the developed numerical procedure for modeling hydraulic fractures in isotropic and anisotropic solids, we employ four specific numerical examples. The used material parameters are given in Table~\ref{material-parameters} (according to \cite{MieheMauthe2015,xia2017phase}). To obtain the solution of the coupled system of equations bilinear quadrilateral $Q1$ finite elements are used, and the consistent linearization, including finite element discretization, is further explained in detail in Appendix A. 
\renewcommand{\tablename}{Table}
\setcounter{table}{0}
\begin{table}[!ht]
	\caption{Material parameters employed in the numerical experiments according to \cite{MieheMauthe2015,xia2017phase}.}	\vspace{1mm}
	\centering
	\begin{tabular}{cclll}
		No.  &parameter & name                   & value    & unit            \\[2mm]\hline 
		1.   &$\mu$        & shear modulus       & $6.65$    & $\mathrm{GPa}$ \\[2mm]
		2.   &$K$      & bulk modulus       & $11$    & $\mathrm{GPa}$                 \\[2mm]
		3.   &$M$        & Biot's modulus        & $12.5$  & $\mathrm{GPa}$ \\[2mm]
		4.   &$B$        & Biot's coefficient     & $0.79$   & -- \\[2mm]
		5.   &$K_D$& Intrinsic permeability  & $2 \times 10^{-14} $  & $\mathrm{m^2}$ \\[2mm]
		6.   &$\zeta$& Permeability transition exponent  & $50 $  & -- \\[2mm]
		7.   &$\eta_F$  & Dynamic fluid viscosity   & $0.001$  & $\mathrm{kg/(m.s)}$ \\[2mm]
		8.   &$G_c$     & Griffith's energy release rate  & $0.00265$ & $\mathrm{GPa}$ \\[2mm]
		9.  &$\eta$    & Crack viscosity     & $10^{-14}$ & $\mathrm{N/m^{2}s}$\\[2mm] 
		10.  &$\kappa$    & Stabilization parameter     & $10^{-8}$ & --\\[2mm]
		\hline
		\label{material-parameters}
	\end{tabular}
\end{table}

Here, we introduce four different numerical experiments. Then, we employ the DRAM technique to identify the influential parameters. The first two examples cover only isotropic materials, where in the next two problems we consider transversely isotropic and orthotropy anisotropic fractures. In the DRAM algorithm, we employ the fidelity parameter $\sigma=10^{-3}$ and we replicate the Bayesian algorithm for $N=10\,000$ number of samples. In all examples, $h=1/200$ is used for the simulations and $h=1/215$ is employed to obtain the reference observation (using the given values in Table \ref{material-parameters}).  A length scale of $l=2h$ in addition to a negligible $\kappa$ (here is $10^{-8}$) is used as well. Regarding the stabilization parameter, we refer the reader to \cite{khodadadian2019bayesian} for a  discussion.
Finally, for all examples, the prior distribution of all desired parameters are listed in Table \ref{prior}. 

%
%
%
%

\begin{table}[!]
	\caption{The prior distribution of the model parameters for different test problems.}
	\vspace{1mm}
	\centering
	\begin{tabular}{cclll}
		&parameter & prior distribution& true value      & test problems                \\\hline 
		&$B$        & $\mathcal{N} (0.8,0.1)$ & 0.79 & all\\[2mm]
		&$\eta_F$  &$\mathcal{N} (0.001,0.0001)$   & 0.001 &all\\[2mm]
		&$G_c$      &$\mathcal{N} (0.0027,0.0003)$  & $0.00265$ &all\\[2mm]
		&$\mu$        & $\mathcal{N} (6.5,0.5)$     & $6.65$    &   all\\[2mm]
		&$K$           & $\mathcal{N} (11,1)$    & 11            &  all \\[2mm]
		&$M$        & $\mathcal{N} (12,1)$       & $12$  &  all\\[1mm]
		&$\beta_a$        & $\mathcal{N} (50,60)$       & 55& Example 3    \\[2mm]
		&$\beta_a$~\text{or}$~\beta_g$        & $\mathcal{U} (0,2)$       & 0.5 & Example 4, Case b    \\[2mm]
		&$\beta_a$~\text{or}$~\beta_g$        & $\mathcal{U} (0,20)$       & 10 & Example 4, Case b    \\[2mm]
		&$\beta_a$~\text{or}$~\beta_g$        & $\mathcal{U} (0,20)$       & 10 & Example 4, Case d    \\[2mm]
		&$\beta_a$~\text{or}$~\beta_g$        & $\mathcal{U} (150,250)$       & 200 & Example 4, Case d    \\[2mm]
		\hline
		\label{prior}
	\end{tabular}
\end{table}


\subsection{Hydraulically induced crack driven by fluid volume injection}\label{Example1}

In the following numerical example, a BVP is applied to the square plate shown in Figure \ref{example1-a}(a). We set $A=40~m$ hence $\calB=(0,80)^2~m^2$ that includes a predefined single notch $\calC_1$ of length $8~m$ in the body center with $a =(36,40)\;m$ and $b =(44,40)\;m$, as depicted in Figure \ref{example1-a}(a). 
A constant fluid flow of $\bar{f} = 0.003\; m^2/s$ is injected in $\calC_1$. At the boundary $\partial_D\calB$, all displacements are fixed in both directions and the fluid pressure is set to zero. Fluid injection $\bar{f}$ continues until failure for $T = 60$ second  with time step $\Delta t = 0.1$ second during the simulation. In the next two examples, we deal with isotropic hydraulic fracture and hence we fixed and set $\beta_a=\chi_a=\beta_g=\chi_g=0$ to recover isotropic formulation.

\begin{figure}[!b]
	\centering
	{\includegraphics[clip,trim=3cm 6cm 6cm 7cm, width=15cm]{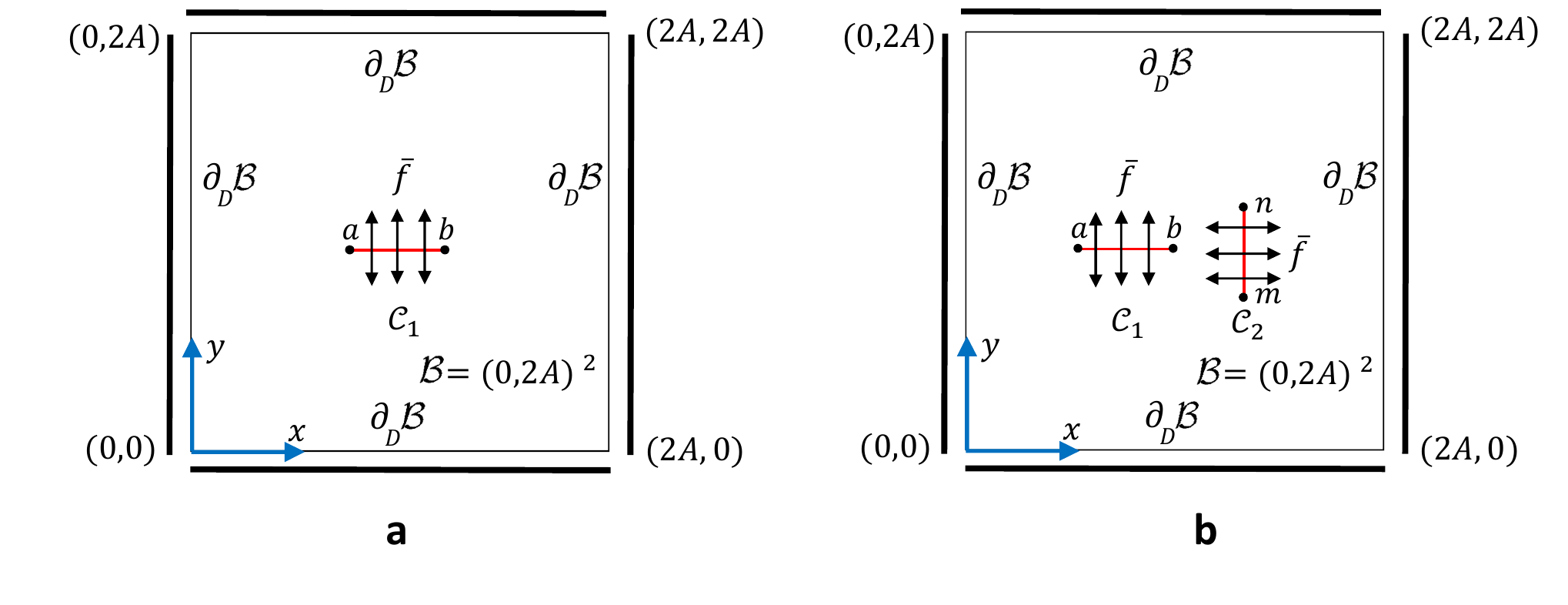}}  
	\caption{Geometry and boundary conditions (a)  Example 1. Hydraulically induced crack driven by fluid volume injection, and (b) Example 2. Joining of two cracks driven by fluid volume injection.
	}
	\label{example1-a}
\end{figure}

\begin{figure}[!]
	\centering
	{\includegraphics[clip,trim=3cm 5.3cm 3.5cm 6cm, width=16.3cm]{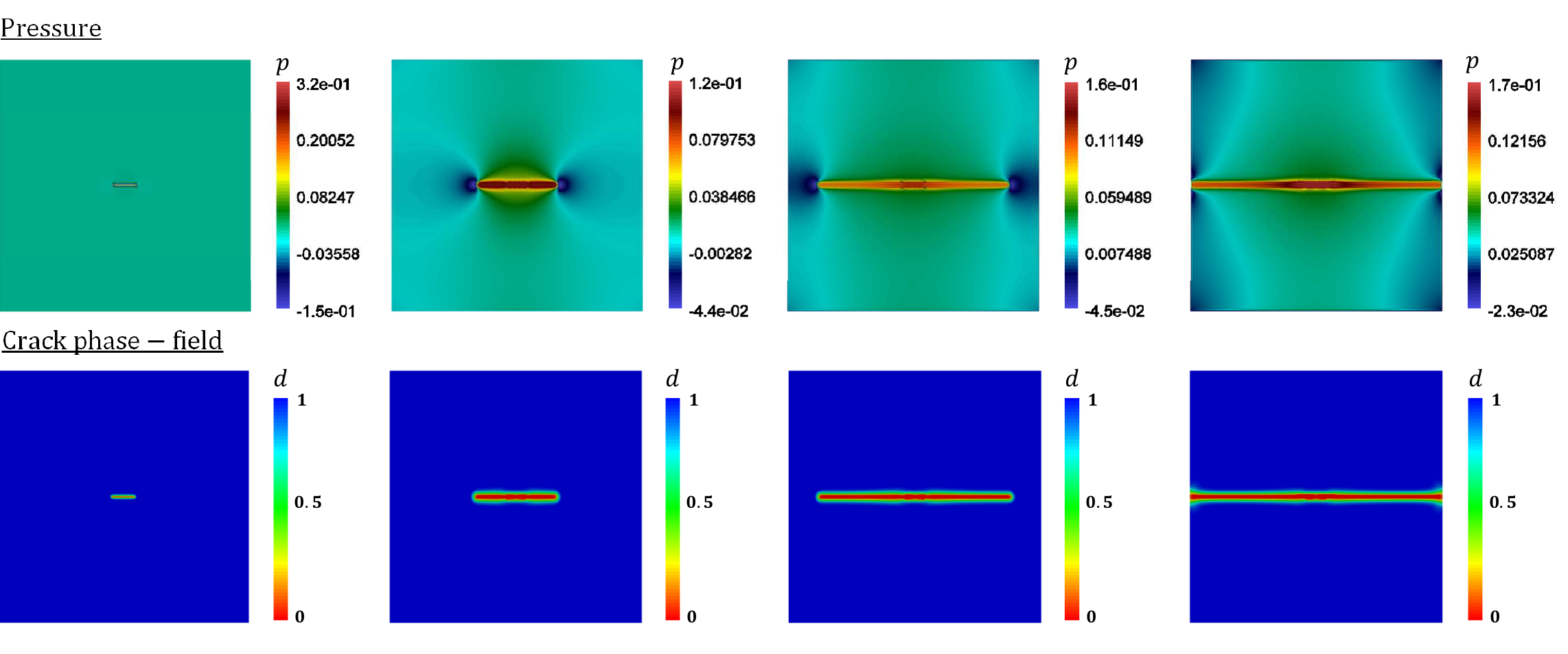}}  
	\caption{Example 1. Reference results of the hydraulically induced crack driven by fluid volume injection. Evolution of the fluid pressure $p$ (first row) and crack phase-field $d$ (second row) for different deformation stages up to the final failure at $t=0.1,\,6.5,\,30,\,49.5$ seconds.}
	\label{example1}
\end{figure}

\begin{figure}[!b]
	\subfloat{\includegraphics[width=0.35\textwidth]{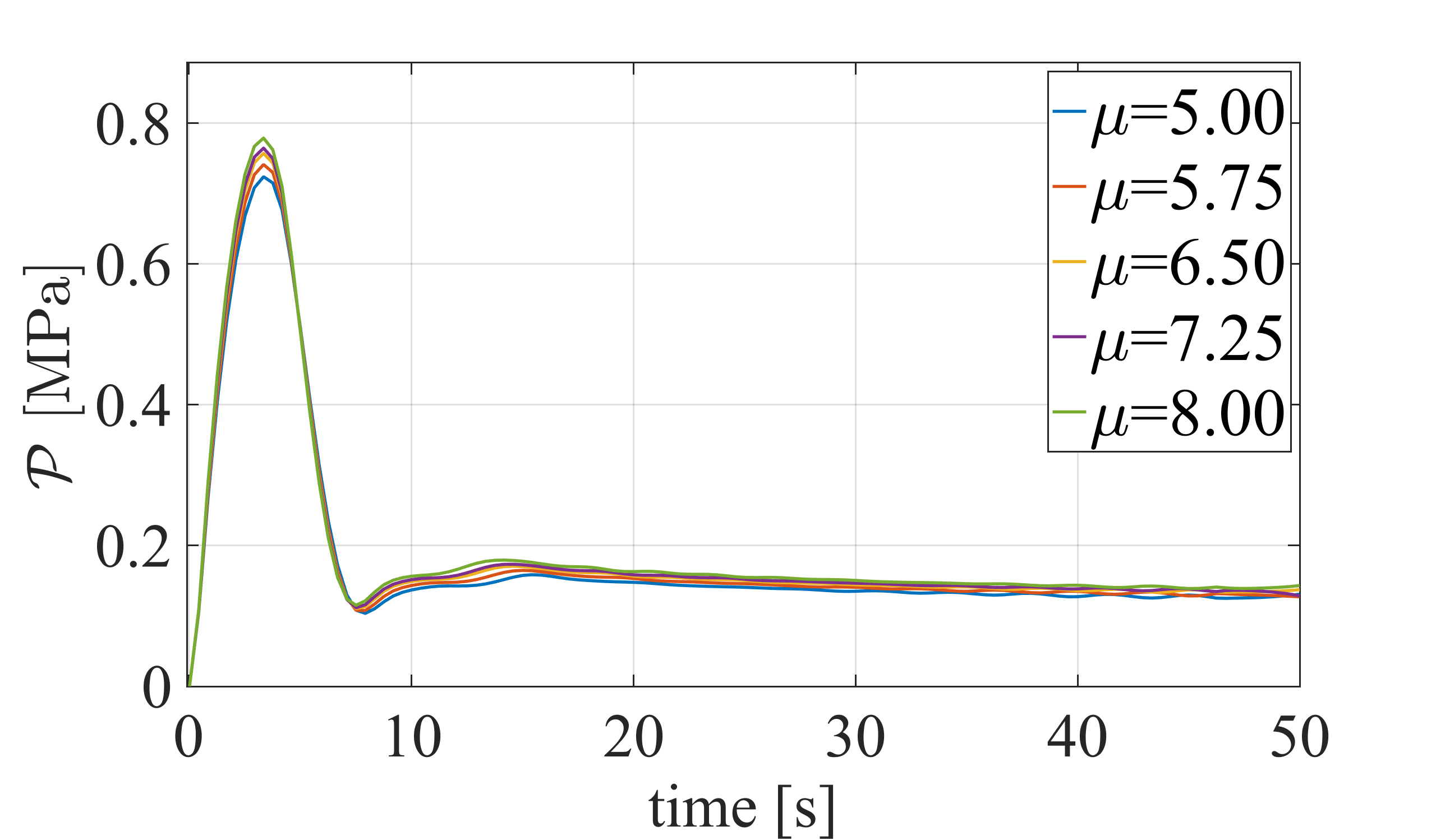}} 
	\subfloat{\includegraphics[width=0.35\textwidth]{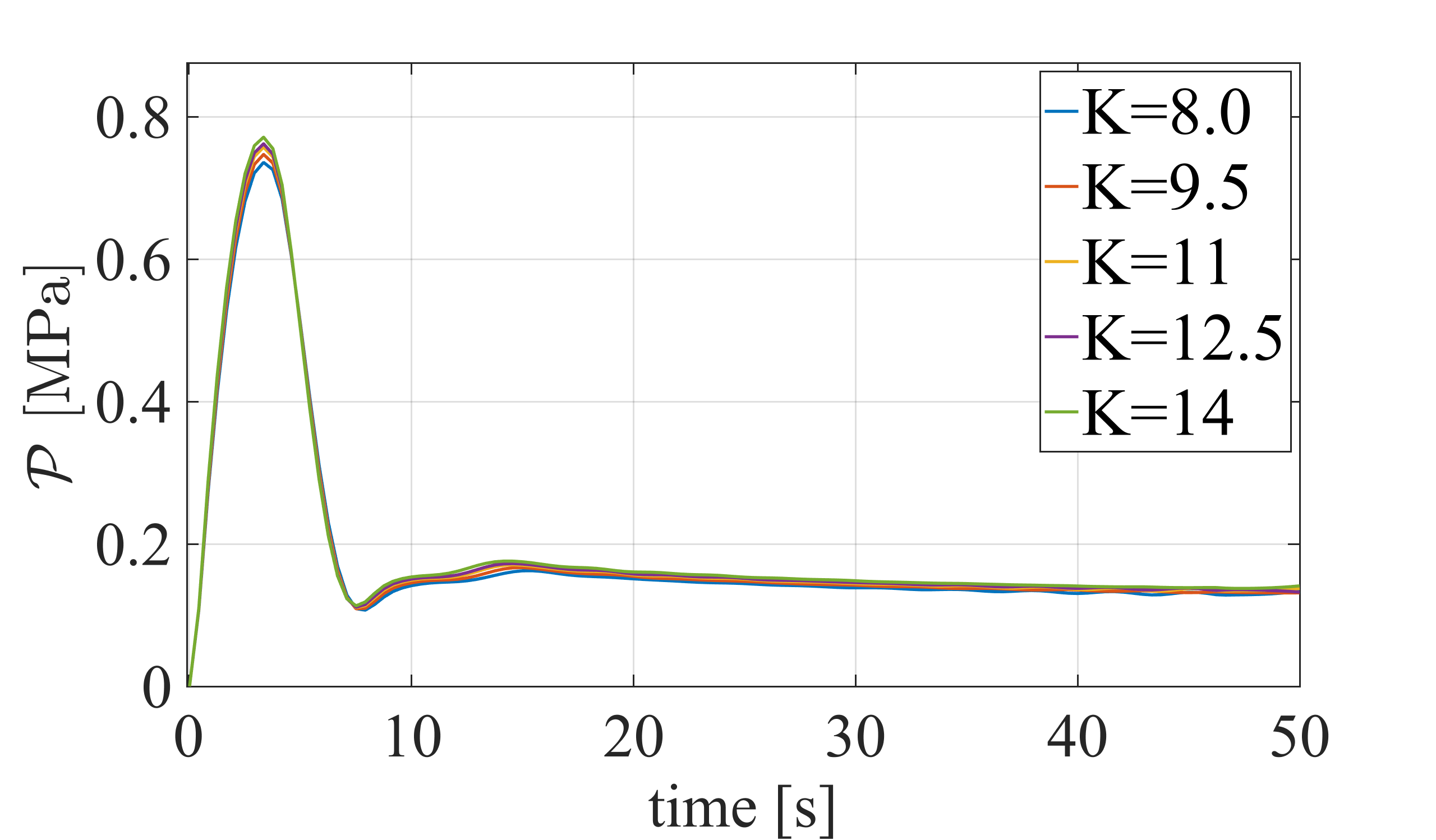}}
	\subfloat{\includegraphics[width=0.35\textwidth]{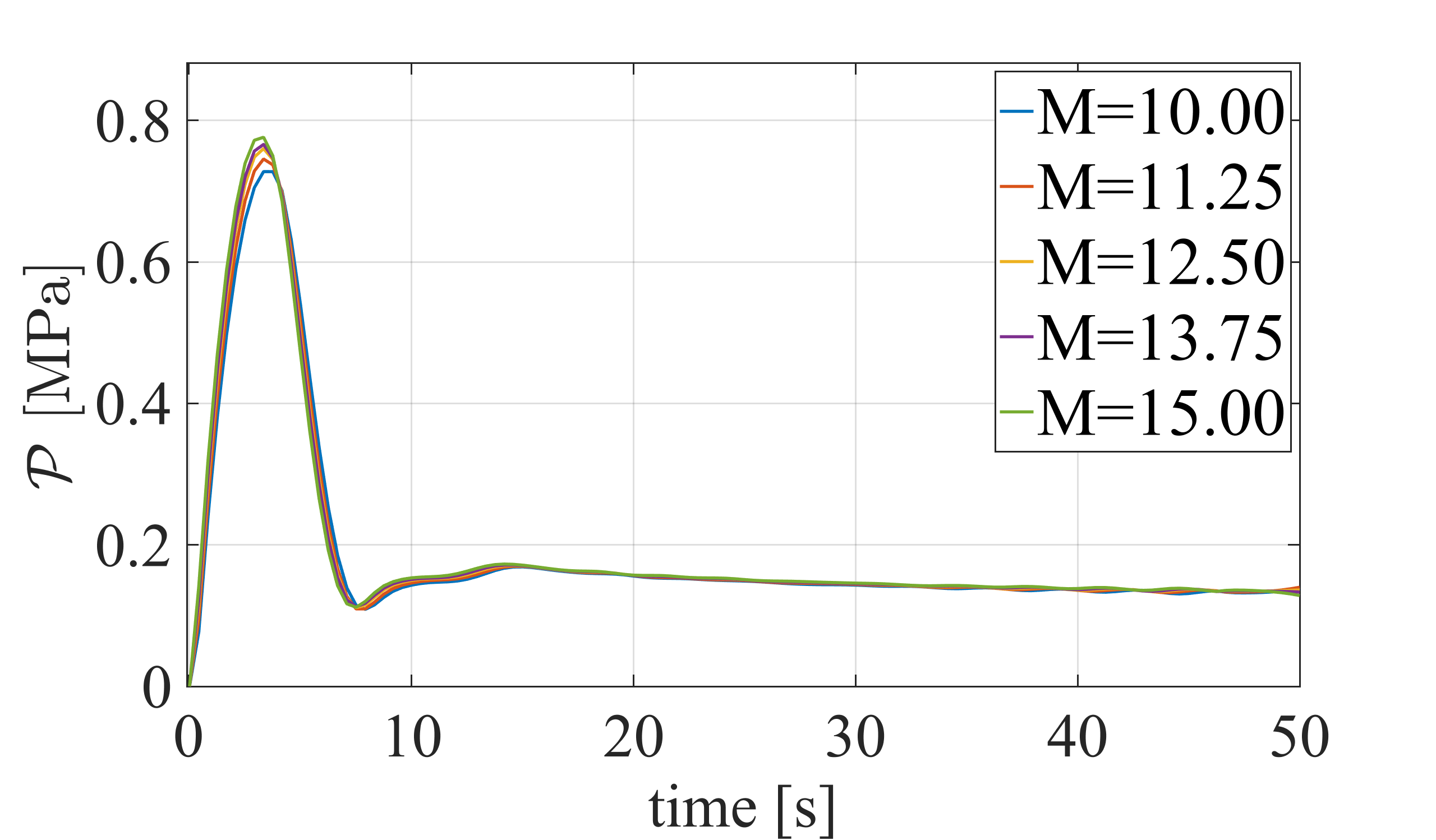}}%
	\newline
	\subfloat{\includegraphics[width=0.35\textwidth]{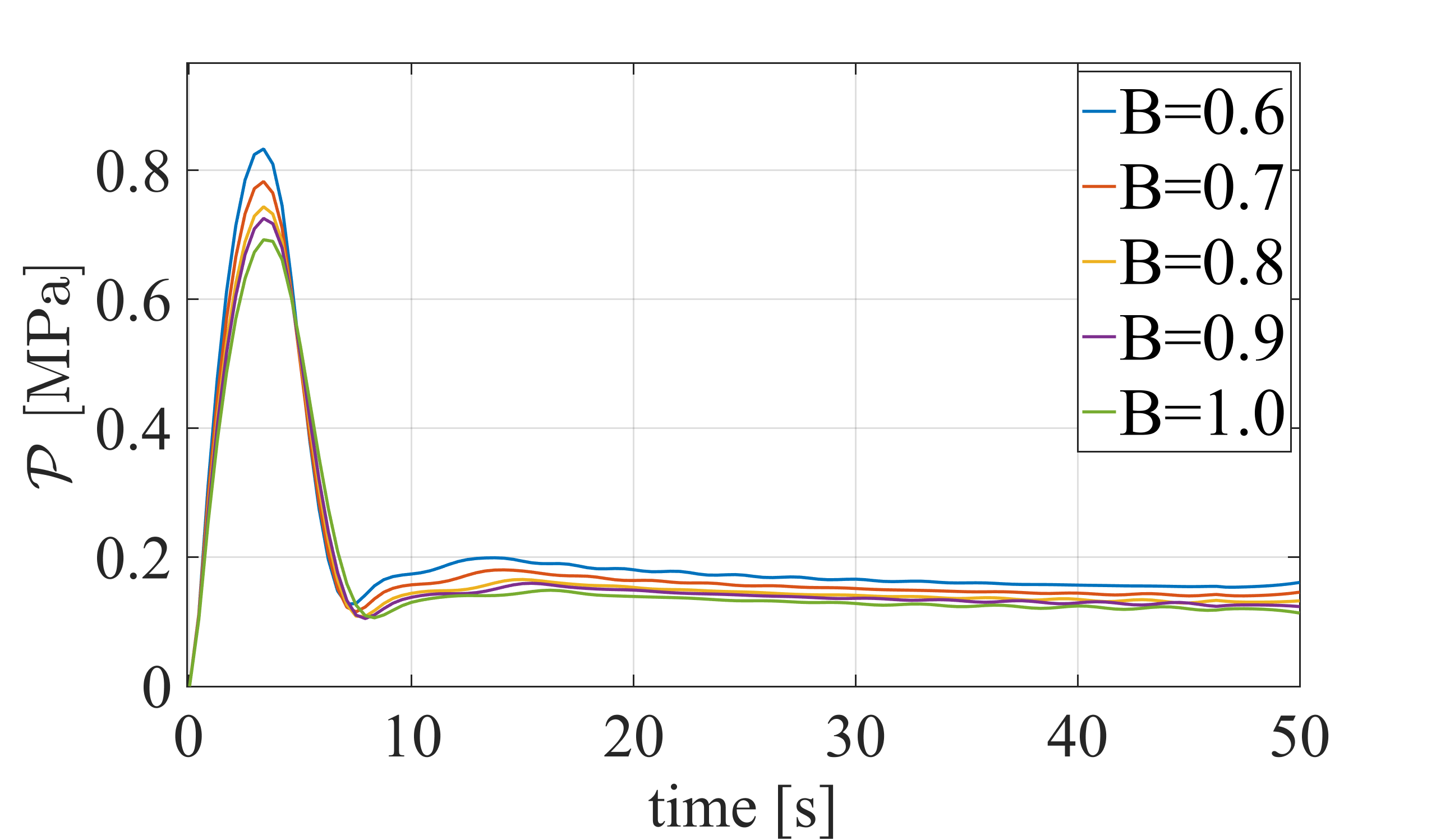}} 
	\subfloat{\includegraphics[width=0.35\textwidth]{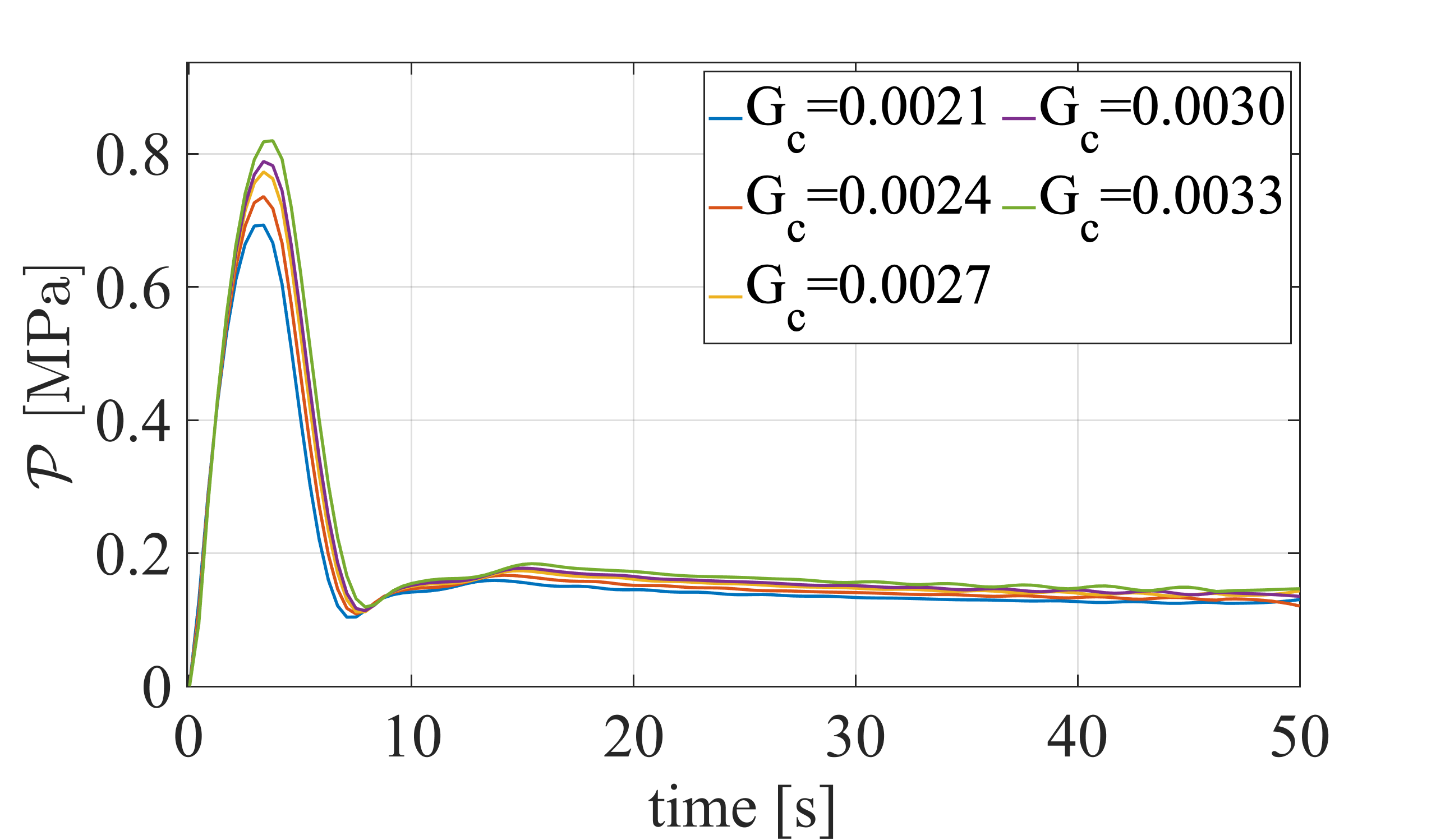}}  
	\subfloat{\includegraphics[width=0.35\textwidth]{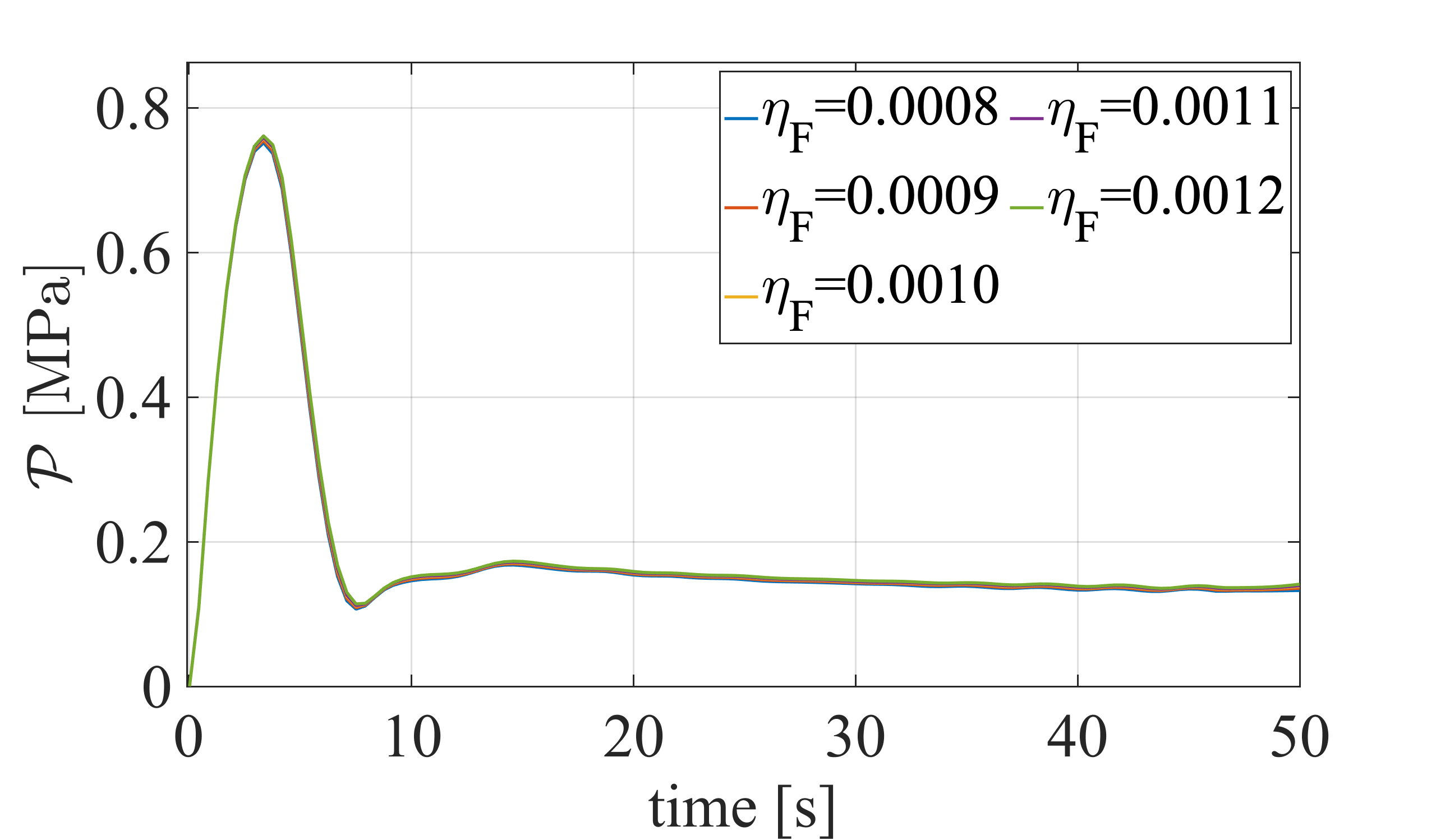}} 
	\caption{Example 1.  The maximum pressure value $\PP$  for different values of $\mu$, $K$, $M$, $B$, $G_c$, and $\eta_F$.}
	\label{fig:exam1_curv}
\end{figure}

We start our analysis by illustrating the reference results for different fluid injection time up to final failure related to Figure \ref{example1-a}a. The  fluid pressure $p$ (first row) and crack phase-field $d$ (second row) evolutions are demonstrated in Figure \ref{example1} for four-time steps, i.e., $t=0.1,\,6.5,\,20,\,34.5$ seconds. The crack initiates at the notch-tips due to fluid pressure increase. Thereafter, the crack propagates horizontally in two directions towards the boundaries. In the fractured zone, $p$ is almost constant due to the increased permeability inside the crack. Whereas, low fluid pressure in the surrounding is observed due to the chosen small time-step in comparison with the permeability of the porous medium, as outlined in \cite{MieheMauthe2015}. The fluid pressure drops down while the crack propagates further as shown in Figure \ref{example1}(b) (second row, middle states). Then, $p$ increases again due to the prescribed fixed boundary conditions $\partial_D\calB$, see Figure \ref{example1} (first row, last state). 

Figure  \ref{fig:exam1_curv} shows the pressure curve during the injection time for different values for six influential parameters. As the figure shows, an increase in the Biot's coefficient raises the pressure peak point; however, for the rest of the parameters, it gives rise to a decline. We continued the pressure estimation until the crack reached the boundary (here $T=60$ seconds is used). 

Now the Bayesian inversion (the DRAM technique) is employed to identify the parameters.  Figure \ref{fig:exam1_hist} depicts the histogram of posterior density of the values. Due to the correlation of the parameters, the joint probability of the elastic modulus and Biot's coefficient/modulus are estimated. As shown, a wide probability density for $\eta_F$ indicates its low impact on the pressure; however, the narrow curve for $G_c$ points out its high effect on the pressure during the injection time. The results are compatible with the obtained curves in Figure \ref{fig:exam1_curv}.

\begin{figure}[!ht]
	\centering
	{\includegraphics[clip,trim=0cm 23.3cm 0cm 17cm, width=16.3cm,height=3.6cm]{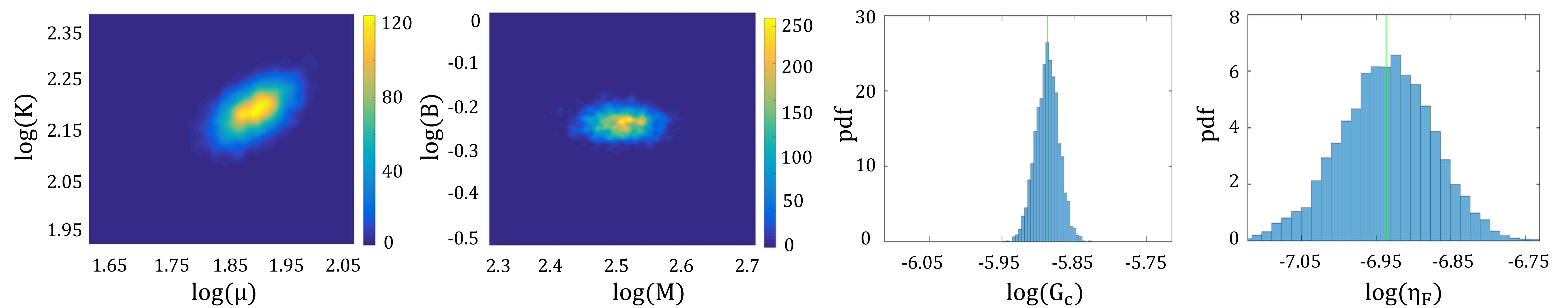}}  
	\caption{Example 1. From left to right: the posterior density  of  the mechanical parameters, Biot's coefficients/modulus, $G_c$, and $\eta_F$. The green lines are the mean values.}
	\label{fig:exam1_hist}
\end{figure}

The main advantage of the DRAM algorithm compared to the Metropolis-Hastings algorithm in phase-field fracture \cite{khodadadian2019bayesian} is a significantly higher acceptance rate. As we already mentioned, the proposal adaptation and the new adjusted candidate improves the reliability/efficiency of the parameter identification.  In order to verify the obtained values, we solved the system with the posterior knowledge and estimated  $\PP$. Figure  \ref{fig:exam1_post} illustrates the pressure diagram obtained by the prior and posterior values and the chosen reference observation. The technique efficiency in the precise estimation of the peak point and curve behavior can be observed here.


\begin{figure}[!]
	\centering
	\includegraphics[width=11cm,height=5.5cm]{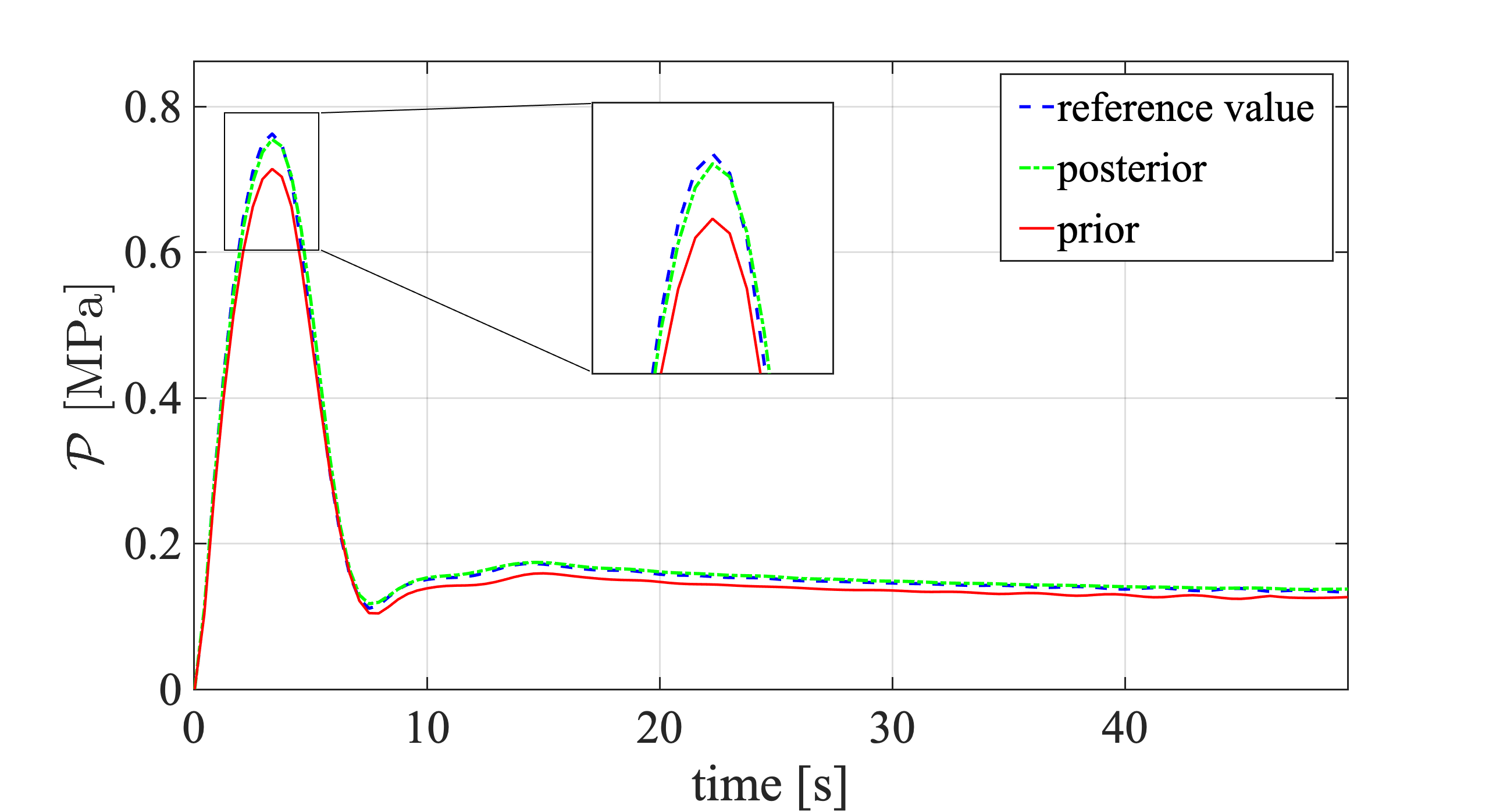}
	\caption{Example 1. A comparison between the maximum pressure (during the injection time) with prior values (red line) and the posterior values (green line). The reference diagram is depicted with a blue line.}
	\label{fig:exam1_post}
\end{figure}

\subsection{Joining of two cracks driven by fluid volume injection}\label{Example2}
The second example is given for handling \textit{coalescence} and \textit{merging} of \textit{crack paths} for the hydraulic fracturing material. 
Crack-initiation and curved-crack-propagation, representing a mixed-mode fracture, are predicted with a phase-field formulation.

The boundary value problem is similar to the benchmark problem of \cite{WickLagrange2014} and depicted in Figure \ref{example1-a}(b). We keep all parameters and loading as in the previous example. The first crack $\calC_1$ is located near the middle of the domain with coordinates $a =(28,40)$ and $b =(36,40)$. The second crack $\calC_2$ is vertically-oriented at $n=(50,44)$ and $m=(50,36)$ with a distance of $14\;m$ from $\calC_1$. A constant fluid flow of $\bar{f} = 0.003\; m^2/s$ is injected in $\calC_1$ and $\calC_2$ as sketched in Figure  \ref{example1-a}(b). At the boundary $\partial_D\calB$, all the displacements are fixed in both directions and the fluid pressure is set to zero. Fluid injection $\bar{f}$ continues until failure for $T = 28$ second with time step $\Delta t = 0.1$ second during the simulation.

Figure \ref{example2_ref} shows the evolutions of the fluid pressure $p$ (first row) and the crack phase-field $d$ (second row) for the reference problem at different times  $t=1.8,\,6.5,\,13.5,\,27.7$ seconds. Here the crack propagates from the notches. We again observe nearly constant fluid pressure in the fractured area ($d=0$), whereas outside the crack zone $p$ is much lower, see Figure \ref{example2_ref} (first row).
\begin{figure}[!t]
	\centering
	{\includegraphics[clip,trim=3cm 5.3cm 3.5cm 6cm, width=16.3cm]{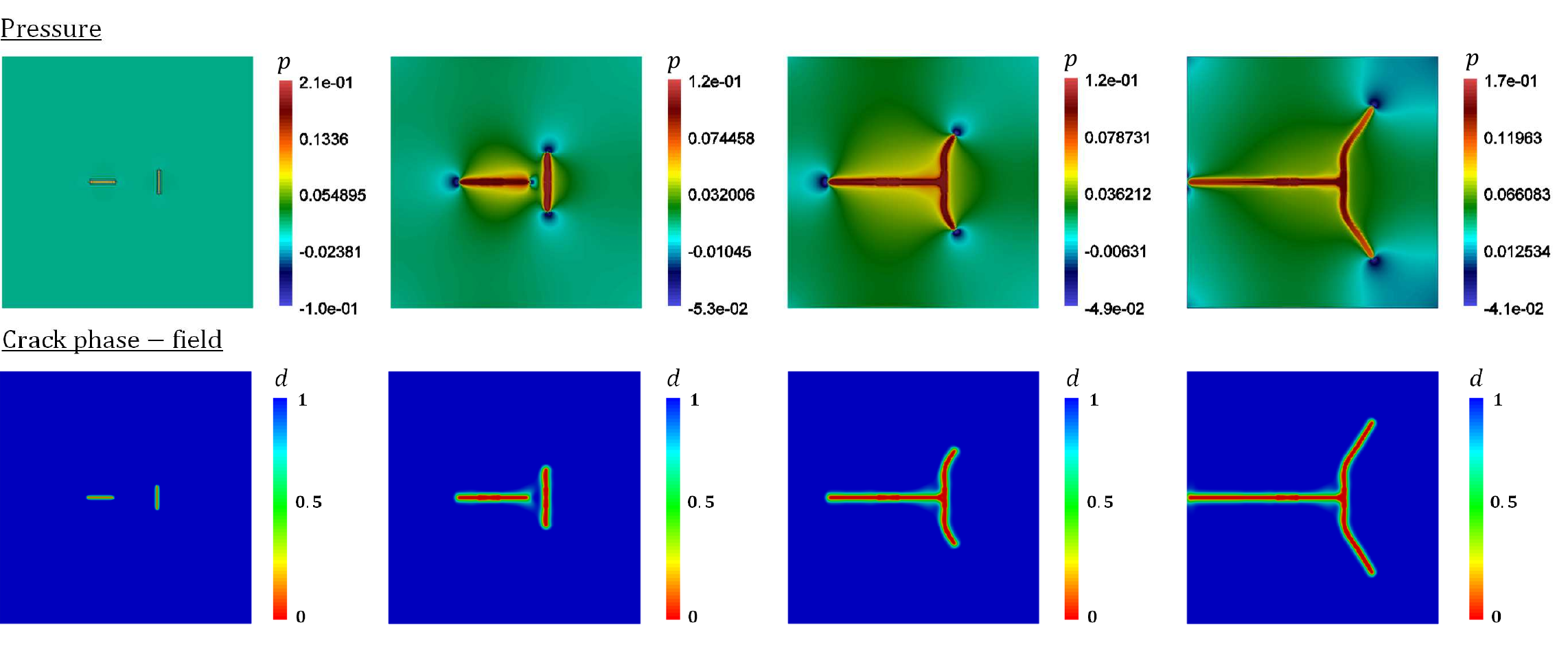}}  
	\caption{Example 2. Reference results of the joining cracks driven by fluid volume injection. Evolution of the vertical displacement $u_y$ (first row), fluid pressure $p$ (second row) and crack phase-field $d$ (third row) for different deformation stages up to final failure at $t=1.8,\,6.5,\,13.5,\,27$ seconds.}
	\label{example2_ref}
\end{figure}

In order to study the parameter effect, we observe the pressure curve with different values of the effective parameters. Figure \ref{fig:exam2_curv} show the influence of on $\PP$   diagram during different injection time. The obtained information from the DRAM algorithm (the posterior distribution) is shown in Figure \ref{fig:exam2_post}.  As depicted, $G_c$ shows a Gaussian distribution and $\eta_F$ has a skewed distribution. 
Finally, the pressure diagram for prior/posterior and the used reference observation (with finite element mesh size  ${h=0.24}$) is shown in Figure  \ref{fig:exam2_post}. Similar to Example 1, employing Bayesian inference enables us to have a more exact model, i.e., the peak point and pressure behavior are predicted more precisely.


\begin{figure}[!b]
	\subfloat{\includegraphics[width=0.35\textwidth]{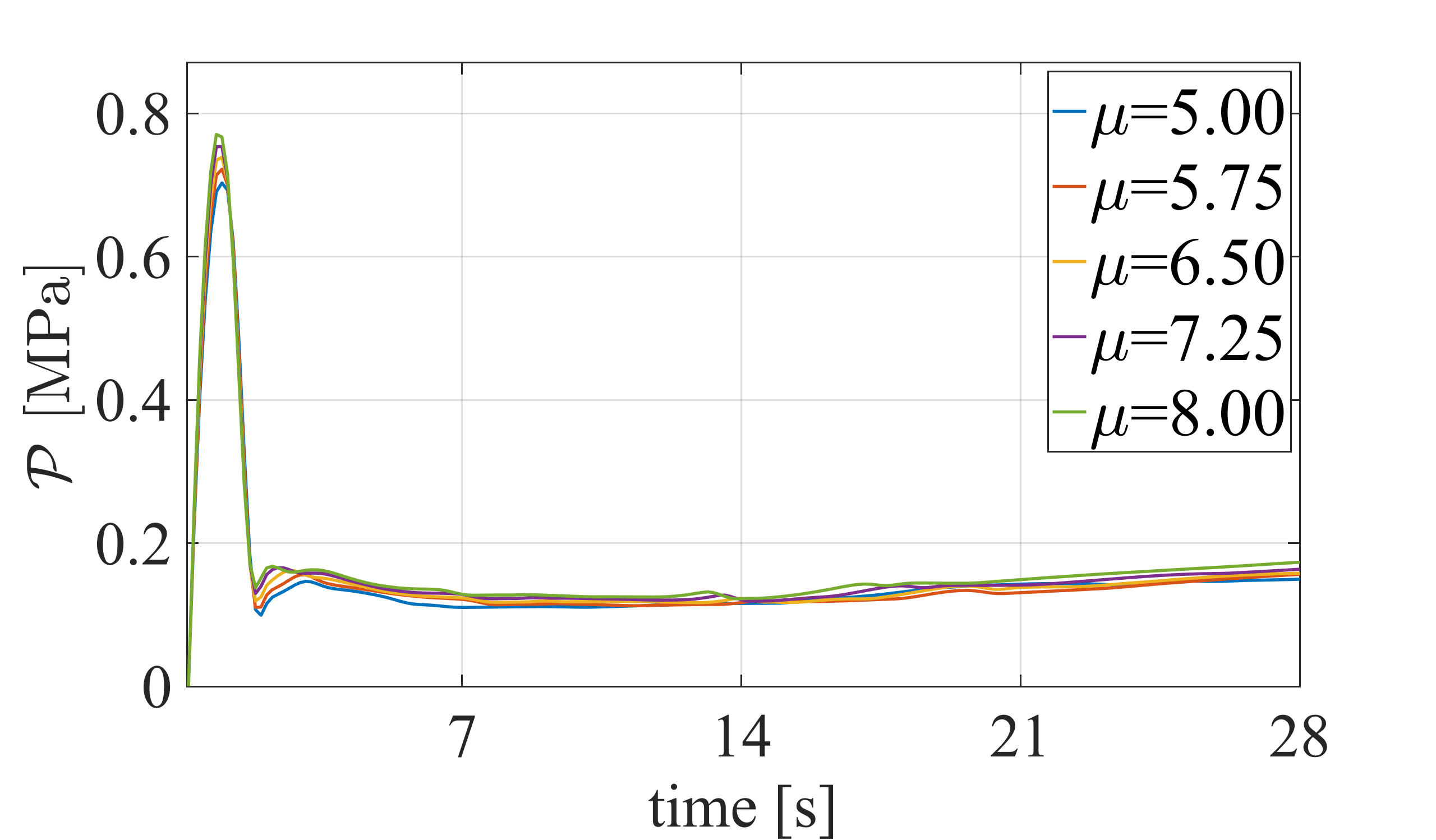}}
	\subfloat{\includegraphics[width=0.35\textwidth]{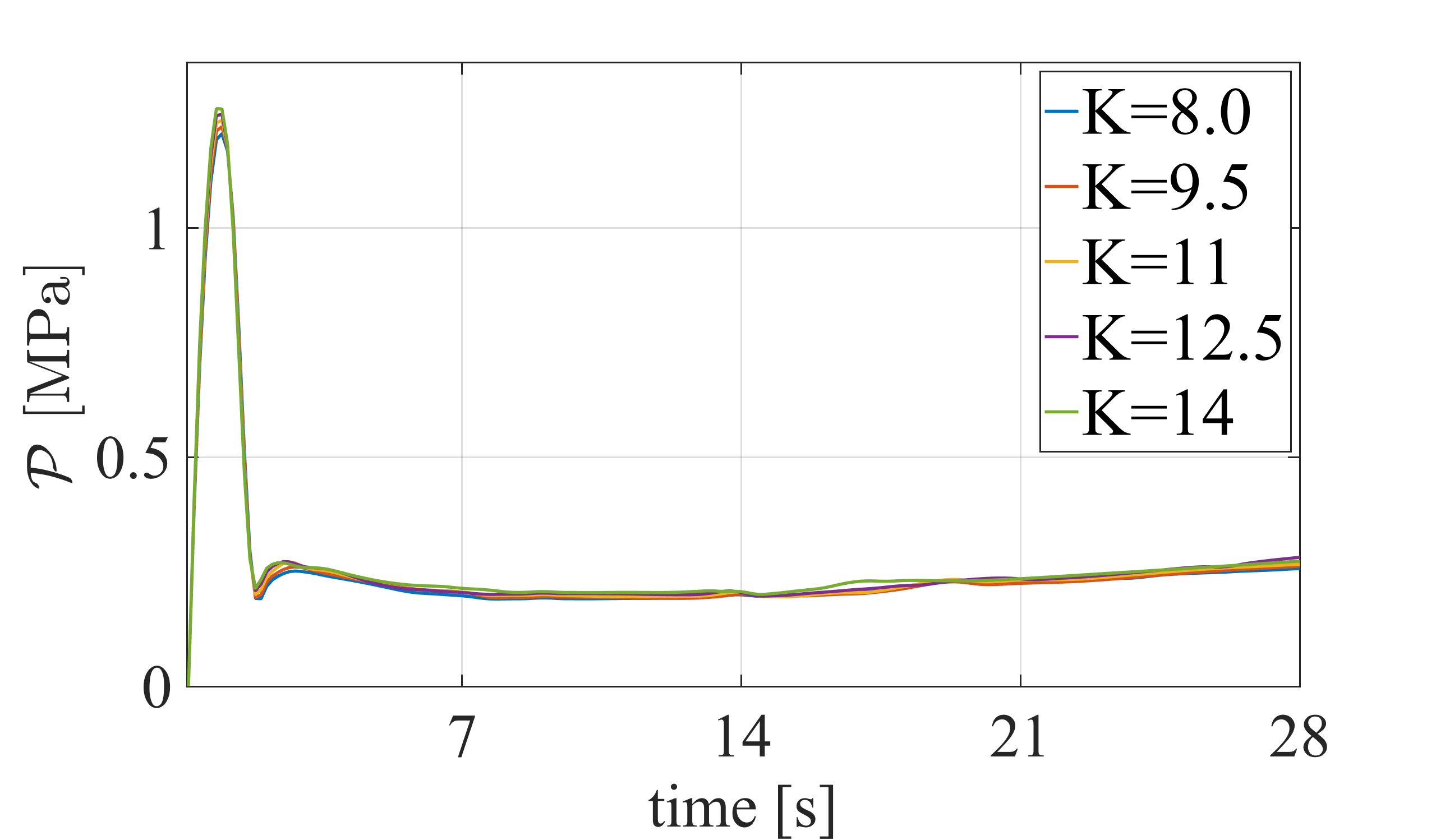}}
	\subfloat{\includegraphics[width=0.35\textwidth]{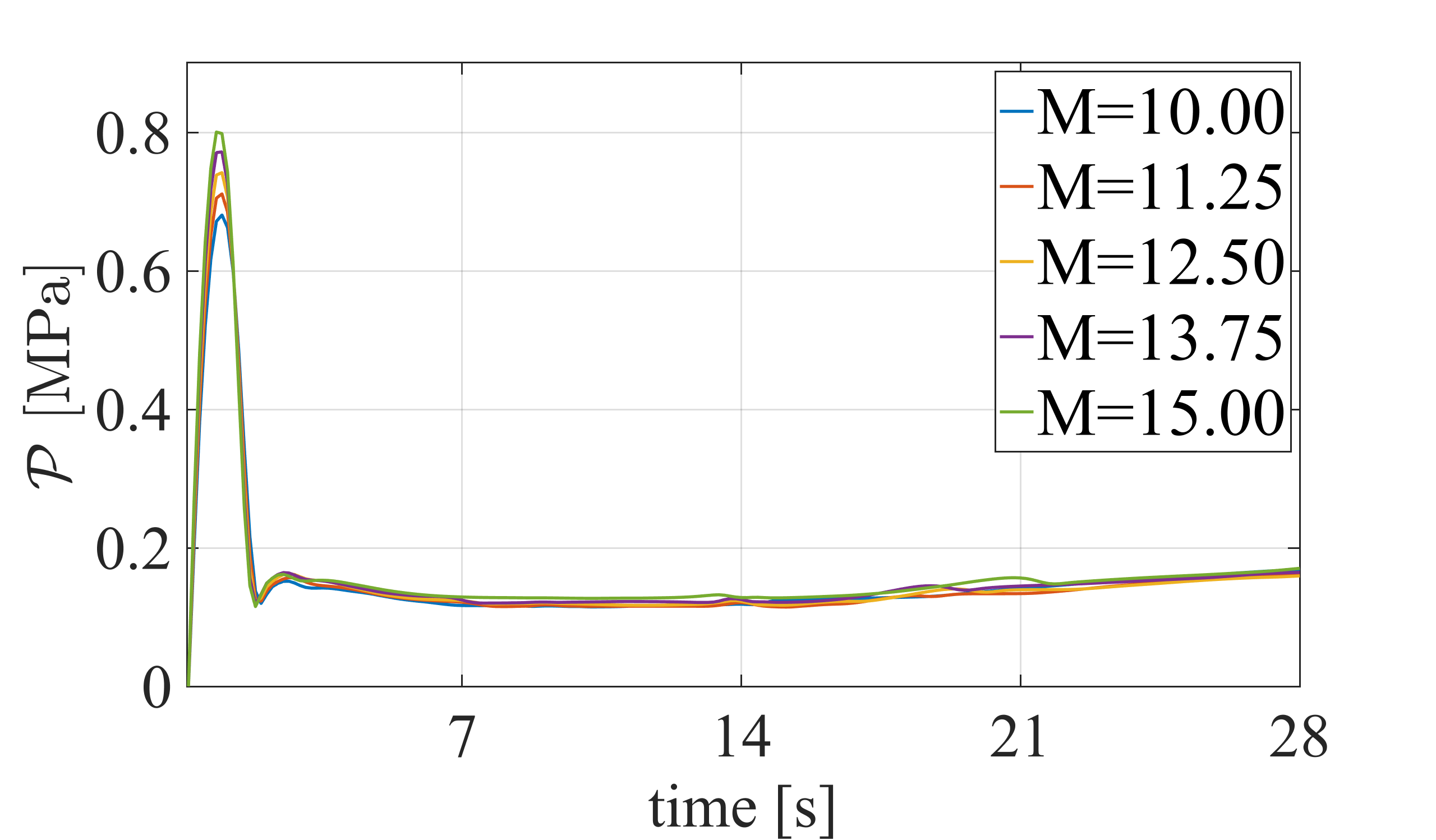}}
	\newline	\subfloat{\includegraphics[width=0.35\textwidth]{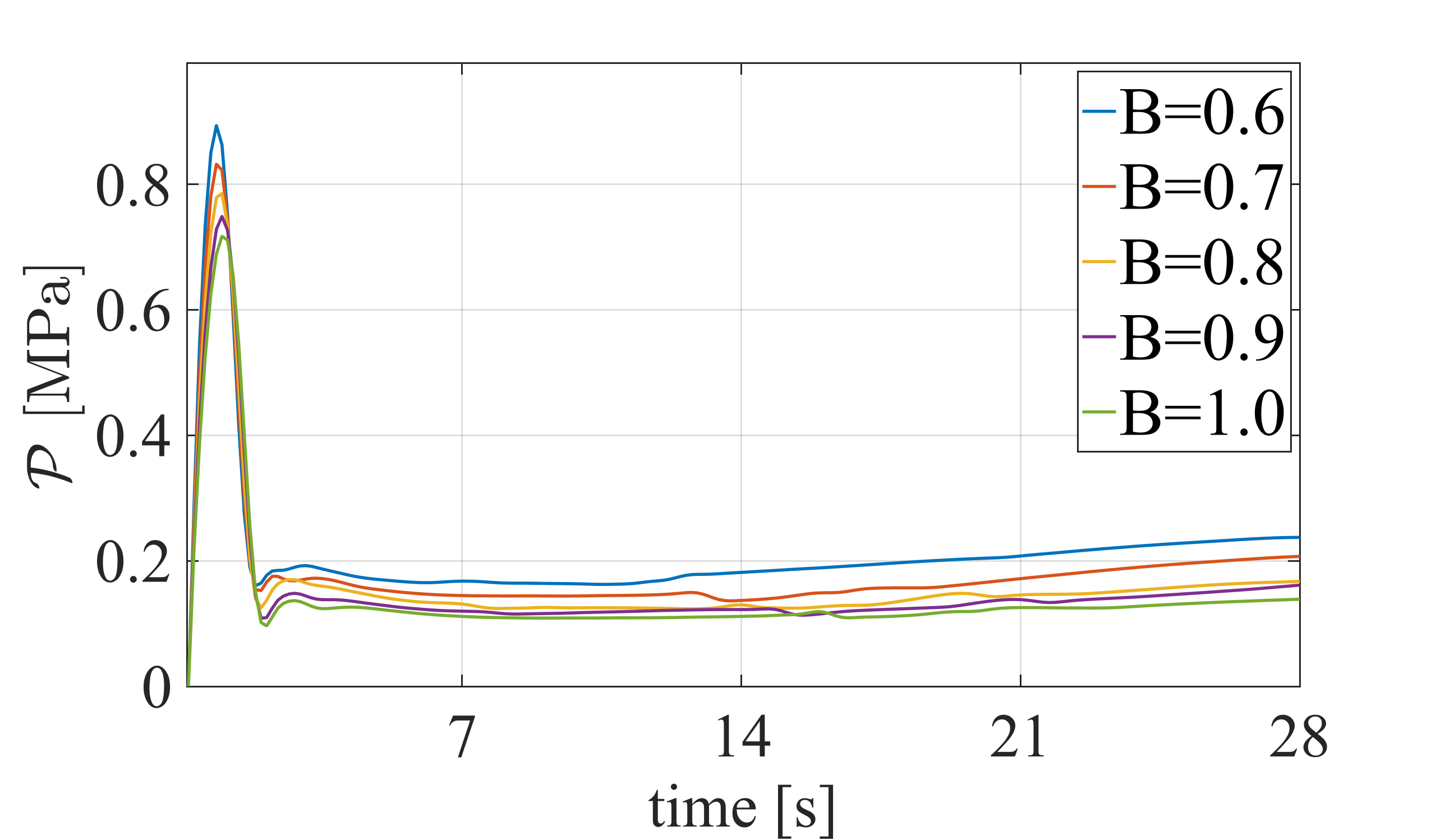}} 
	\subfloat{\includegraphics[width=0.35\textwidth]{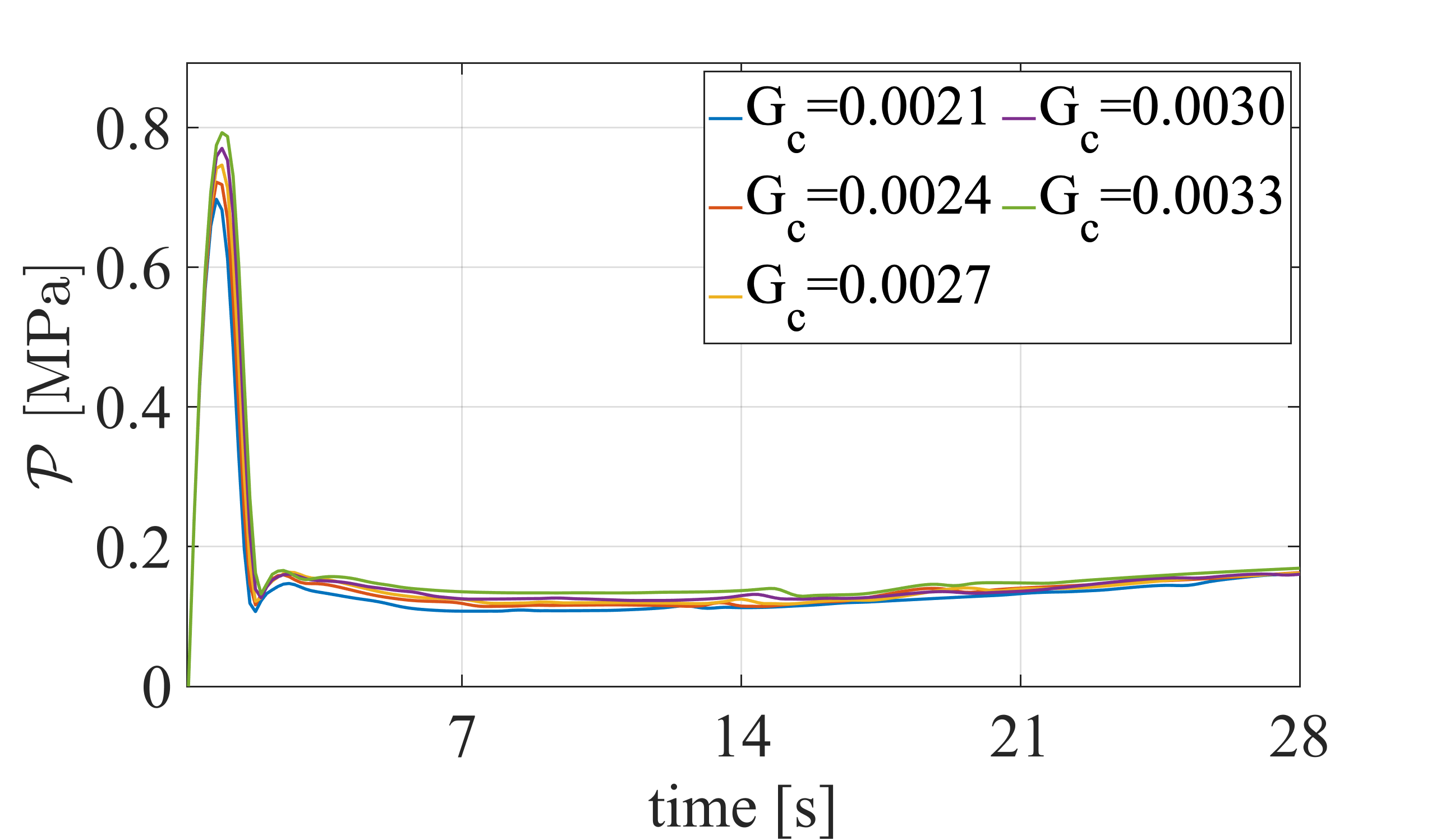}}  
	\subfloat{\includegraphics[width=0.35\textwidth]{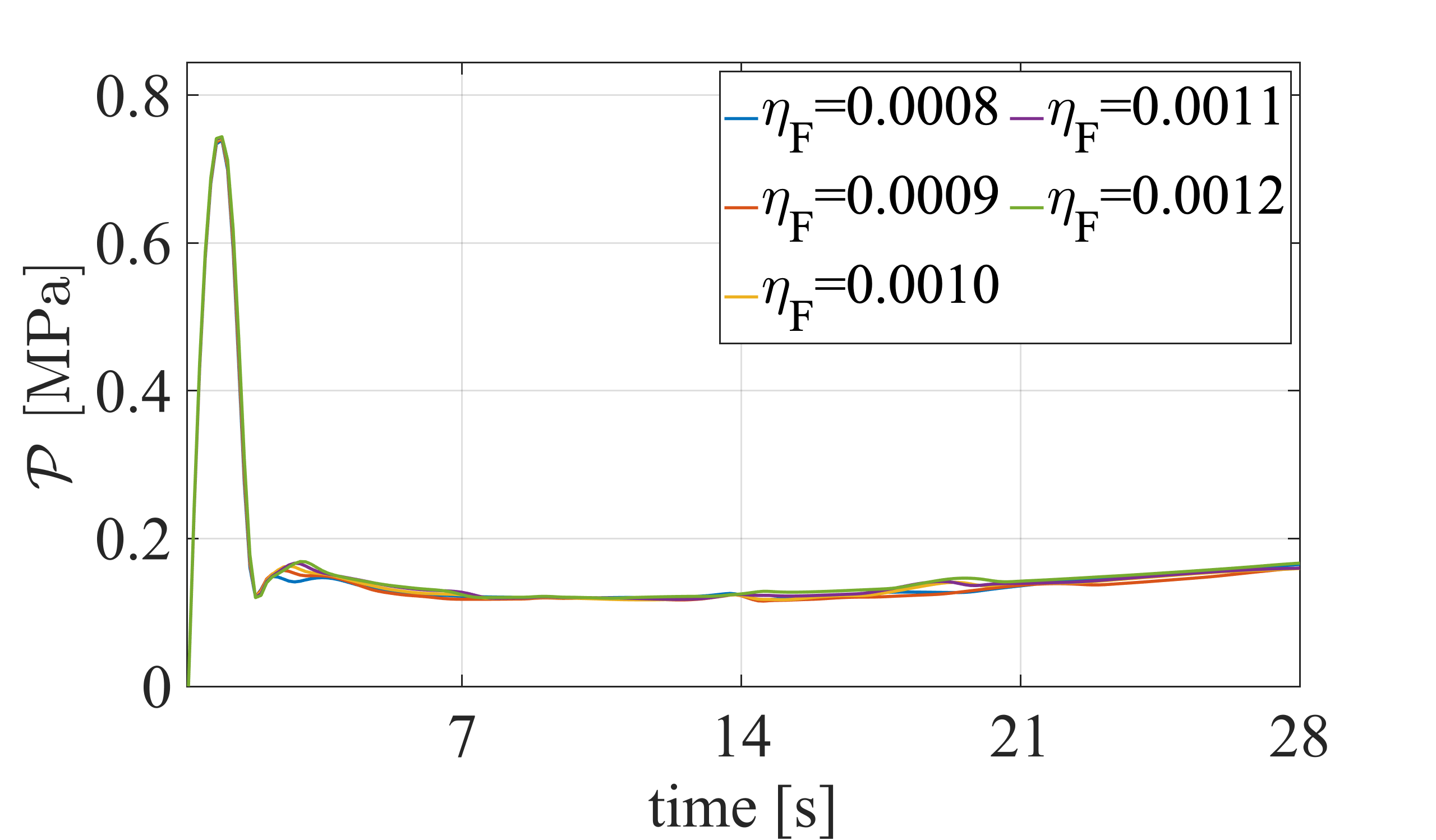}} 
	\caption{Example 2. The maximum pressure value $\PP$  for different values of $\mu$, $K$, $M$, $B$, $G_c$, and $\eta_F$.}
	\label{fig:exam2_curv}
\end{figure}

\begin{figure}[!ht]
	{\includegraphics[clip,trim=0cm 23.3cm 0cm 17cm, width=16.3cm,height=3.6cm]{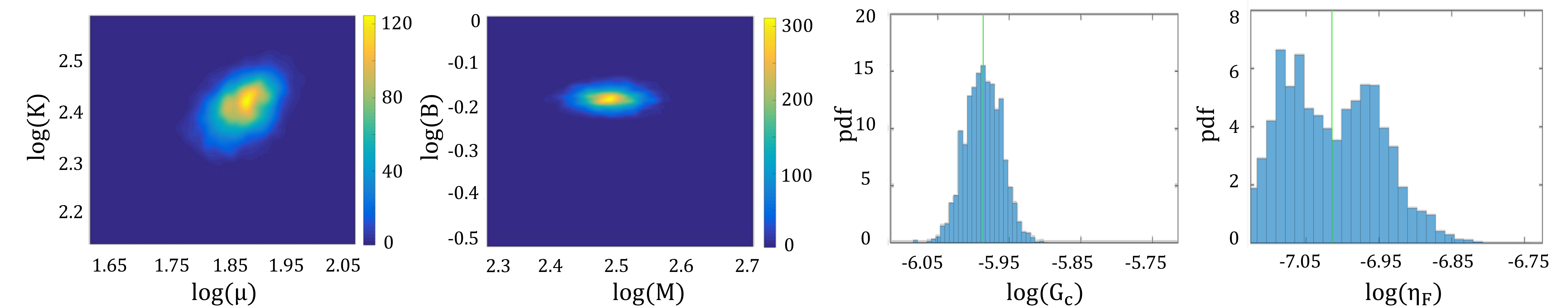}}  
	\caption{Example 2.  From left to right: the posterior density  of  the mechanical parameters, Biot's coefficients/modulus, $G_c$, and $\eta_F$. The green lines are the mean values.}
	\label{fig:exam2_hist}
\end{figure} 

\begin{figure}[ht!]
	\centering
	\includegraphics[width=11cm,height=5.5cm]{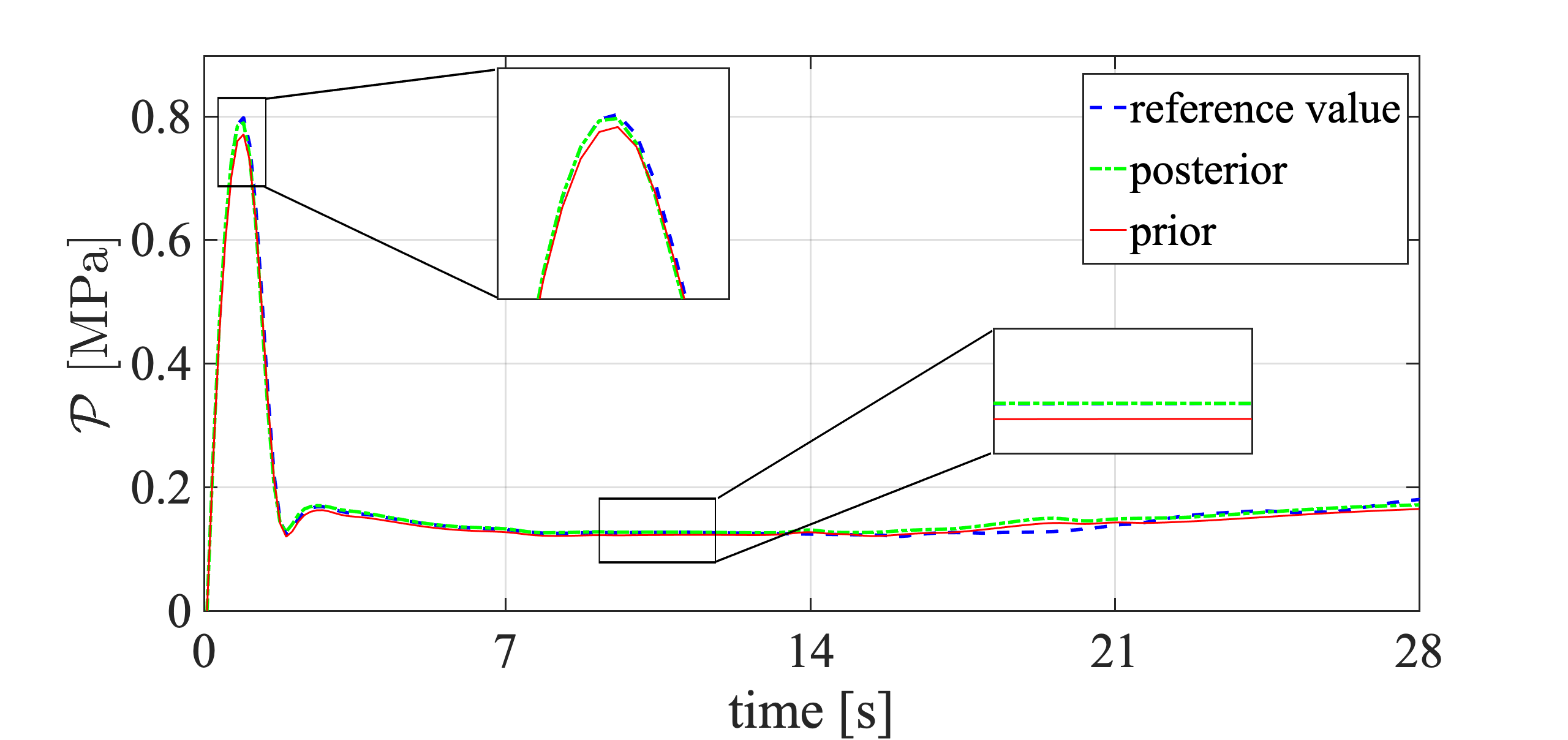}
	\caption{Example 2. A comparison between the maximum pressure (during the injection time) with prior values (red line) and the posterior values (green line). The reference diagram is depicted with a blue line.}
	\label{fig:exam2_post}
\end{figure}


\subsection{Example 3: Transversely isotropic fracture for the poroelastic layered material induced by fluid volume injection }\label{Example3}
The following example deals with transversely isotropic material responses induced by the fluid volume injection. The boundary value problem is given in Figure \ref{example3-a}(a). 
We use identical poroelastic material parameters to observe hydraulic fracture response that is given in Table \ref{material-parameters}.
Here, the domain is divided identically into two vertical layers (see Figure \ref{example3-a}(a)), with thickness $A=40~m$ such that layer 1 and layer 2 are enforced with unidirectional fibers with different orientations which are inclined under an angle  $\phi=+60^\circ$ and $\phi=-60^\circ$ with respect to the $x$-axis
of a fixed Cartesian coordinate system. We set penalty-like parameter by $\beta^{i}_a=\chi^{i}_a:=200$ with $i=(1,2)$ and letting $\beta_g=\chi_g=0$.

Similar as before, at the boundary $\partial_D\calB$, all the displacements are fixed in both directions and the fluid pressure is set to zero. A constant fluid flow of $\bar{f} = 0.004\; m^2/s$ is injected in $\calC$. Fluid injection $\bar{f}$ continues until failure for $T = 45$ second with time step $\Delta t = 0.1$ second during the simulation.

\begin{figure}[!]
	\centering
	{\includegraphics[clip,trim=3cm 7cm 6cm 7cm, width=15cm]{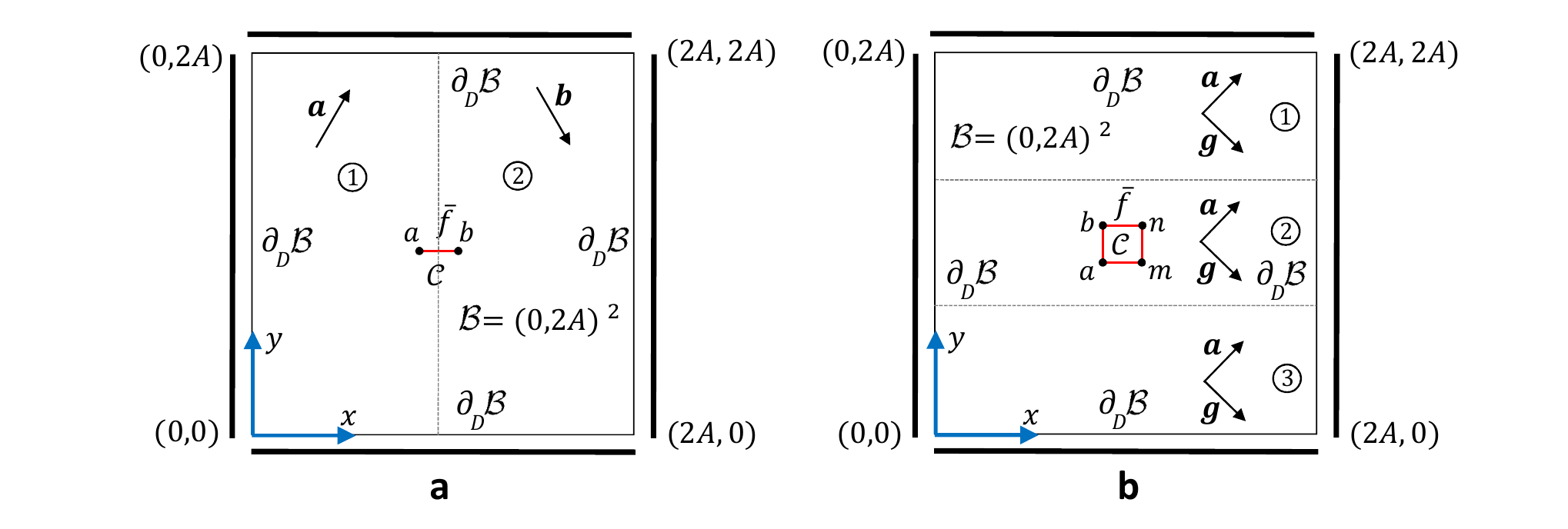}}  
	\caption{Hydraulically induced crack driven by fluid volume injection. (a) Example 3. Layered transversely isotropic poroelastic material, and (b) Example 4. layered orthotropy anisotropic poroelastic material.
	}
	\label{example3-a}
\end{figure}
\subsubsection{Estimation of the penalty parameter}
In this example, we first assume that the penalty parameter is a random field. Therefore, we strive to study the effect of its randomness on each element.\\

The Karhunen-Lo\'eve expansion (KLE) expansion technique is a useful computational method used  to reduce the dimensionality of the random field. Here
the field
$\Lambda$ indicates the penalty parameter (here $\beta_a$ while $\beta_g=0$ is fixed)
and can be decomposed by its mean value  and variation. Denoting the probability density function $\mathbb{P}$ and the random variable $\omega\in\Omega$ belongs the probability space $\Omega$, the covariance function has the form
\begin{align}
	\operatorname{Cov}_{\Lambda}(\bs{x},\bs{y})=\int_{\Omega}\left(\Lambda(\bs{x},\omega)-\Lambda(\bs{x})\right)\left(\Lambda(\bs{y},\omega)-\Lambda(\bs{y})\right)\,\text{d}\mathbb{P}(\omega).
\end{align}
Therefore, the the KL-expansion reads
\begin{align}\label{KL}
	\Lambda (\xx,\omega)=\bar{\Lambda}(\xx)+\sum_{i=1}^{\infty}\sqrt{\psi_i }k_i(\xx) \xi_i (\omega).
\end{align}
The first term indicates the expectation, $k_i$ are the orthogonal eigenfunctions, $\psi_i$ are the corresponding eigenvalues of the eigenvalue problem 
\begin{eqnarray}
\int_{\mathcal{B}}\text{Cov}_{\Lambda} (\xx,\yy)k_i(\yy)~ d\yy=\psi_i k_i(\xx),
\end{eqnarray}
and the $\lbrace \xi_i(\omega)\rbrace$ are mutually uncorrelated random variables satisfy the following condition
\begin{eqnarray}
\mathbb{E}[\xi_i]=0,\hspace{1em}\mathbb{E}[\xi_i \xi_j]=\delta_{ij}.
\end{eqnarray}
Also $\mathbb{E}$ denotes the expected value of the random
variables, and $\delta_{ij}$ denotes the Kronecker product. For the Gaussian random field, we use a Gaussian covariance kernel defined by
\begin{align}
	\operatorname{Cov}_{\Lambda} (\xx,\yy)=\sigma^2\exp \left( -\frac{(x_1-y_1)^2}{\zeta_1} -\frac{(x_2-y_2)^2}{\zeta_2}\right),
\end{align}
where $\zeta_1$, and $\zeta_2$ are the anisotropic correlation lengths and $\sigma$ is the standard deviation.
The infinite series can be truncated to a finite series
expansion (i.e., an $N_\mathrm{KL}$-term truncation) by 
\begin{align}\label{KL1}
	\Lambda (\xx,\omega)=\bar{\Lambda}(\xx)+\sum_{i=1}^{N_{_{\text{KL}}}}\sqrt{\psi_i}k_i(\xx)\xi_i(\omega).
\end{align}
In order to define $N_\mathrm{KL}$, we use the following  criterion
\begin{align}
	\frac{\sum_{i=1}^{N_\mathrm{KL}} \psi_i}{\sum_{i=1}^{\infty} \psi_i}=:\varphi,
\end{align}
to preserve the variance. In this work in order to decompose the random field (penalty parameters) we assume that it has the expectation of 55, the correlation lengths are $\zeta_1=0.1$, $\zeta_2=0.1$, the standard deviation is	$\sigma=5$, and $\varphi=0.95$. The values of the random field in the elements is shown in Figure \ref{example3-a1}.

From now onward, due to dealing with the anisotropic solids, we follow the following parameter estimation procedure.
\begin{enumerate}[I]
	\item  Propose $\theta^*\in$($\beta_a,\,\beta_g$) according to the given distribution to determine the posterior density of the penalty parameters, i.e.,
	\begin{align}
		(\bar{\beta_a},\,\bar{\beta_g})=	\text{DRAM}(\theta^*,\upsilon),
	\end{align}
	where other unknowns $(\upsilon=\{  \mu, K, M, B, G_c, \eta_F\})$
	are according to the true values.
	
	\item Then, use the extracted information from the estimated parameters to identify other unknown values, namely
	\begin{align}
		(\bar{\mu},\,\bar{K},\,\bar{M},\,\bar{B},\,\bar{G_c},\,\bar{\eta_F})=	\text{DRAM}(\theta^*,\bar{\beta_a},\,\bar{\beta_g}),
	\end{align}
	where the candidates $\theta^*  \in \{\mu, K, M, B, G_c, \eta_F\}$ are proposed based on the given distribution. 
\end{enumerate}

\begin{figure}[!]
	\centering
	{\includegraphics[width=0.5\textwidth]{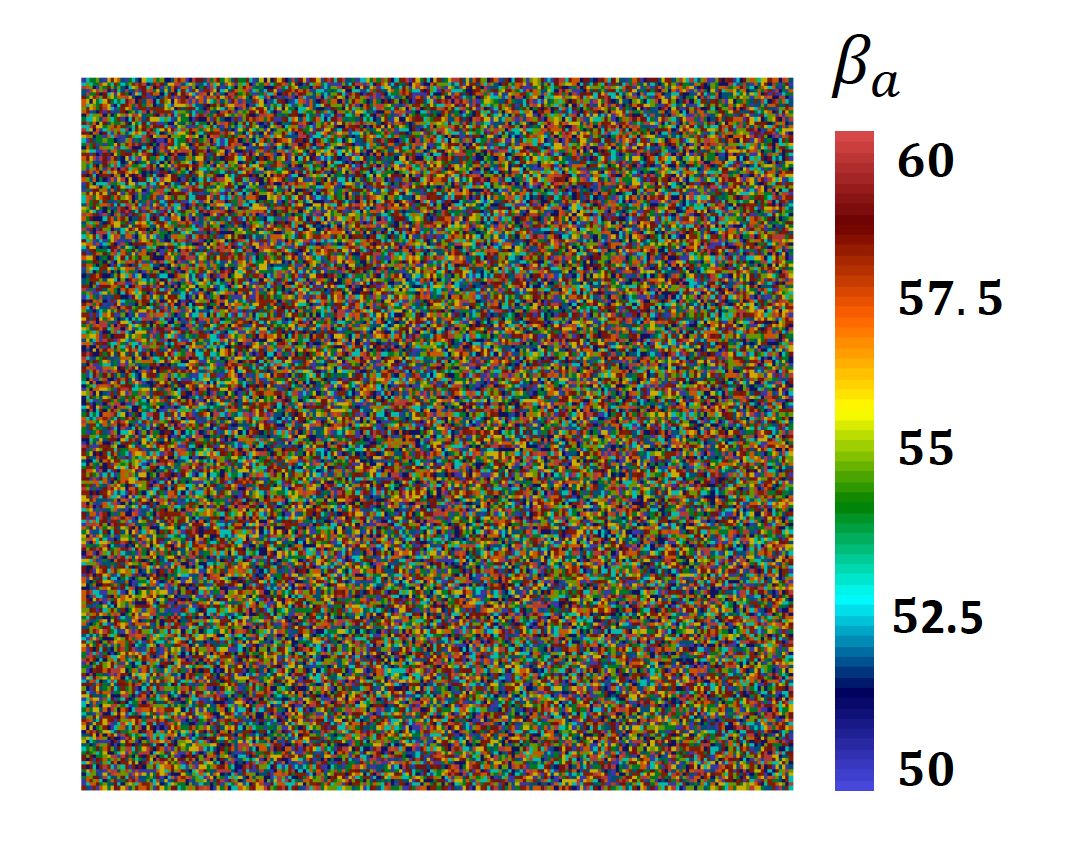}}  
	\caption{The values of the penalty-like parameter $\beta_a$ (prior density) on each element of the domain $\calB$.
	}
	\label{example3-a1}
\end{figure}

\begin{figure}[!]
	\centering
	{\includegraphics[clip,trim=3cm 11cm 3.5cm 0cm, width=16.3cm]{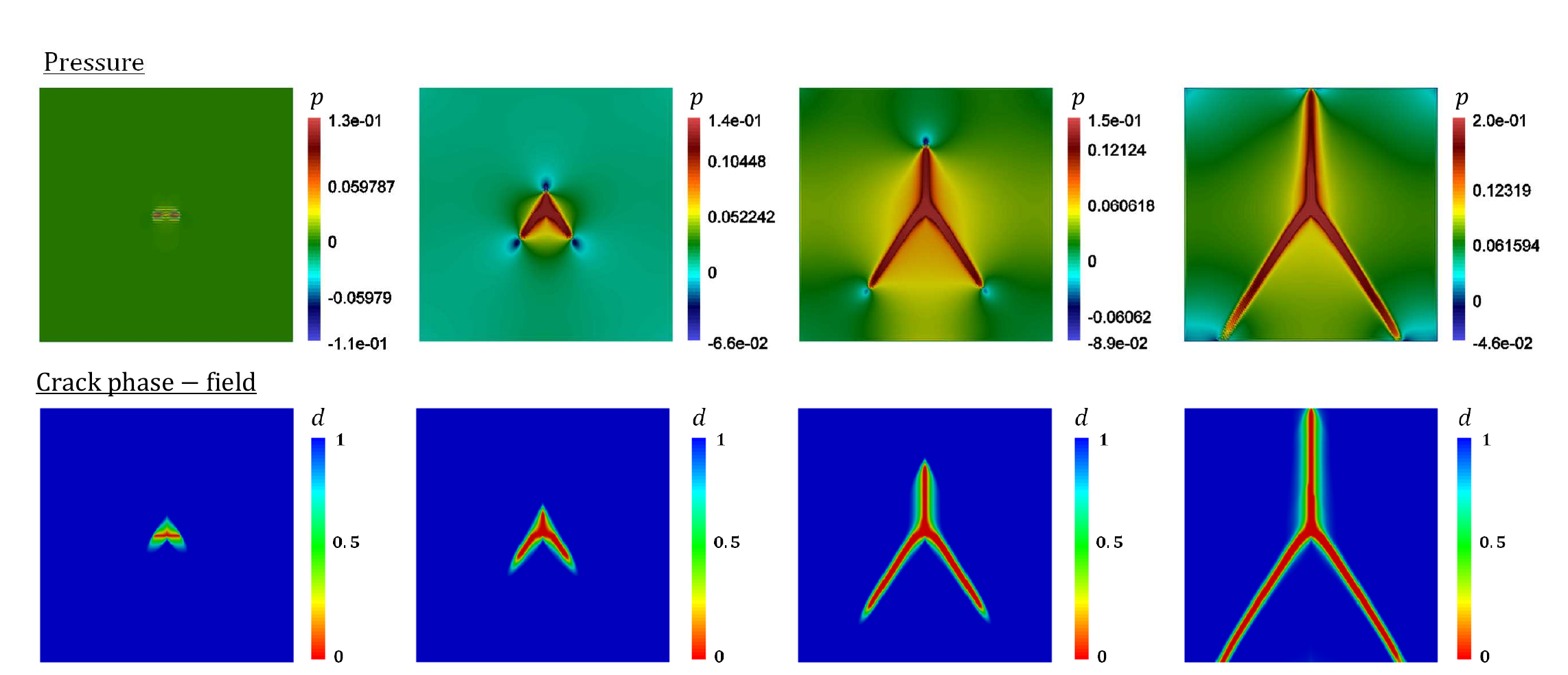}}  
	\caption{Example 3. Reference results of the hydraulically induced crack driven by the fluid volume injection for the layered orthotropy anisotropic poroelastic material. Evolution of the fluid pressure $p$ (first row) and crack phase-field $d$ (second row) for different deformation stages up to final failure at $t=0.1,\,5.1,\,20.1,\,44.5$ seconds.}
	\label{example3_ref}
\end{figure}

Next, we start our analysis by illustrating the computed reference results for different fluid injection time up to final failure related to Figure \ref{example3-a}(a). The fluid pressure $p$ (first row) and crack phase-field $d$ (second row) evolutions are demonstrated in Figure \ref{example3_ref} for four-time steps, i.e., $t=0.1,\,10,\,40,\,80.2$ seconds. The crack initiates at the notch-tips due to fluid pressure increase. The crack profile at first time step (i.e. $t=0.1$ second), evidently intend to the preferential fiber direction within each layer (see diffusivity area in Figure \ref{example3_ref}, second row). Afterward, the crack phase-field propagates toward fiber directions and in some certain time ($t=10$ second), secondary crack initiates through the middle point of the notch induced by fluid injection and then propagates through the interface between two layers. That is an interesting observation (and it is typical for the interface problem) which is shown in  Figure \ref{example3_ref} at $t=40$ second. Primary and secondary crack propagates in three directions towards the boundaries. Same as before, in the fractured zone, $p$ is almost constant due to the increased permeability inside the crack while low fluid pressure in the surrounding is observed. Another impacting factor that should be noted, the highest pressure is aligned with the highest strength direction of the material at each layer, see Figure \ref{example3_ref}, the first row.
\begin{figure}[ht!]
	\centering
	\includegraphics[width=10.78cm,height=5cm]{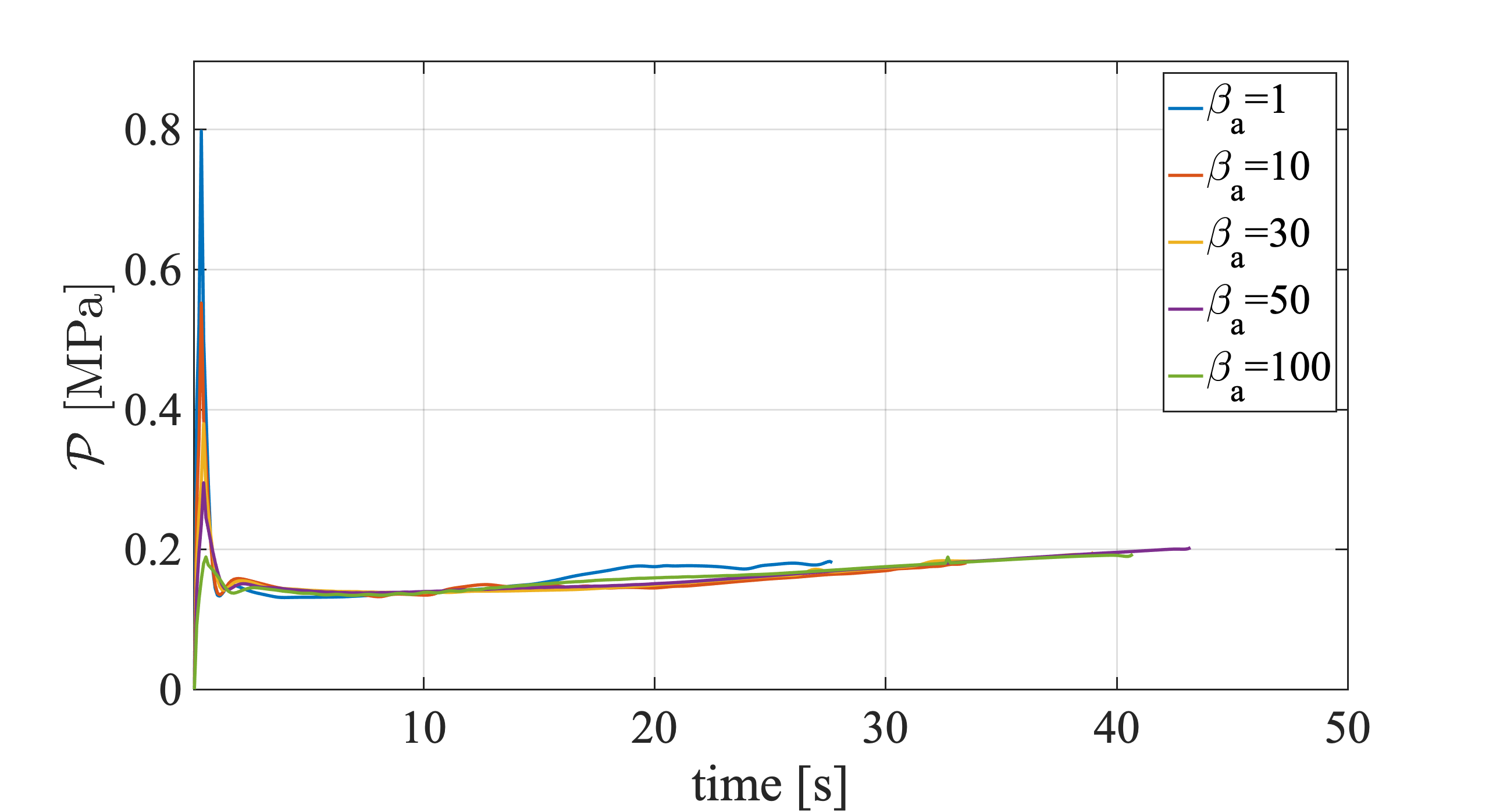}
	\caption{Example 3. The effect of penalty parameter on $\mathcal{P}$.}
	\label{fig:exam3_penalty}
\end{figure}

The effect of different $\beta_a$ on $\mathcal{P}$ is shown in Figure \ref{fig:exam3_penalty}.
As we already mentioned the penalty parameter is assumed a random field, and the KL-expansion used to determine the parameter in the elements. We extract the information to estimate the penalty parameter, where the probability density is shown in Figure \ref{exam3_hist_prnalty}.

\begin{figure}[!]
	\centering
	\includegraphics[width=0.7\textwidth]{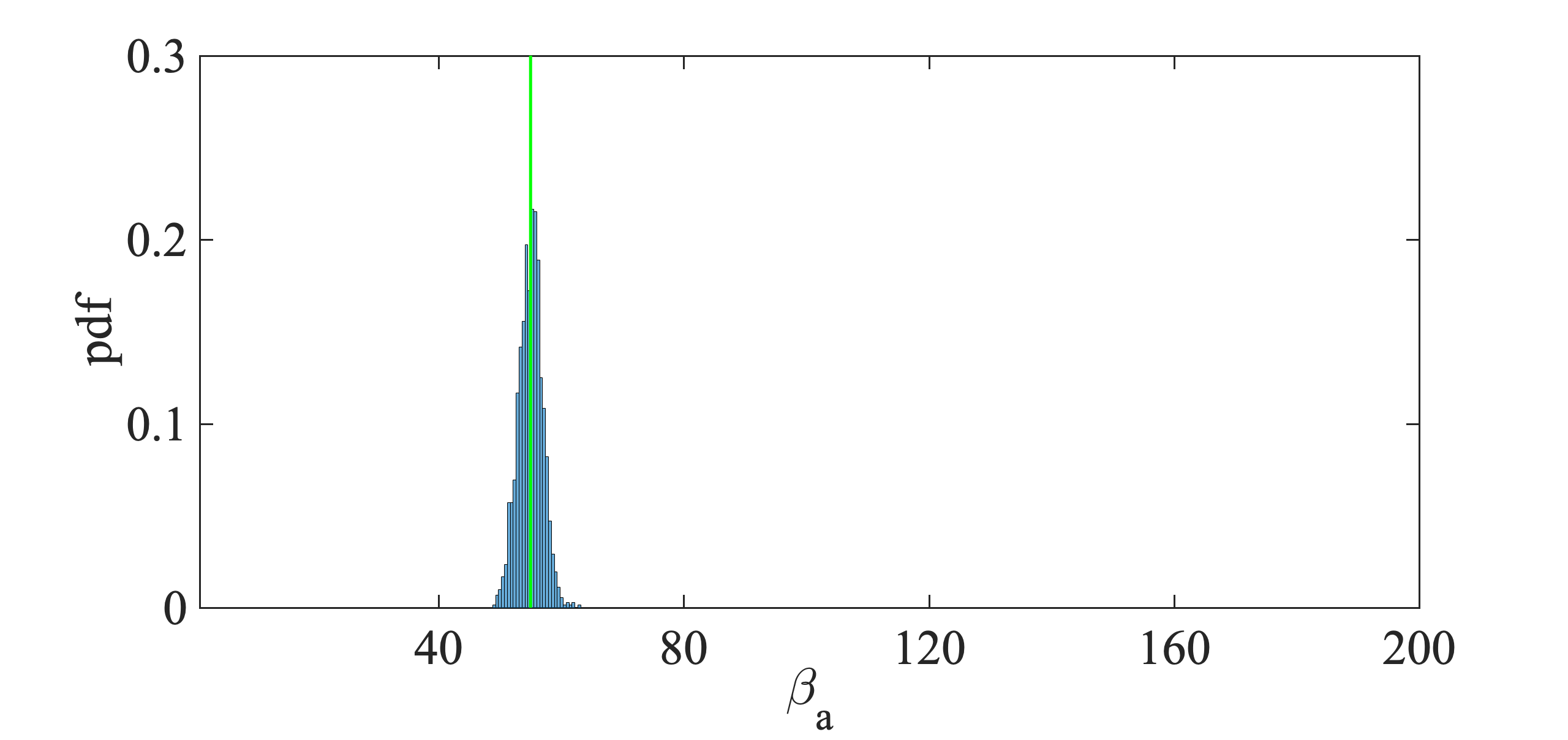}
	\caption{Example 3. The probability density of the penalty parameter.}
	\label{exam3_hist_prnalty}
\end{figure}

Now, we strive to determine the desired values (using the determined $\beta_a$).  The effect of the different values of the parameters on $\mathcal{P}$ is depicted in Figure \ref{fig:exam3_curv} and the obtained posterior densities are shown in Figure \ref{fig:exam3_hist}. 
The different ending point of the curves is due to the impact of the parameters on the crack propagation (reaching the boundary).
Then, we compare the estimated knowledge from the posterior with prior value (see Figure \ref{exam3_post}). Using the posterior information we can estimate the peak point and the pressure ending point precisely.

\begin{figure}[!]
	\subfloat{\includegraphics[width=0.35\textwidth]{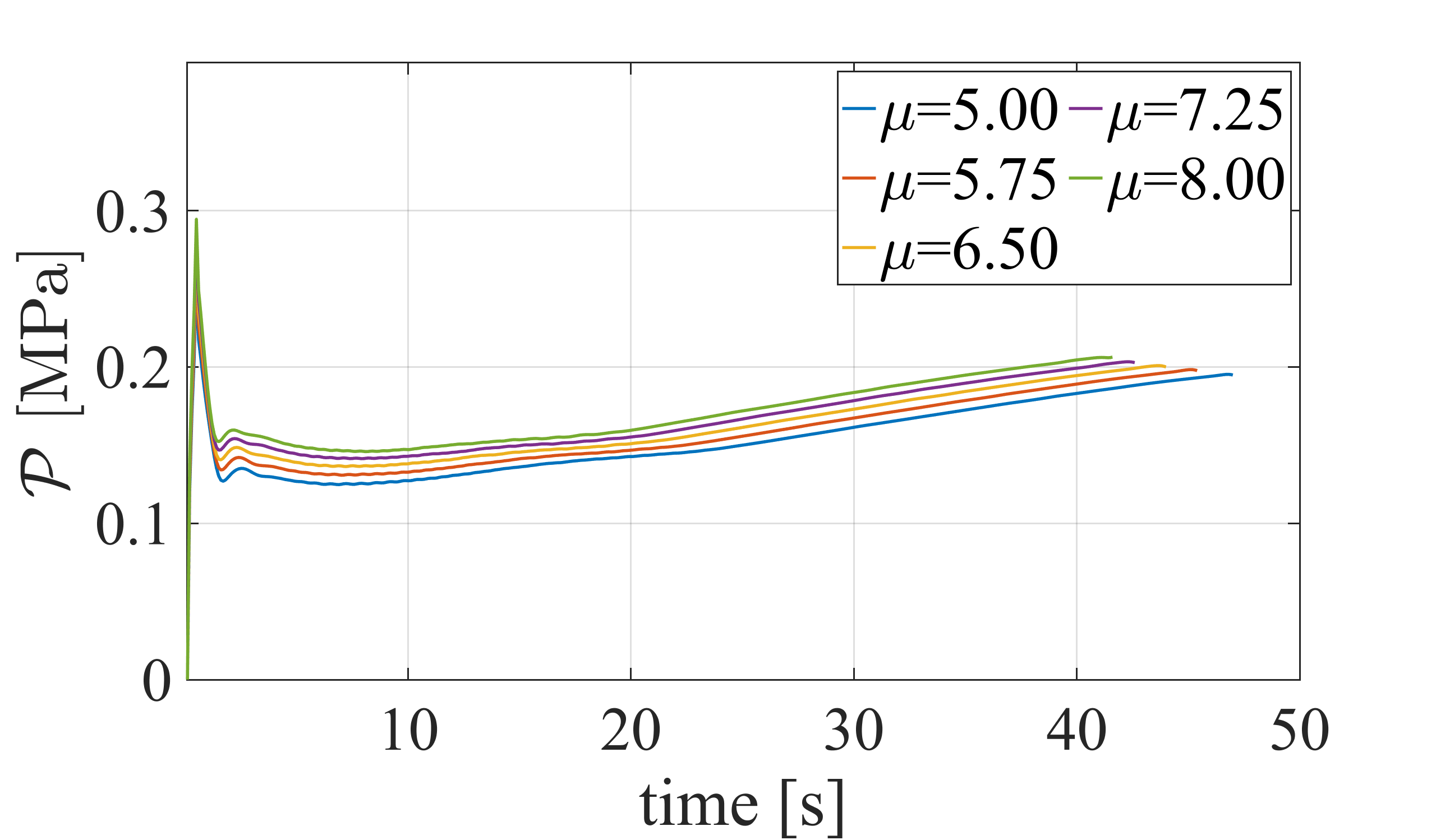}}  
	\subfloat{\includegraphics[width=0.35\textwidth]{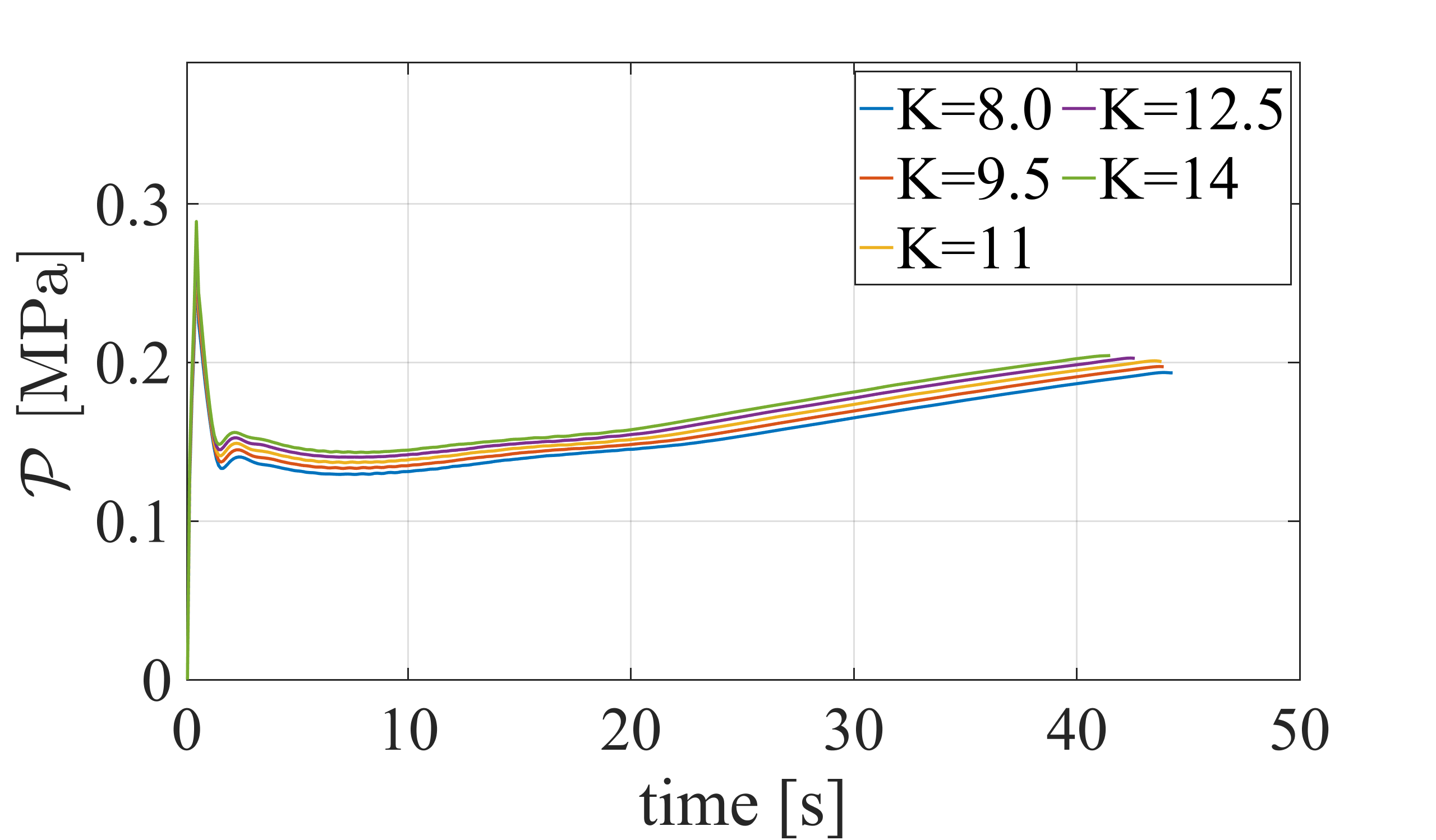}}
	\subfloat{\includegraphics[width=0.35\textwidth]{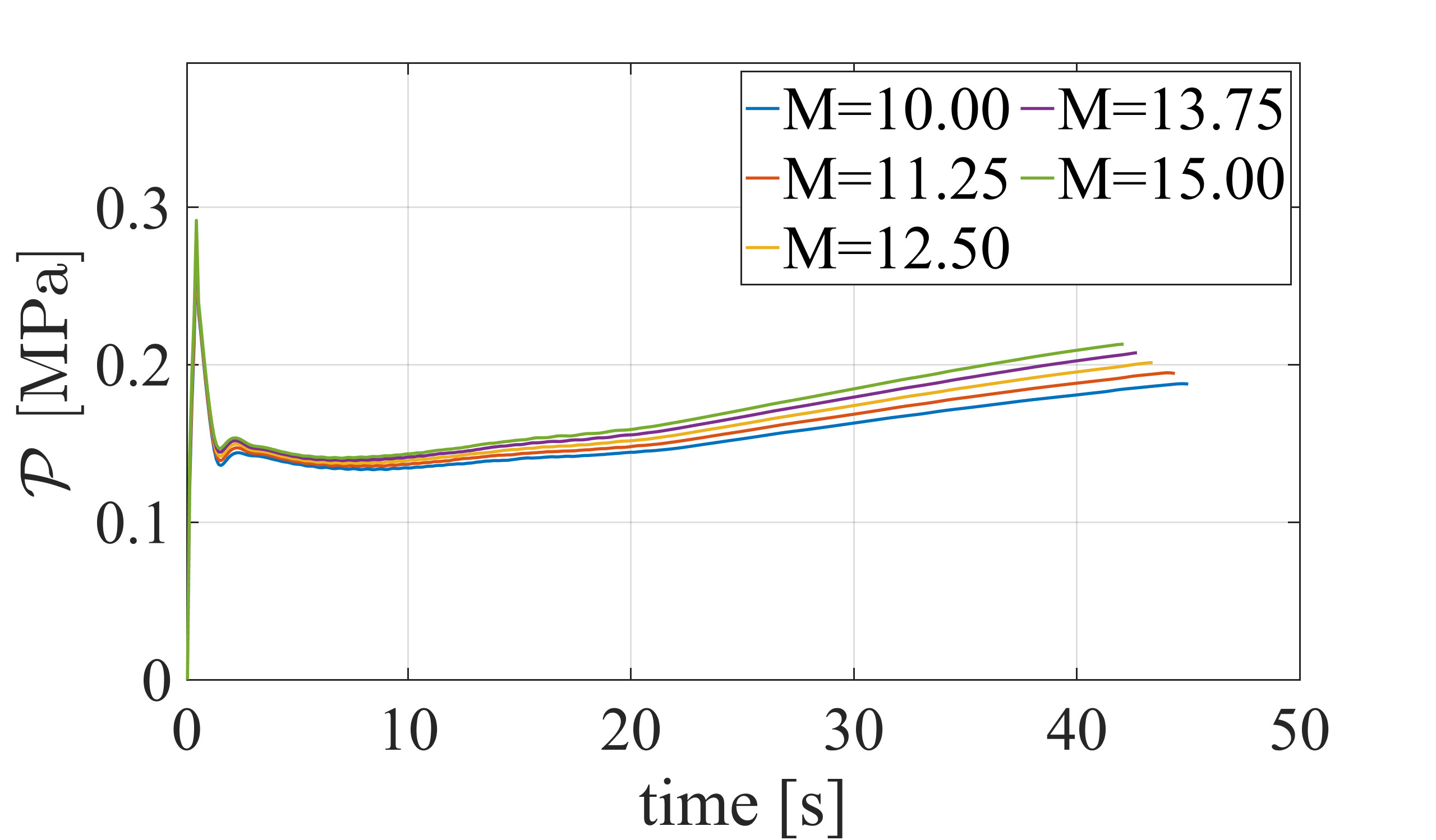}}
	\newline
	\subfloat{\includegraphics[width=0.35\textwidth]{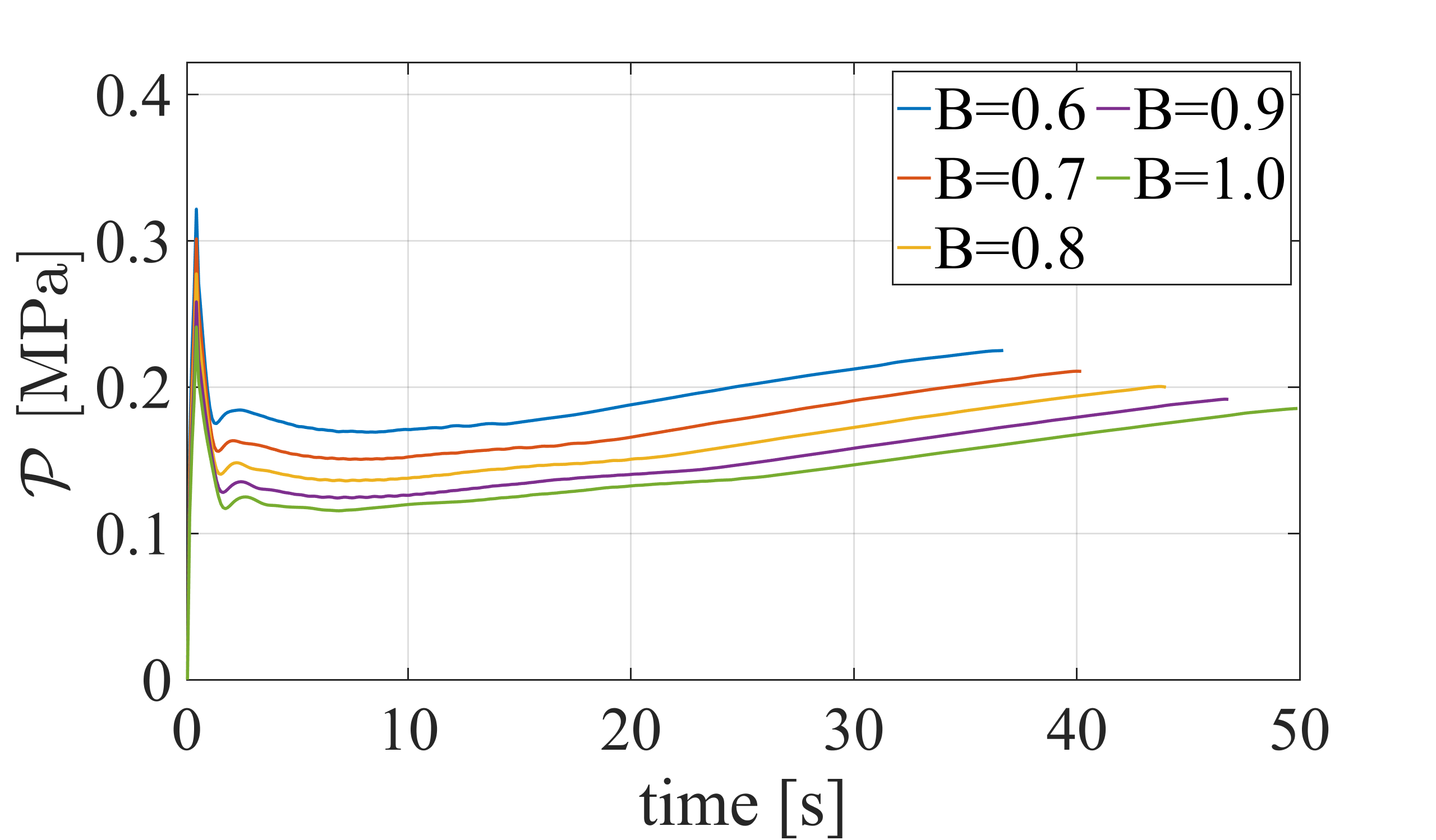}} 
	\subfloat{\includegraphics[width=0.35\textwidth]{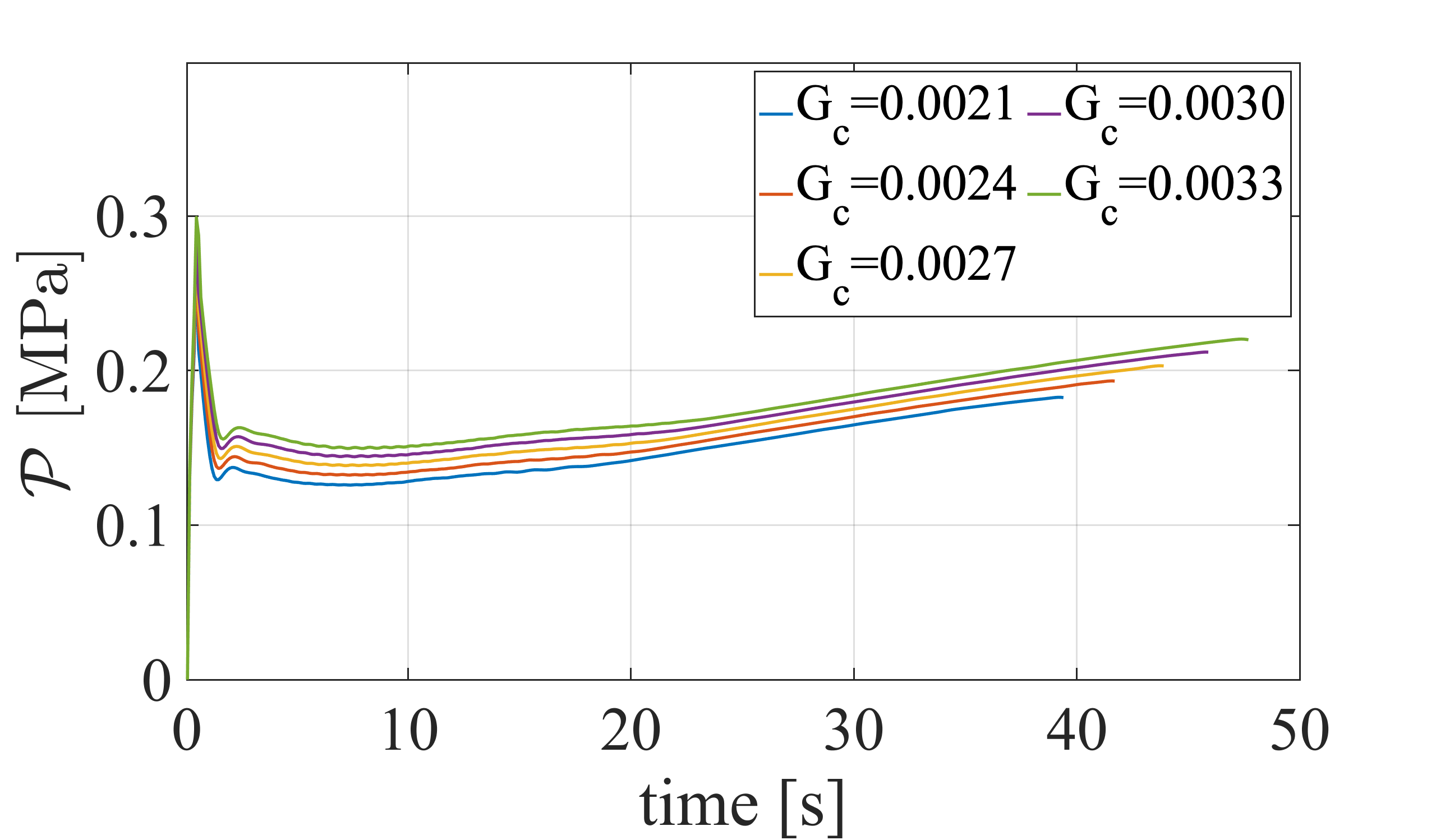}}  	\subfloat{\includegraphics[width=0.35\textwidth]{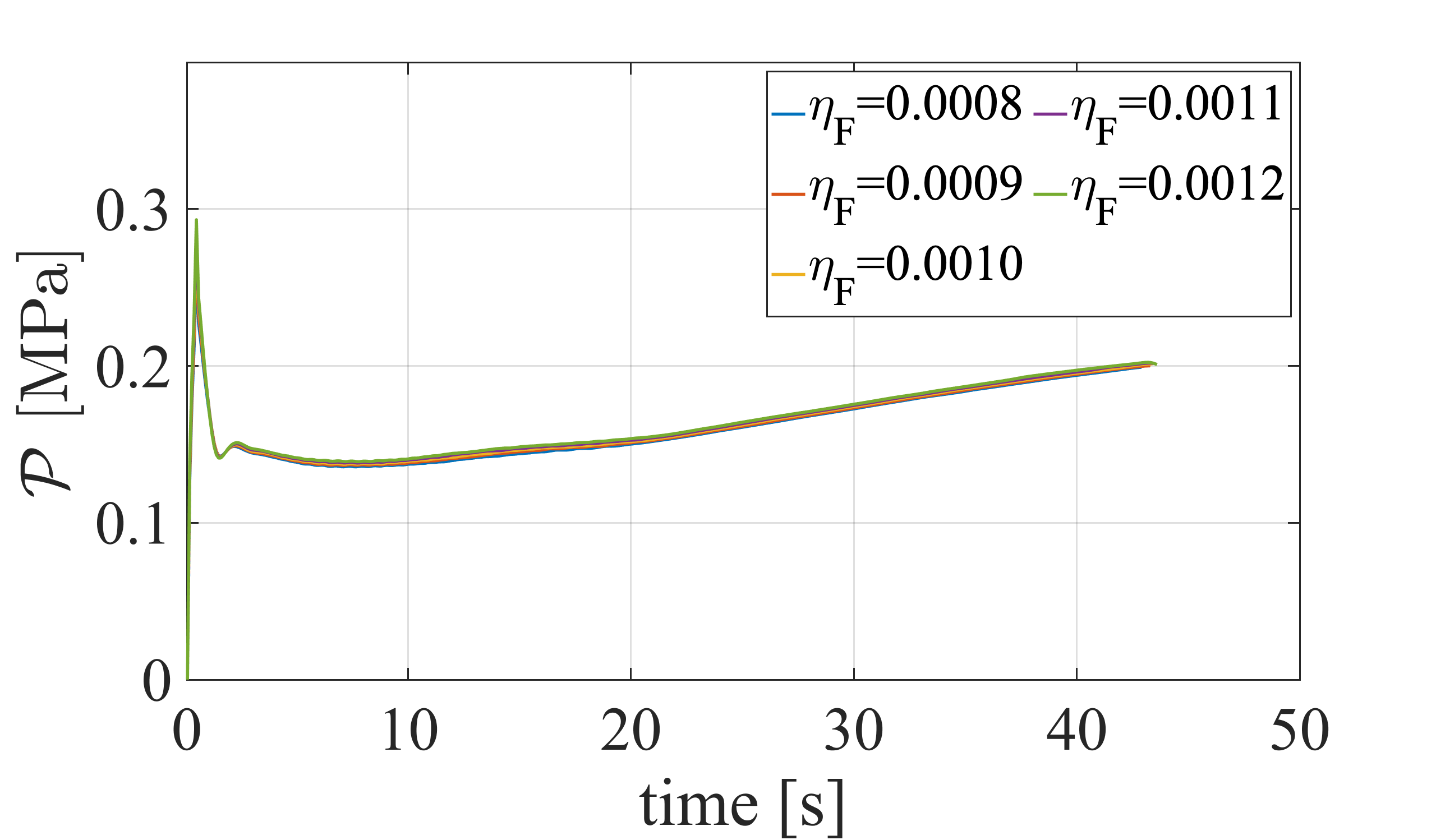}} 
	\caption{Example 3.  The maximum pressure value $\PP$  for different values of $\mu$, $K$, $M$, $B$, $G_c$, and $\eta_F$.}
	\label{fig:exam3_curv}
\end{figure}

\begin{figure}[!ht]
	\centering
	{\includegraphics[clip,trim=0cm 23.3cm 0cm 17cm, width=16.3cm,height=3.6cm]{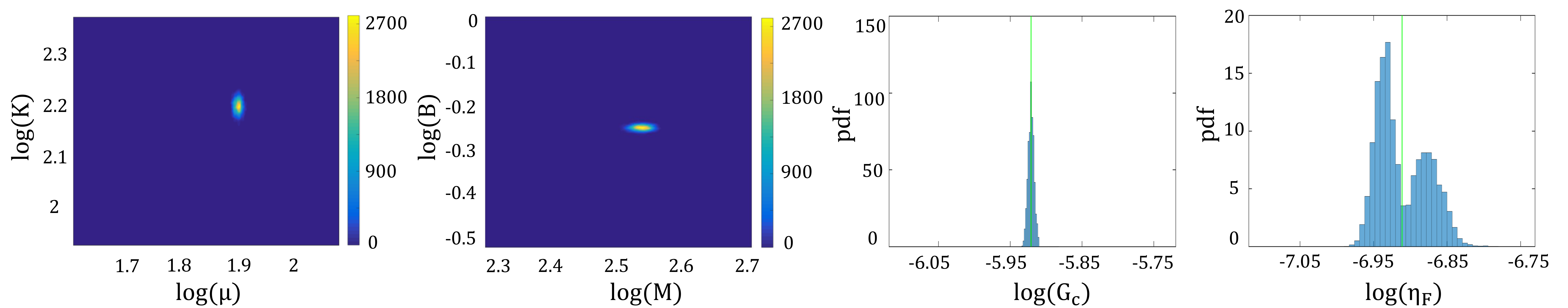}}  
	\caption{Example 3.  From left to right: the posterior density  of  the mechanical parameters, Biot's coefficients/modulus, $G_c$, and $\eta_F$. The green lines are the mean values.}
	\label{fig:exam3_hist}
\end{figure}

\begin{figure}[ht!]
	\centering
	\includegraphics[width=10.78cm,height=5cm]{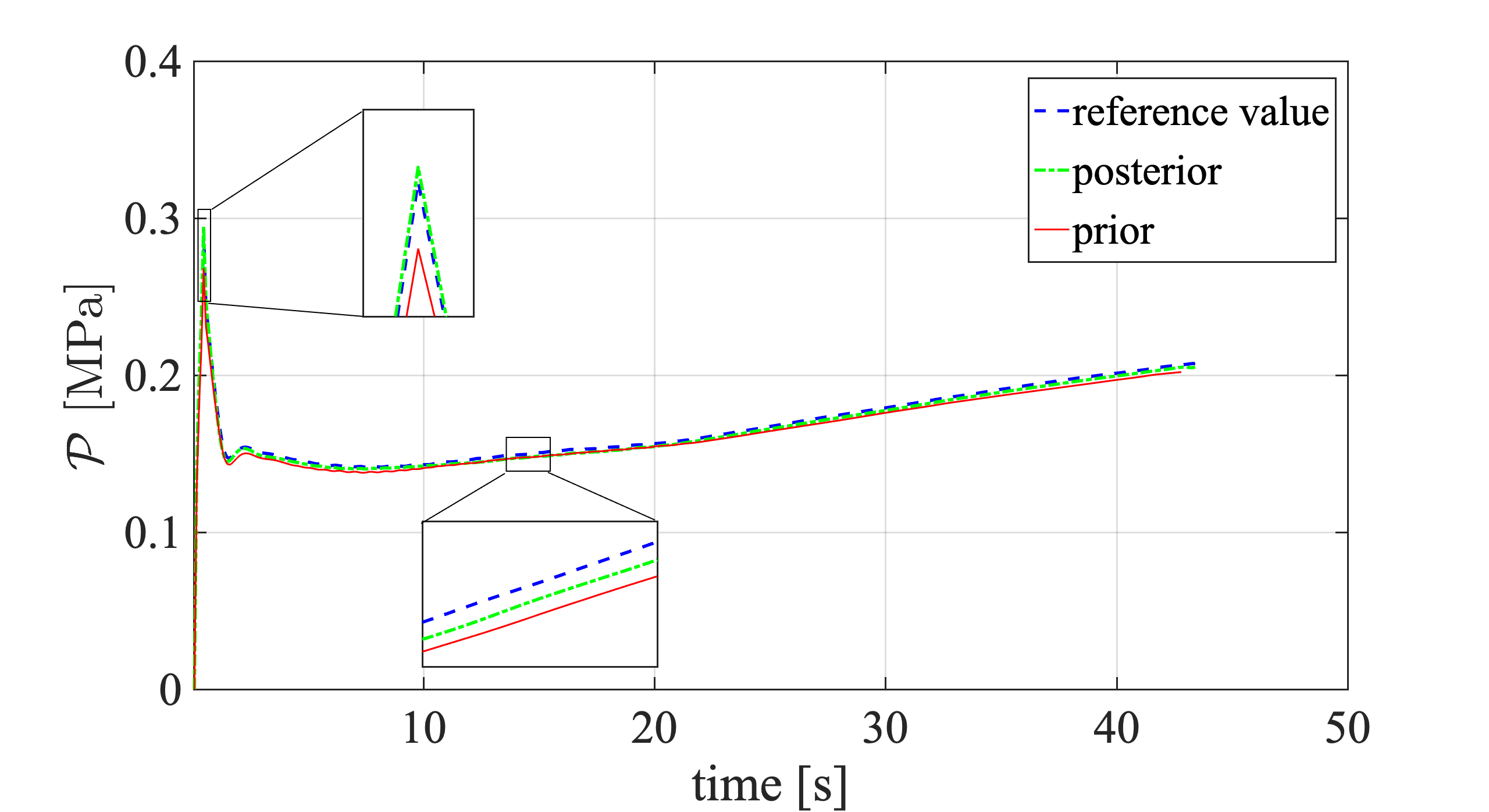}
	\caption{Example 3. A comparison between the maximum pressure (during the injection time) with prior values (red line) and the posterior values (green line). The reference diagram is depicted with a blue line.}
	\label{exam3_post}
\end{figure}
\subsection{Example 4: Orthotropy anisotropic fracture for a poroelastic layered material induced by fluid volume injection}\label{Example4}

The last numerical test is concerned with orthotropic anisotropic poroelastic materials with two families of fibers induced by the fluid volume injection. The layered boundary value problem is given in Figure \ref{example3-a}b. The material properties are used the same as before. A constant fluid flow of $\bar{f} = 0.004\; m^2/s$ is injected in $\calC$ until failure for $T = 13$ second with time step $\Delta t = 0.1$ second during the simulation.

Here, the domain is divided identically into three horizontal layers (see Figure \ref{example3-a}b) with a thickness of $2A/3=26.66~m$. Each layer of the poroelastic material is reinforced with two orthogonal unidirectional fibers embedded in the matrix, namely $\bm{a}$ and $\bm{g}$. The preferential fiber direction in each layer of the laminate is given by the structural director $\bm{a}$ and $\Bg$ which is inclined by $\theta = +30^\circ$ and $\theta = -60^\circ$, respectively,  with respect to the $x$-axis of a fixed Cartesian coordinate system. Here, penalty-like parameters act as a material parameter, hence families of fibers with higher penalty-like parameters respond stiffer, and hence anisotropic response is oriented in that direction. Specifically, we define the mismatched ratio between two families of fibers and denoted by $\xi$ which is given by

\begin{equation}
\xi^{i}:=\frac{\beta^{i}_a}{\beta^{i}_g}=\frac{\chi^{i}_a}{\chi^{i}_g}, \quad \text{with} \quad i=(1,2,3).
\end{equation}
Herein, $i$ refers to the layer number within the domain, see Figure \ref{example3-a}b, with $(\beta_a,\chi_a)$ and $(\beta_g,\chi_g)$ are corresponding to the $\bm{a}$ and $\bm{g}$, respectively, see \req{euler-eq-d} and \req{eq24}. Thus, if $\xi^{i}>1$ means $\bm{a}$ is stiffer than $\bm{g}$ then crack orientation is in direction parallel to $\bm{a}$. Otherwise, if $\xi^{i}<1$ means $\bm{g}$ is stiffer than $\bm{a}$ then crack orientation is in direction parallel to $\bm{g}$. To formulate the fracture process, the stiffer fiber is set with a larger value of the penalty-like parameter in the \req{euler-eq-d} and \req{eq24}. Therefore, in the following, we considered four different cases.

\begin{itemize}
	
	\item \textbf{Case a.} In the first case, we consider the isotropic hydraulic fracture and hence we fixed and set $\beta_a=\chi_a=\beta_g=\chi_g=0$ to recover isotropic formulation. The fluid pressure $p$ and the crack phase-field $d$ evolutions are shown in Figure \ref{example4_ref} and \ref{example41_ref} first row, respectively, for four-time steps, i.e., $t=0.1,\,5.1,\,10.1,\,12.2$ seconds. Here, the crack initiates at the notch-tips (where we have a singularity-like shape) due to fluid pressure increase. Then, it propagates about $45^\circ$ and in the very final stage, see Figure \ref{example41_ref} the first row at $t=12.2$, we observed the crack branching near boundaries induced by the fluid injection. 	
	
	\item \textbf{Case b.} In this and the next two cases, for the first and third layers, we assume fiber $\bm{g}$ is stiffer than $\bm{a}$ while within the second layer $\bm{a}$ is stiffer than $\bm{g}$. Hence, we set $\beta^{1}_g=\beta^{2}_a=\beta^{3}_g=10$ and $\beta^{1}_a=\beta^{2}_g=\beta^{3}_a=0.5$. The same values are also holds for the ($\chi^{i}_a$,$\chi^{i}_g$) with $i=(1,2,3)$. Table \ref{Ex5_Table} summarizes penalty parameters and the mismatch ratio for each layer. The fluid pressure $p$ and crack phase-field $d$ evolutions are shown in Figure \ref{example4_ref} and \ref{example41_ref} second row, respectively, for four-time steps, i.e., $t=0.1,\,3.1,\,5.1,\,11.4$ seconds. Here, the crack initiates at the notch-tips due to fluid pressure increase. The crack profile at first time step (i.e. $t=0.1$ second), propagate toward the preferential fiber direction $\beta_a$ in the second layer. Afterwards, the crack phase-field initiates and then propagates along the interface between layers 2 and 1 and, accordingly, layer 2 and 3, see Figure \ref{example41_ref}, second row. This crack profile occurs mainly because the material is not very stiff in the preferential direction such that crack propagates toward the fibers. Thus, it continuous along the interface between two layers.
	
	\item \textbf{Case c.} In this case, we set $\beta^{1}_g=\beta^{2}_a=\beta^{3}_g=50$ and $\beta^{1}_a=\beta^{2}_g=\beta^{3}_a=2.5$. The same values are also holds for the ($\chi^{i}_a$,$\chi^{i}_g$) with $i=(1,2,3)$. By means of Table \ref{Ex5_Table}, in layer 2 where $\xi^{i}>1$, crack propagate in a direction of $\bm{a}$, otherwise $\bm{g}$, e.g. layers 1 and 3. 	The fluid pressure $p$ and crack phase-field $d$ evolutions are shown in Figure \ref{example4_ref} and \ref{example41_ref} third row, respectively, for four-time steps, i.e., $t=0.1,\,3.1,\,5.1,\,12$ seconds. Here, the crack initiates at the notch-tips due to fluid pressure increase. The crack profile at first time step (i.e. $t=0.1$ second), evidently intend to the preferential fiber direction within each layer (see diffusivity area in Figure \ref{example41_ref}, third row). Afterwards, the crack phase-field propagates toward fiber directions which is inclined under $\theta = +30^\circ$, because we have a situation $\xi>1$, meaning that the stiffer response is observed in $\Ba$ orientation of the poroelastic material ($t=3.1$ second). In some certain time ($t=5.1$ second), the crack direction is changed toward $\Bg$ ($\theta = -60^\circ$) since $\xi<1$. A secondary crack initiates along the interface between two layers which is depicted in Figure \ref{example41_ref} at $t=12$ second. Additionally, it can be grasped the highest pressure is aligned with the highest strength direction of the material at each layer, see Figure \ref{example4_ref}, the third row.	
	
	\item \textbf{Case d.} Here, we set $\beta^{1}_g=\beta^{2}_a=\beta^{3}_g=200$ and $\beta^{1}_a=\beta^{2}_g=\beta^{3}_a=10$. The same values are also holds for the ($\chi^{i}_a$,$\chi^{i}_g$) with $i=(1,2,3)$. The fluid pressure $p$ and crack phase-field $d$ evolutions are shown in Figure \ref{example4_ref} and \ref{example41_ref} last row, respectively, for four-time steps, i.e., $t=0.1,\,3.1,\,5.1,\,10.4$ seconds. The first important observation is that the crack surface, precisely, follows the mismatch ratio $\xi^{i}$ criteria indicated in Table \ref{Ex5_Table}. Another impacting factor that should be noted that the crack surface is very similar with Case c, except in this case, a secondary crack is not anymore observed. This is mainly because the material behaves much stiffer in each fiber direction compared to Case c.
	
\end{itemize}

\begin{figure}[!]
	{\includegraphics[clip,trim=3.3cm 0cm 3.3cm 0cm, width=17.0cm]{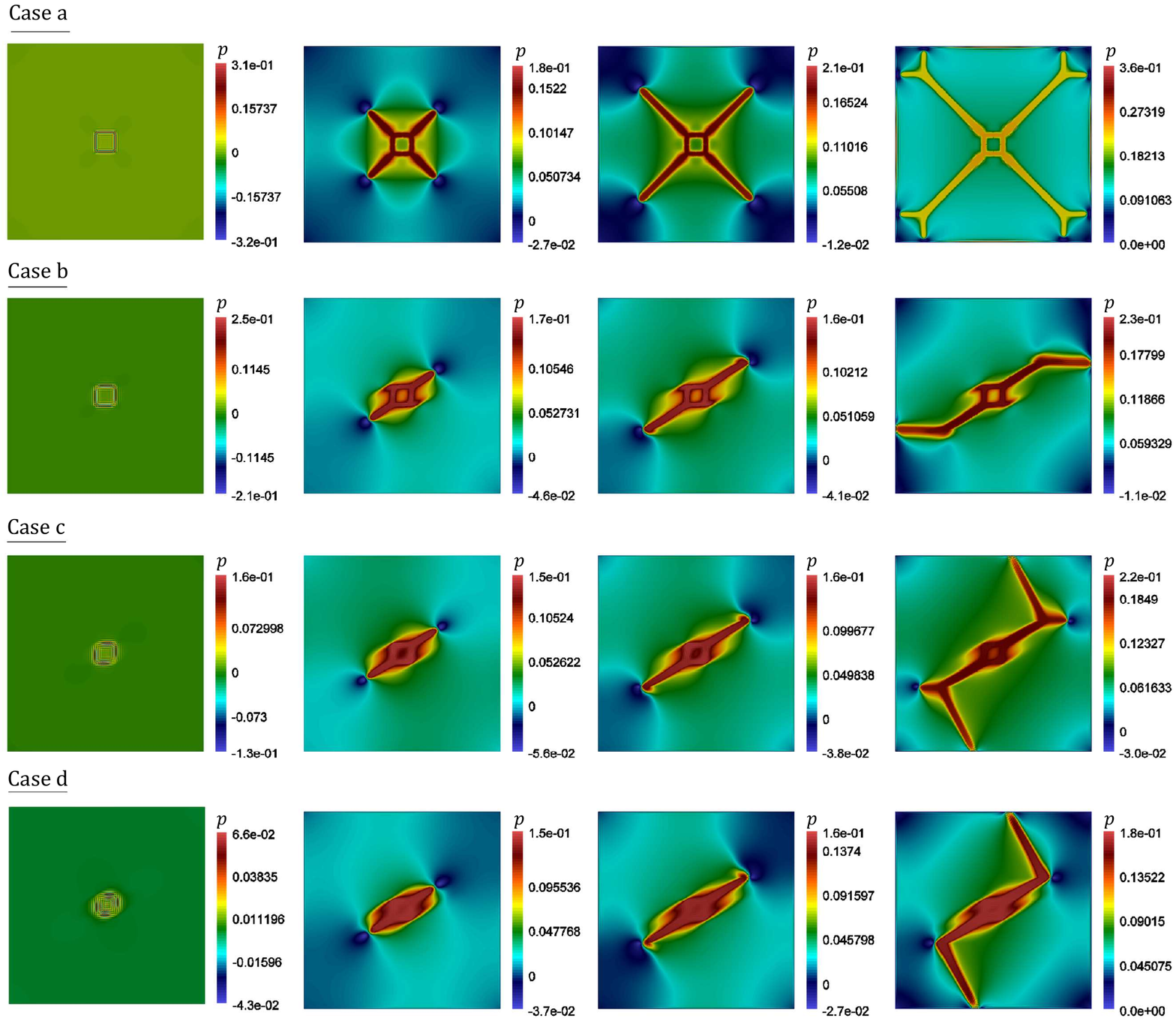}}   
	\caption{Example 4. Reference results of the hydraulically induced crack driven by the fluid volume injection for the layered anisotropic poroelastic material. Evolution of the fluid pressure $p$ for the Case a (first row) that is isotropic setting for different deformation stages up to final failure at $t=0.1,\,5.1,\,10.1,\,12.2$ seconds; Case b (second row) at $t=0.1,\,3.1,\,5.1,\,11.4$ seconds; Case c (third row) at $t=0.1,\,3.1,\,5.1,\,12$ seconds and Case d (fourth row) at $t=0.1,\,3.1,\,5.1,\,10.4$ seconds.}
	\label{example4_ref}
\end{figure}

\begin{figure}[!]
	\centering
	{\includegraphics[clip,trim=3.3cm 0cm 3.3cm 0cm, width=17.5cm]{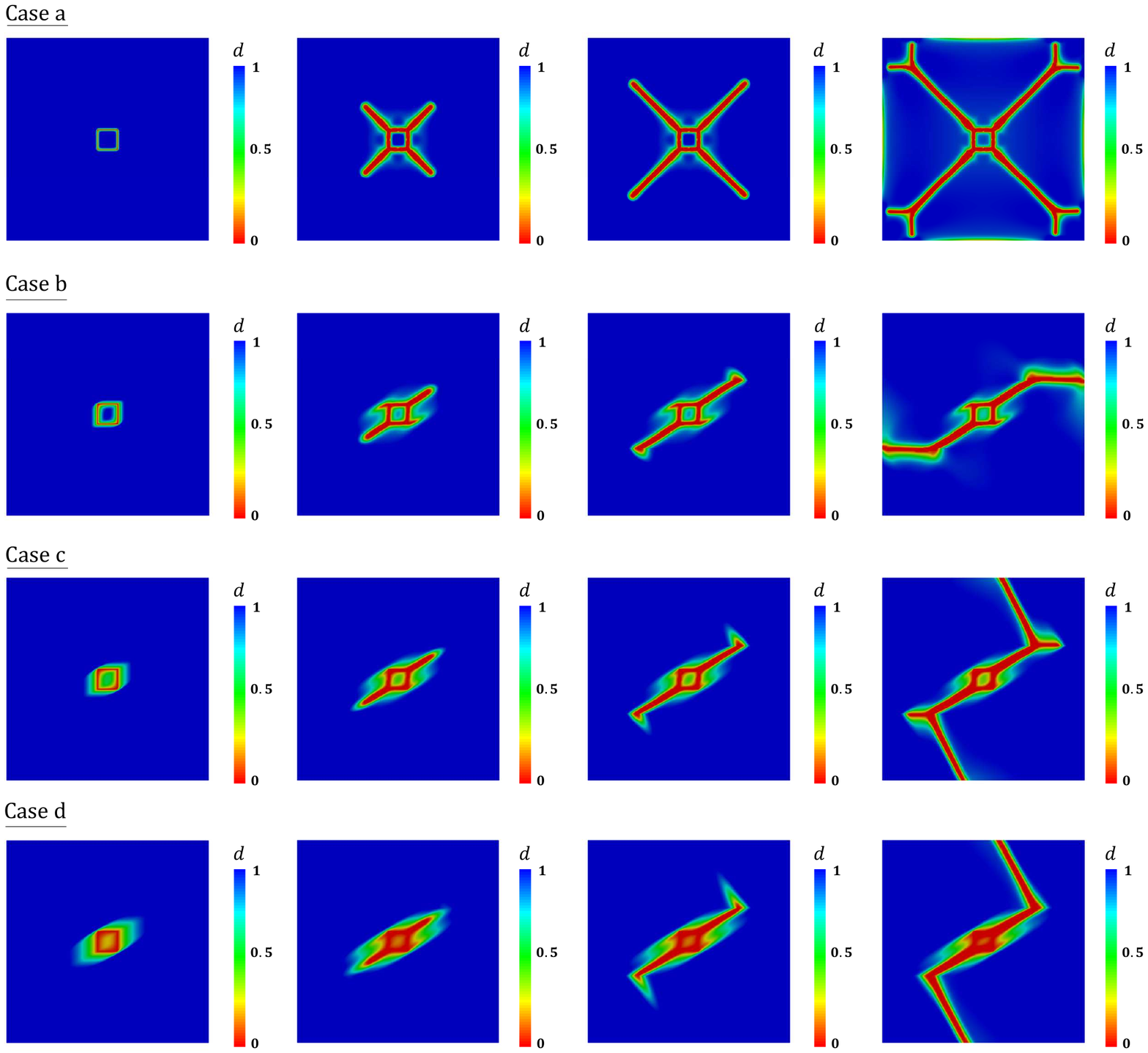}}   
	\caption{Example 4. Reference results of the hydraulically induced crack driven by the fluid volume injection for the layered anisotropic poroelastic material. Evolution of the crack phase-field $d$ for the Case a (first row) that is isotropic setting for different deformation stages up to final failure at $t=0.1,\,5.1,\,10.1,\,12.2$ seconds; Case b (second row) at $t=0.1,\,3.1,\,5.1,\,11.4$ seconds; Case c (third row) at $t=0.1,\,3.1,\,5.1,\,12$ seconds and Case d (fourth row) at $t=0.1,\,3.1,\,5.1,\,10.4$ seconds. }
	\label{example41_ref}
\end{figure}

\begin{table}[!ht]
	\centering
	\begin{tabular}{lllll}
		\hline
		$(\beta_a,\beta_g,\xi)$\; & $layer\;1$ &   $layer\;2$ & $layer\;3$ \\[2mm]
		\hline
		\textbf{Case a}       &  (0,\,0,\,--)         & (0,\,0,\,--)      &  (0,\,0,\,--)   \\[2mm]
		\textbf{Case b}       &  (0.5,\,10,\,0.05)   & (10,\,0.5,\,20)  &  (0.5,\,10,\,0.05)   \\[2mm]
		\textbf{Case c}       &  (2.5,\,50,\,0.05)   & (50,\,2.5,\,20)  &  (2.5,\,50,\,0.05)   \\[2mm]
		\textbf{Case d}       &  (10,\,200,\,0.05)   & (200,\,10,\,20)  &  (10,\,200,\,0.05)   \\[2mm]
		\hline
	\end{tabular}
	\caption{Example 4. The mismatched ratio $\xi$ between two families of fibers $(\bm{a},\bm{g})$. Large value for $\xi>1$, e.g. $\xi=20$, results stiffer response in direction of structural director $\bm{a}$ otherwise $\bm{g}$.} 
	\label{Ex5_Table}
\end{table}

Next, in order to investigate the accuracy of the Bayesian framework, for this example, we consider Case b and also Case d. The first main goal is to identify the penalty parameter in all layers. As we already mentioned, in each region, different $\beta_a$ and $\beta_g$ are employed where the prior densities and the true values are shown in Table \ref{prior}. Figure \ref{fig:exam41_penalty} shows the joint probability density of both penalty parameters. As shown for both parameters a narrow distribution is obtained.

Using the estimated penalty parameters (the posterior densities), we present the effect of the unknown parameters on $\mathcal{P}$ in Figure \ref{fig:exam41_curv} and the posterior densities (joint/marginal) are depicted in Figure \ref{fig:exam41_hist}. Finally, we used the obtained information to compared the posterior and prior densities as shown in Figure \ref{fig:exam41_post}(a).

We use the same Bayesian framework for Case d and strive to estimate the penalty parameters. Again, the prior densities and the true values are shown in Table \ref{prior}. The effect of the parameters in different layers on $\boldsymbol{u}$ and $p$ can be observed in the last line of Figure \ref{example4_ref} and Figure \ref{example41_ref}, respectively. Figure \ref{fig:exam42_penalty} shows the posterior density for the three-layer, were compared to Case b a wider distribution is obtained. Using the extracted mean values (of the posterior density), we solve the system of equations to study the effect of the penalty parameter on the pressure curve as shown in Figure \ref{fig:exam42_penalty}.
We consider which parameter is more influential on the pressure pattern in Figure \ref{fig:exam42_curv}. Here, Biot's coefficient is the most effective, although $\eta_F$ does not have a noticeable impression. We employ the Bayesian inversion to estimate the posterior density of the parameters and show the results in Figure \ref{fig:exam42_hist}. Finally, a comparison between the prior and posterior values with the reference observation is drawn in Figure \ref{fig:exam41_post}(b).

\begin{figure}[!]
	\centering
	\subfloat{\includegraphics[clip,trim=2cm 0cm 1.5cm 1cm, width=5cm,height=4.2cm]{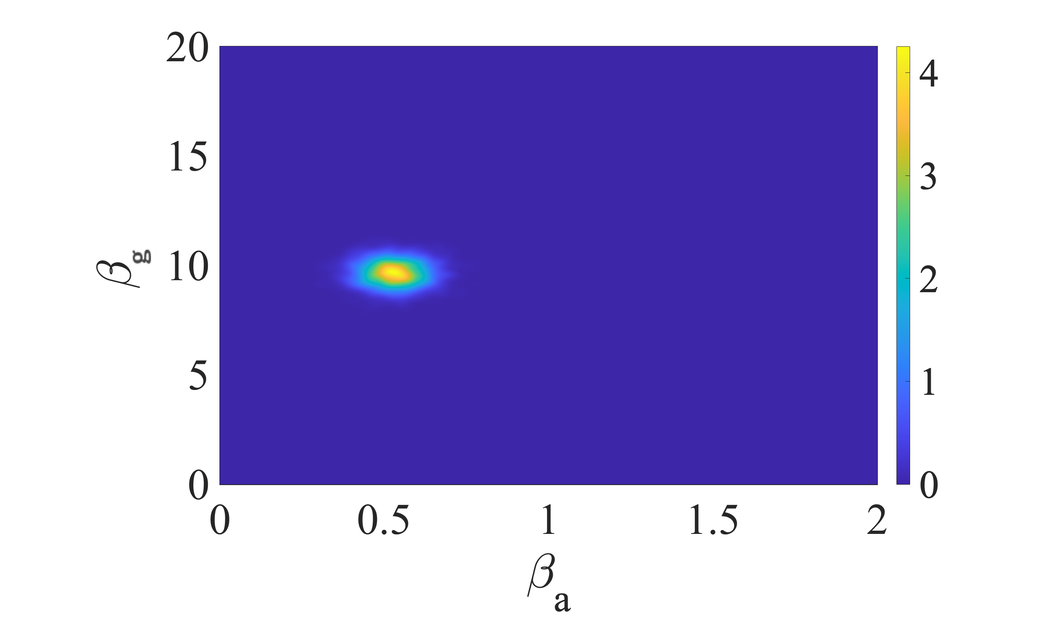}}
	\hspace{-0.2cm}
	\subfloat{\includegraphics[clip,trim=2cm 0cm 1.5cm 1cm, width=5cm,height=4.2cm]{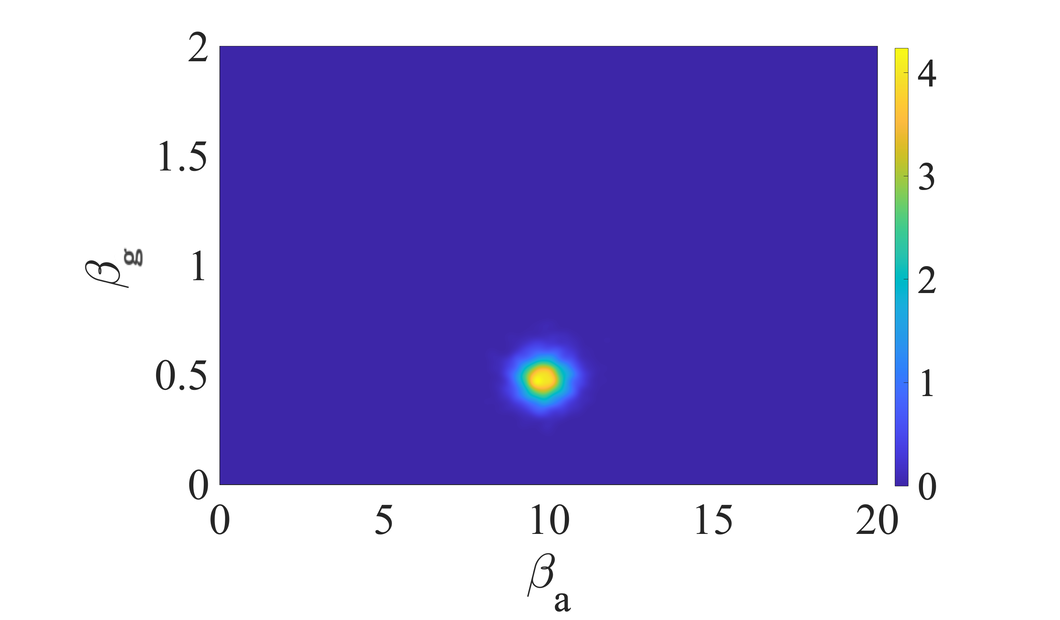}}
	\subfloat{\includegraphics[clip,trim=2cm 0cm 1.5cm 1cm, width=5cm,height=4.2cm]{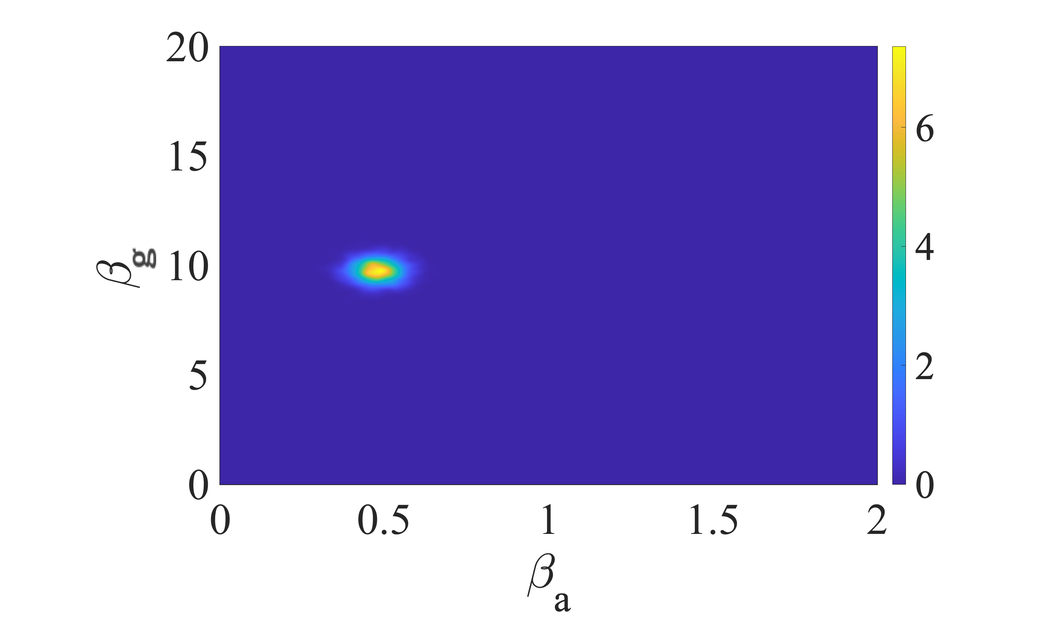}}
	\caption{Example 4 (Case b).  The joint probability density of the penalty parameters ($\beta_a$ and $\beta_g$) in layer 1 (left), layer 2 (middle), and layer 3 (right).}
	\label{fig:exam41_penalty}
\end{figure}

\begin{figure}[!b]
	\subfloat{\includegraphics[width=0.35\textwidth]{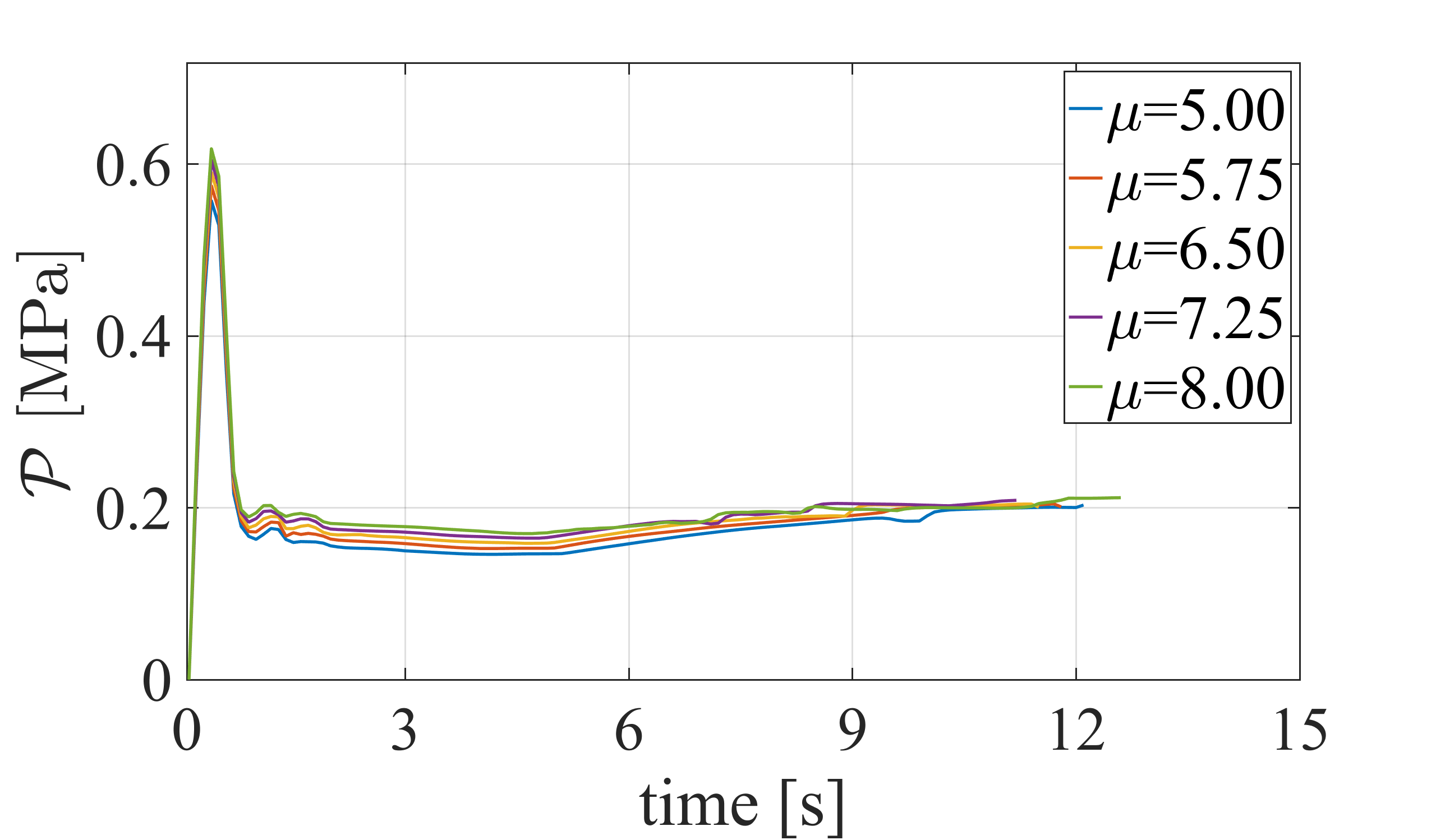}} 
	\subfloat{\includegraphics[width=0.35\textwidth]{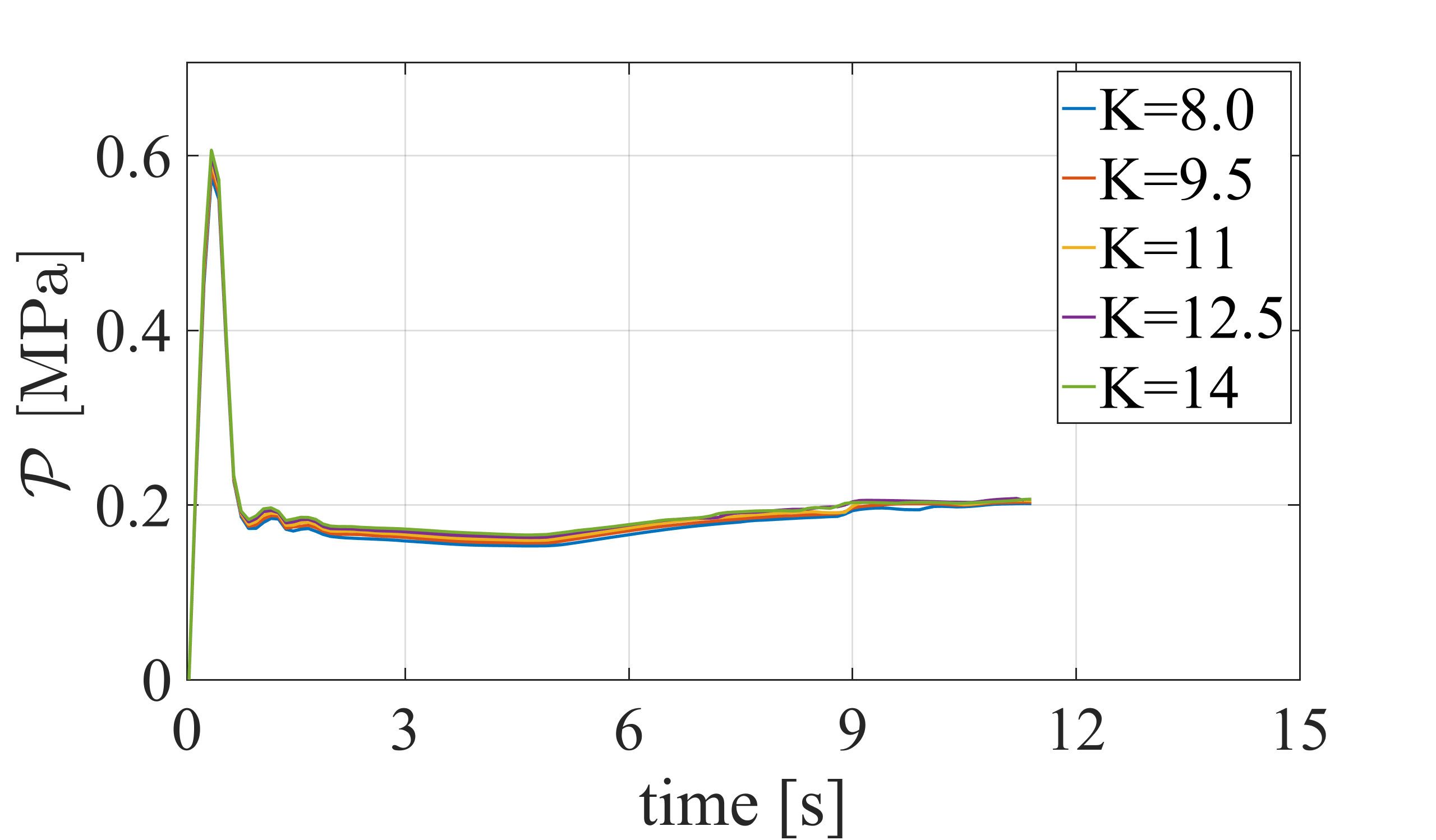}}
	\subfloat{\includegraphics[width=0.35\textwidth]{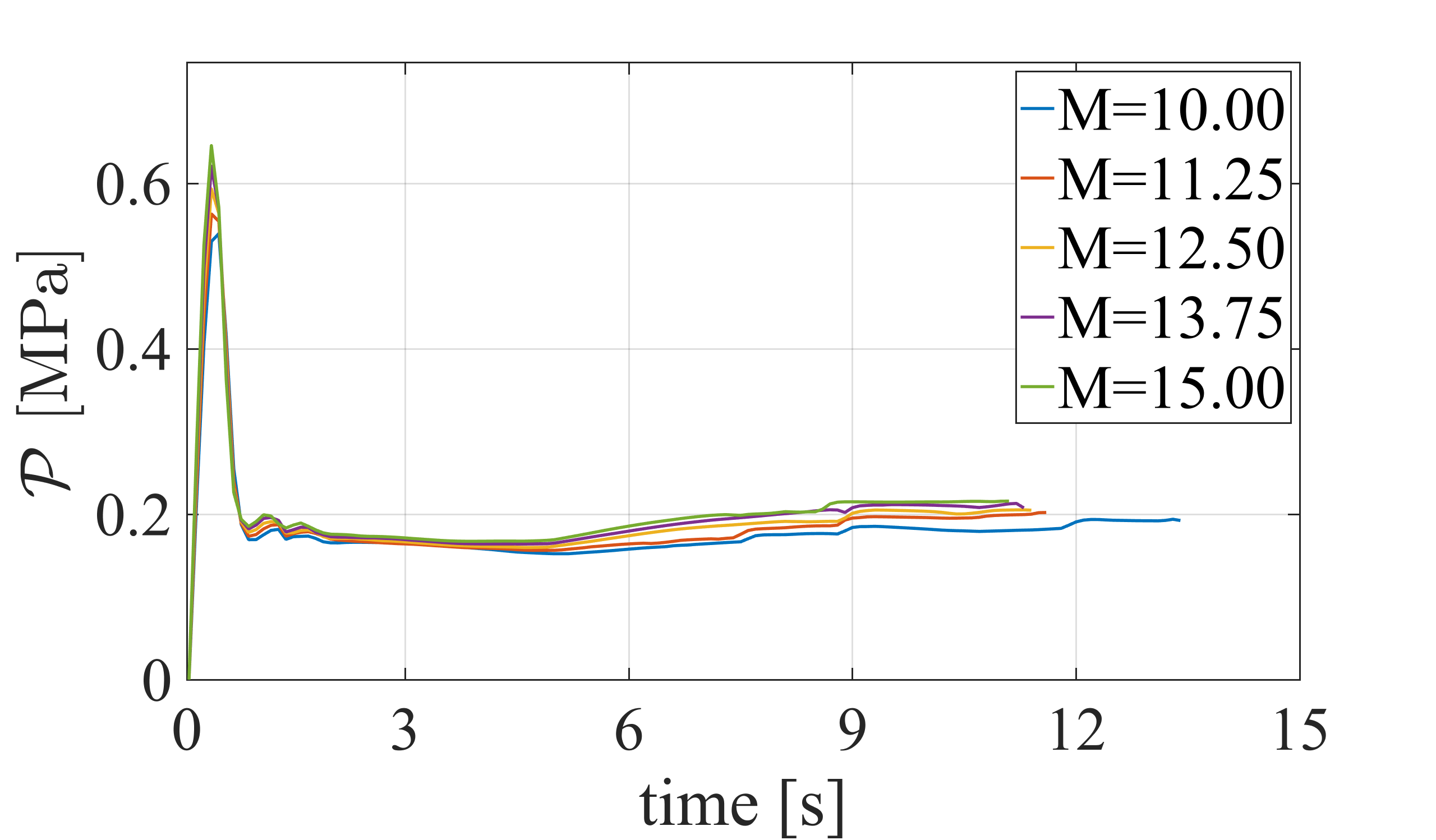}}%
	\newline
	\subfloat{\includegraphics[width=0.35\textwidth]{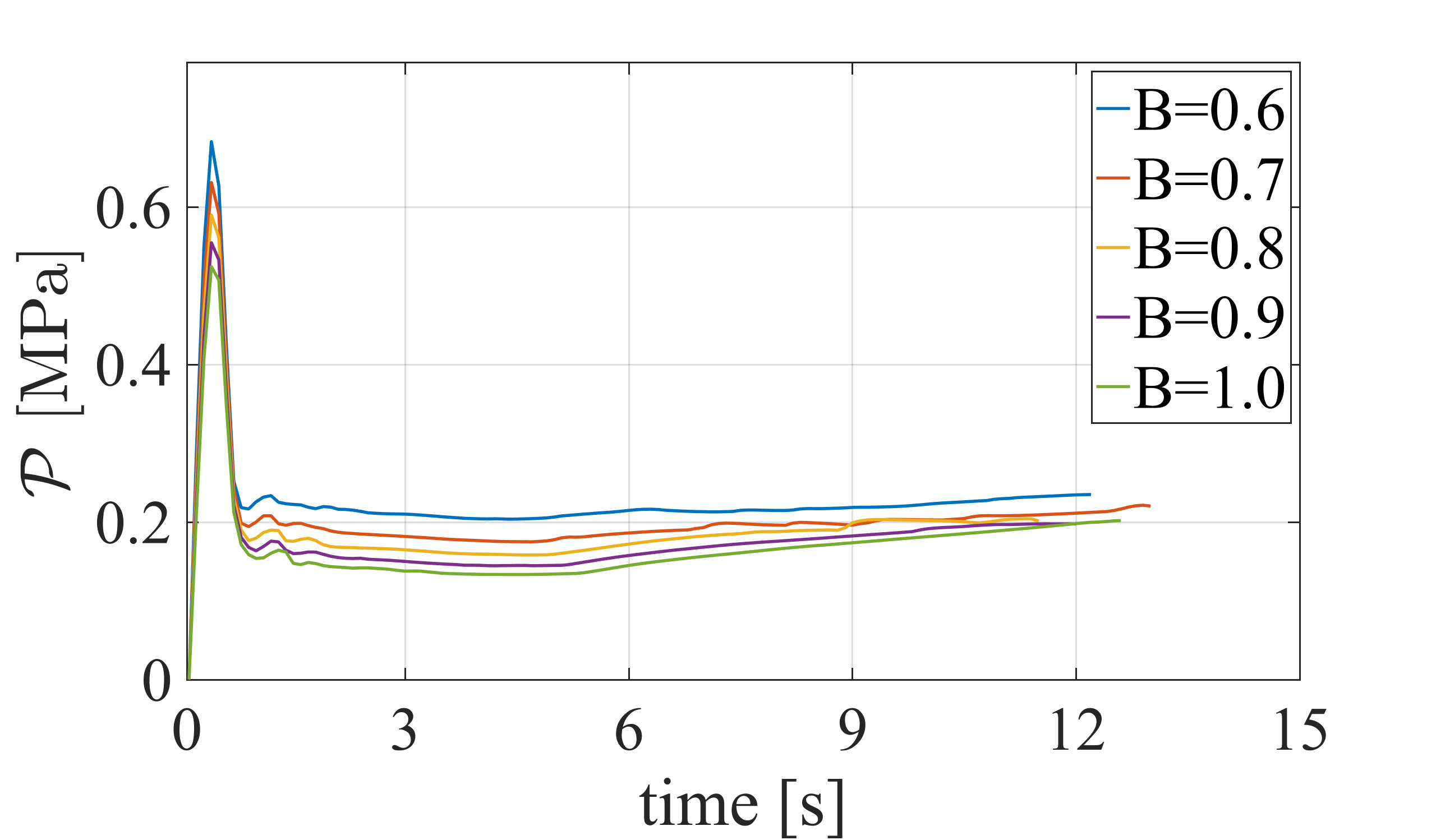}} 
	\subfloat{\includegraphics[width=0.35\textwidth]{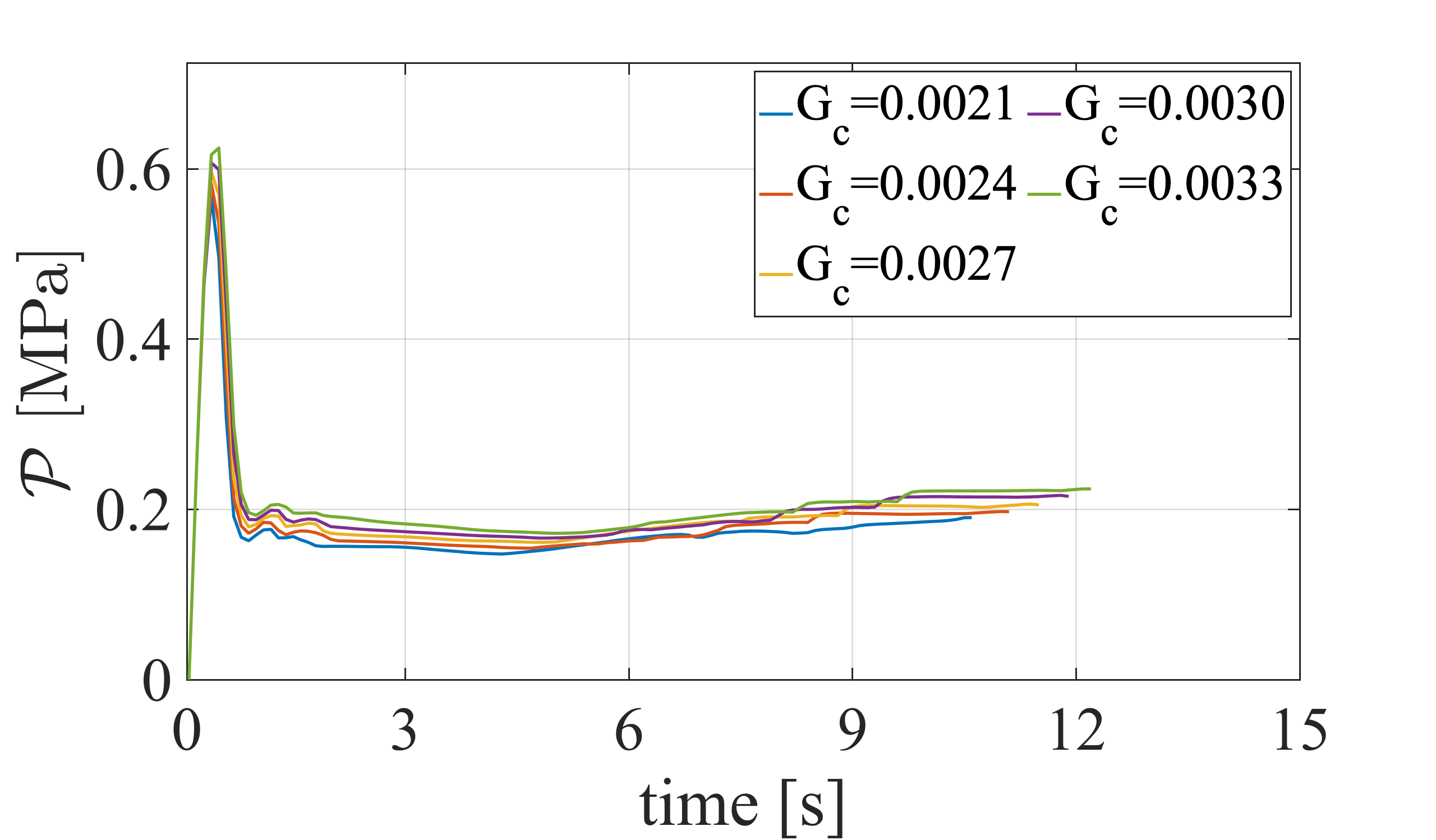}}  
	\subfloat{\includegraphics[width=0.35\textwidth]{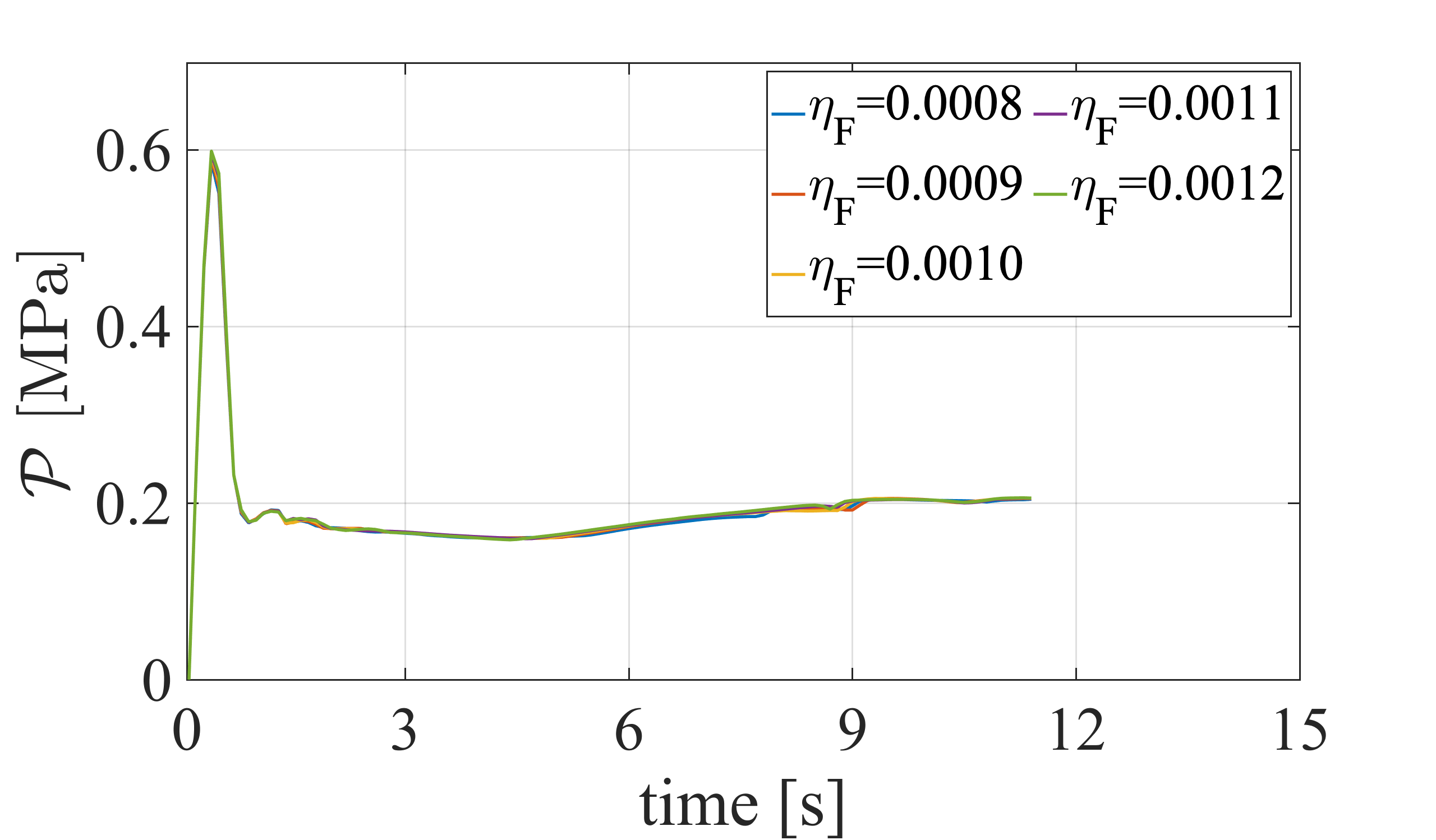}} 
	\caption{Example 4 (Case b).  The maximum pressure value $\PP$  for different values of $\mu$, $K$, $M$, $B$, $G_c$, and $\eta_F$.}
	\label{fig:exam41_curv}
\end{figure}

\begin{figure}[!ht]
	\centering
	{\includegraphics[clip,trim=0cm 23.3cm 0cm 17cm, width=16.3cm,height=3.6cm]{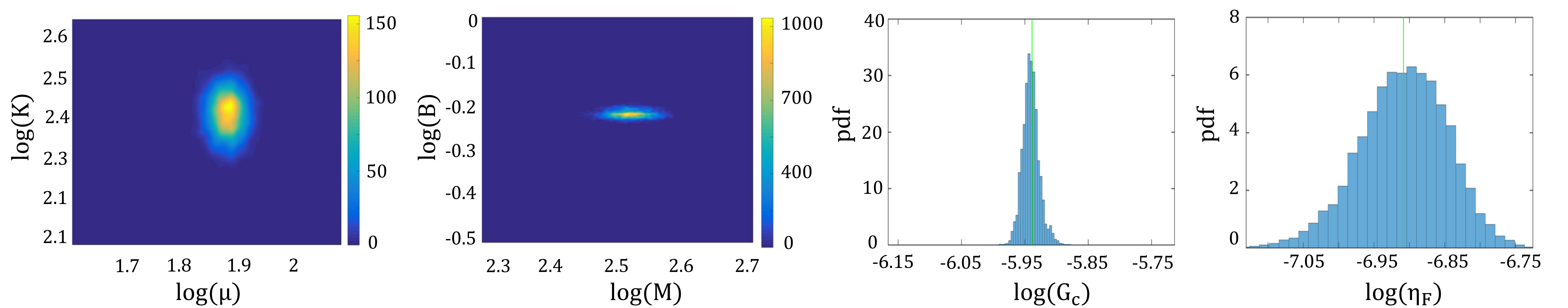}}  
	\caption{Example 4 (Case b). From left to right: the posterior density  of  the mechanical parameters, Biot's coefficients/modulus, $G_c$, and $\eta_F$. The green lines are the mean values.}
	\label{fig:exam41_hist}
\end{figure}  

\begin{figure}[ht!]
	\centering
	\subfloat{\includegraphics[width=8cm,height=4.2cm]{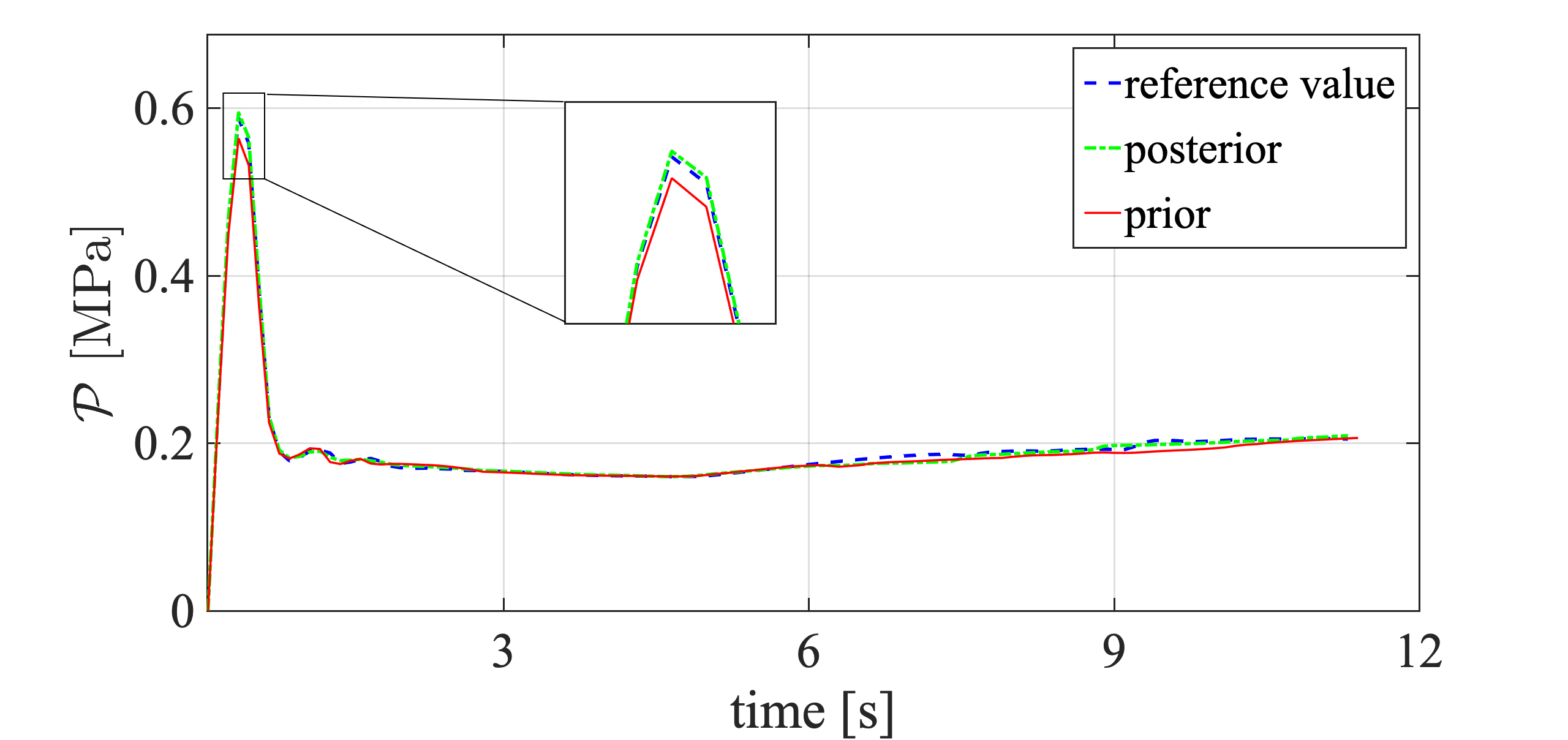}} 
	\subfloat{\includegraphics[width=8cm,height=4.2cm]{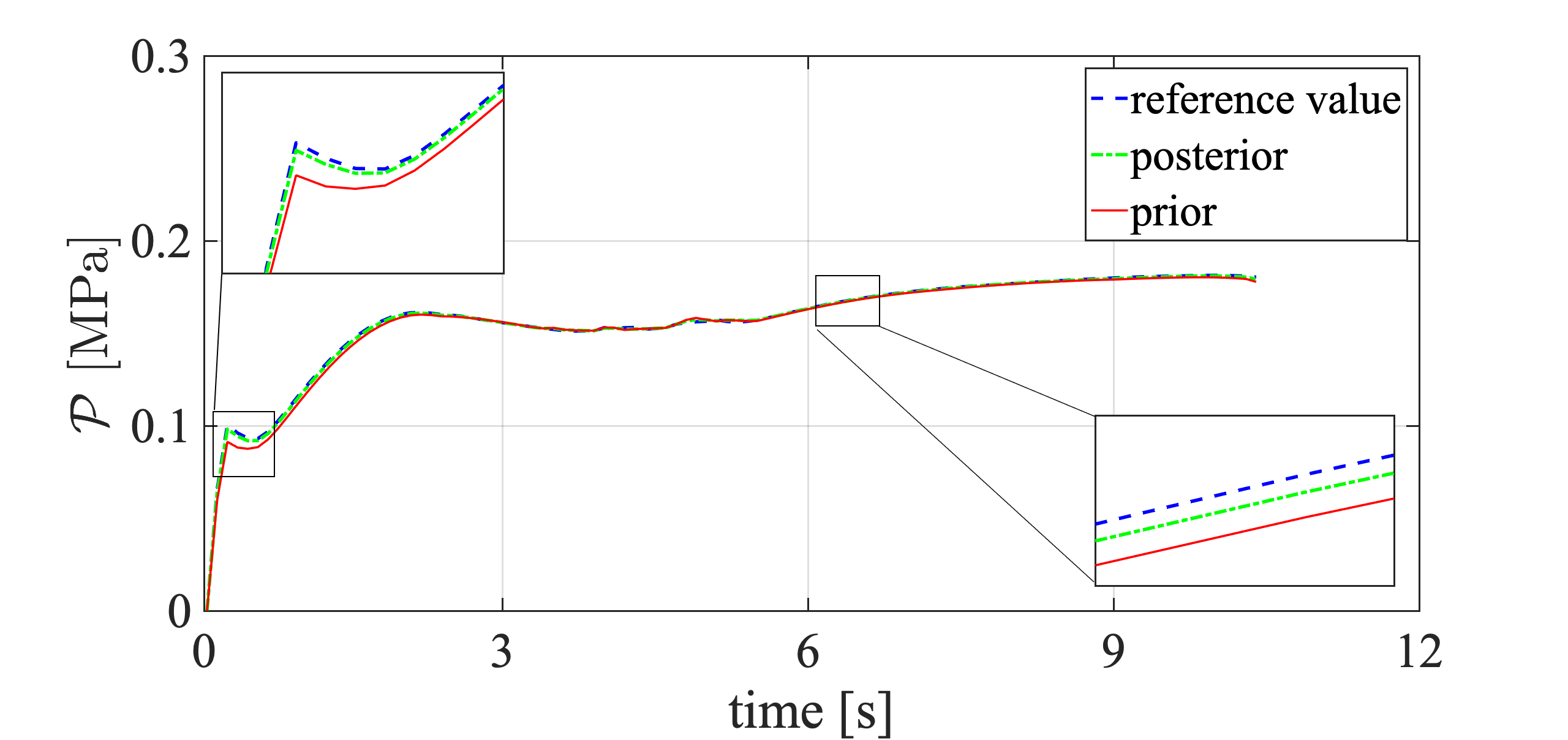}}
	\caption{Example 4. A comparison between the maximum pressure (during the injection time) with prior values (red line), the posterior values (green line), and the reference diagram in Case b (left) and Case d (right).}
	\label{fig:exam41_post}
\end{figure}

\begin{figure}[!]
	\centering
	\subfloat{\includegraphics[clip,trim=2cm 0cm 1.5cm 1cm, width=5cm,height=4.2cm]{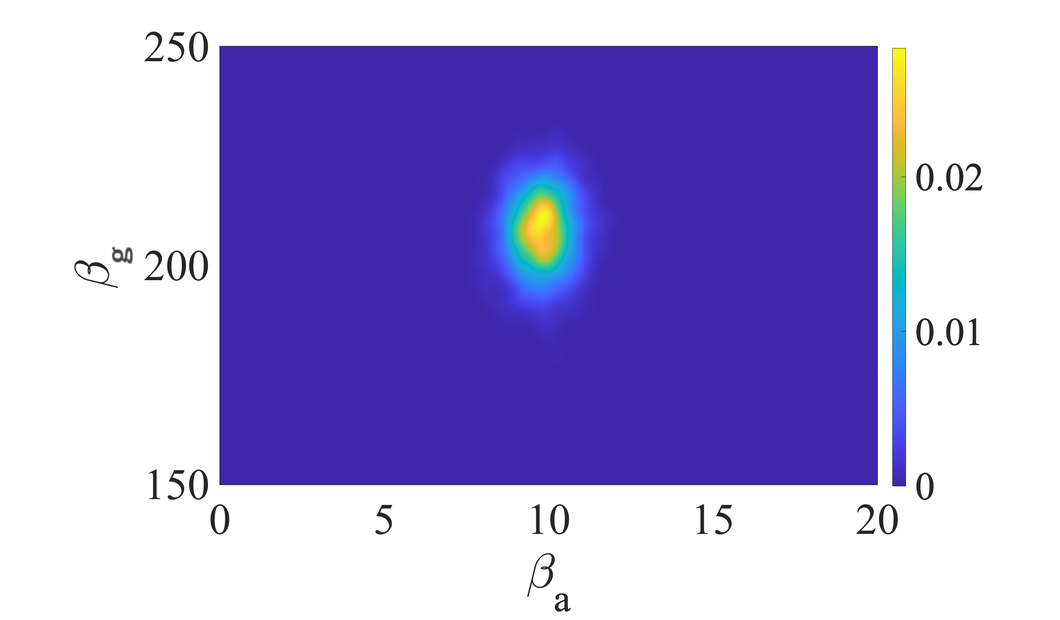}}	\hspace{-0.2cm}
	\subfloat{\includegraphics[clip,trim=2cm 0cm 1.5cm 1cm, width=5cm,height=4.2cm]{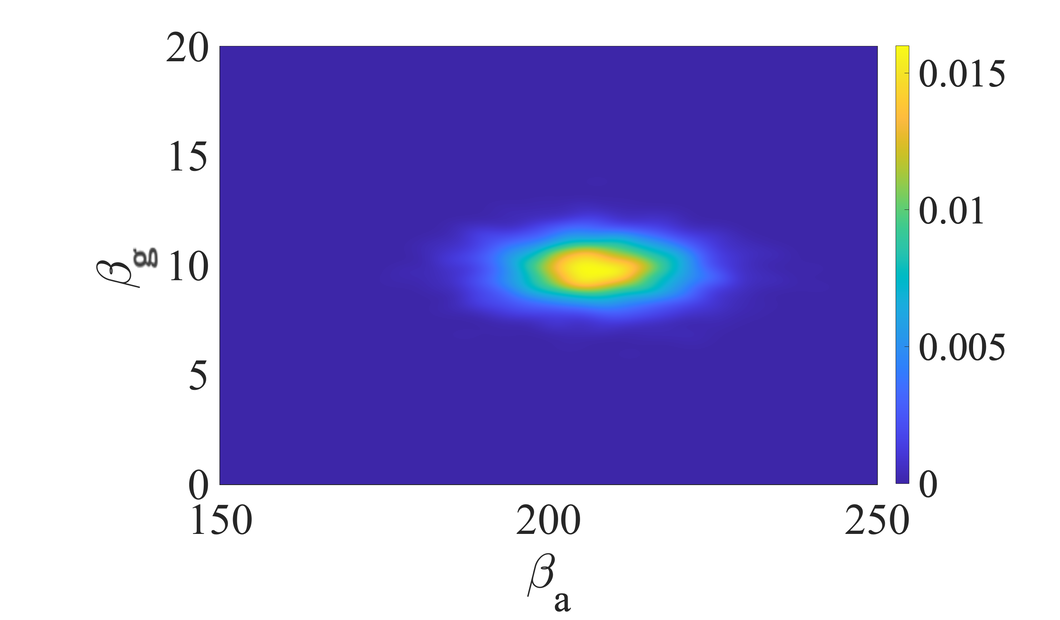}}
	\subfloat{\includegraphics[clip,trim=2cm 0cm 1.5cm 1cm, width=5cm,height=4.2cm]{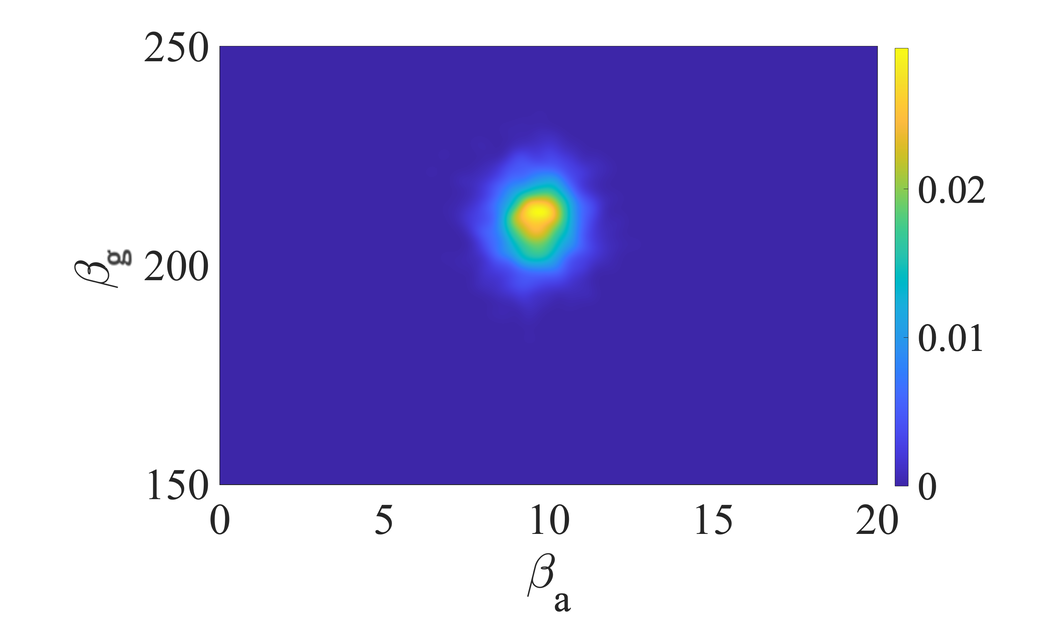}}	
	\caption{Example 4 (Case d).  The joint probability density of the penalty parameters ($\beta_a$ and $\beta_g$) in layer 1 (left), layer 2 (middle), and layer 3 (right).}
	\label{fig:exam42_penalty}
\end{figure}

\begin{figure}[!]
	\subfloat{\includegraphics[width=0.35\textwidth]{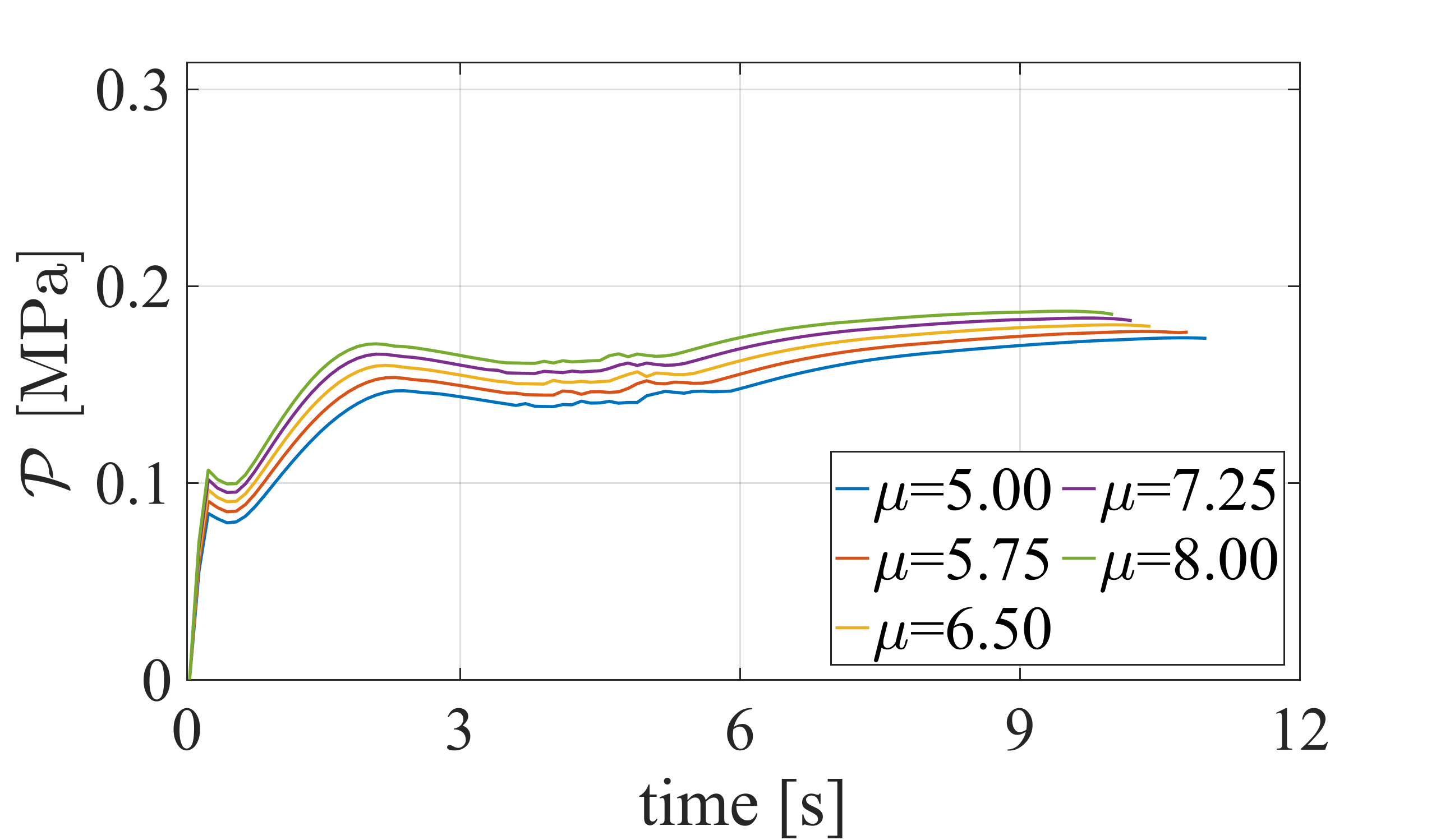}} 
	\subfloat{\includegraphics[width=0.35\textwidth]{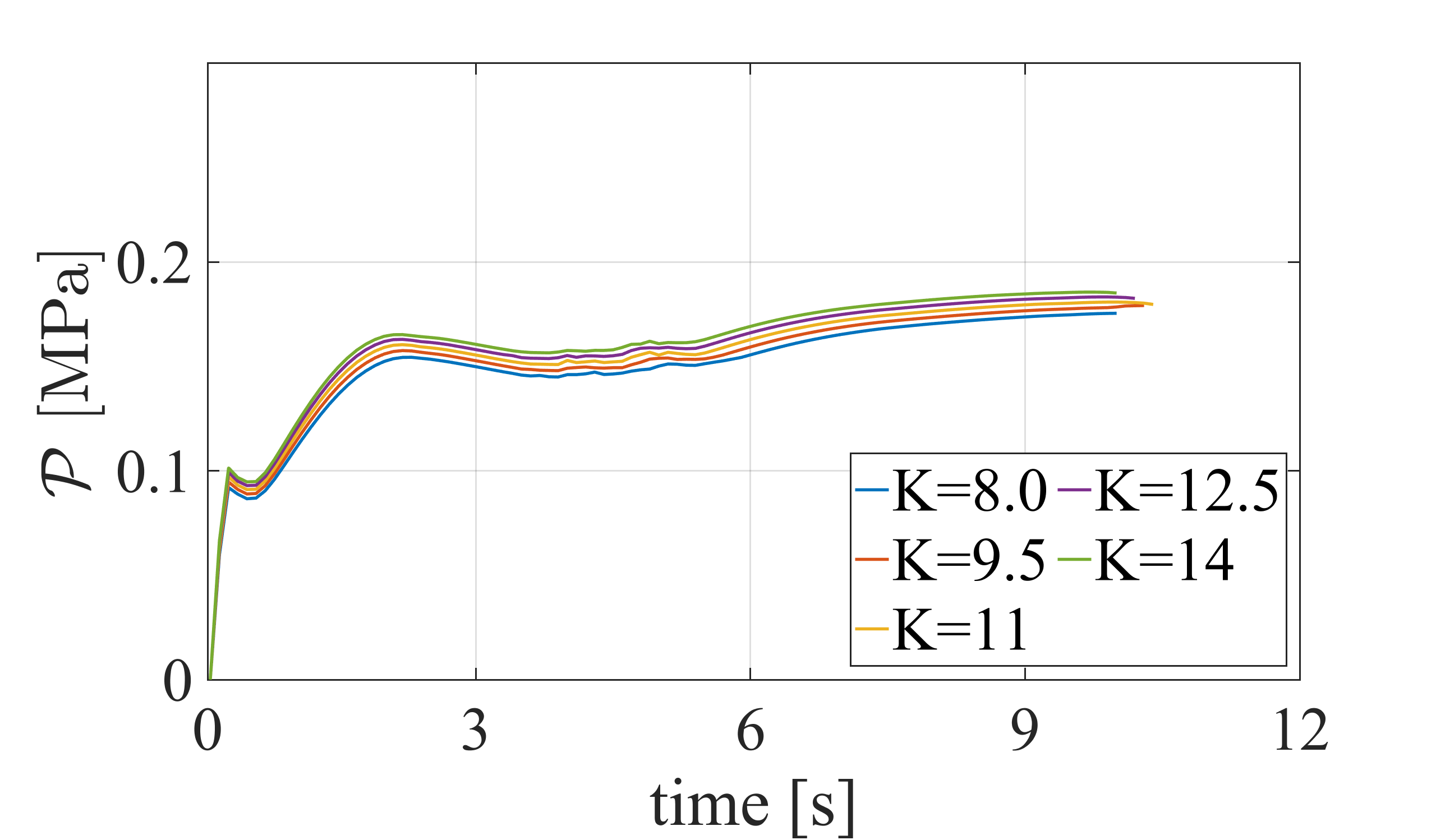}}  
	\subfloat{\includegraphics[width=0.35\textwidth]{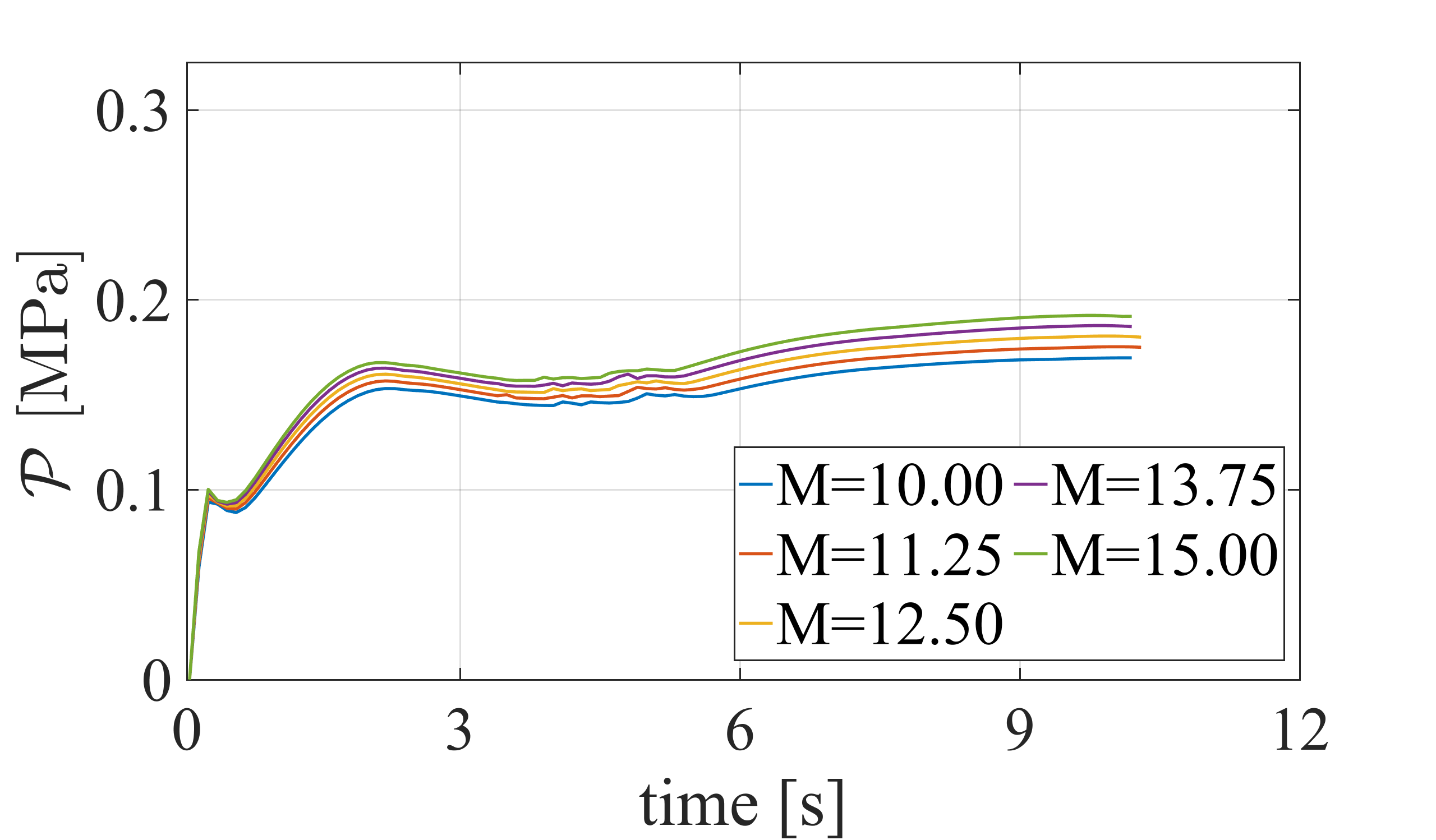}} 
	\newline
	\subfloat{\includegraphics[width=0.35\textwidth]{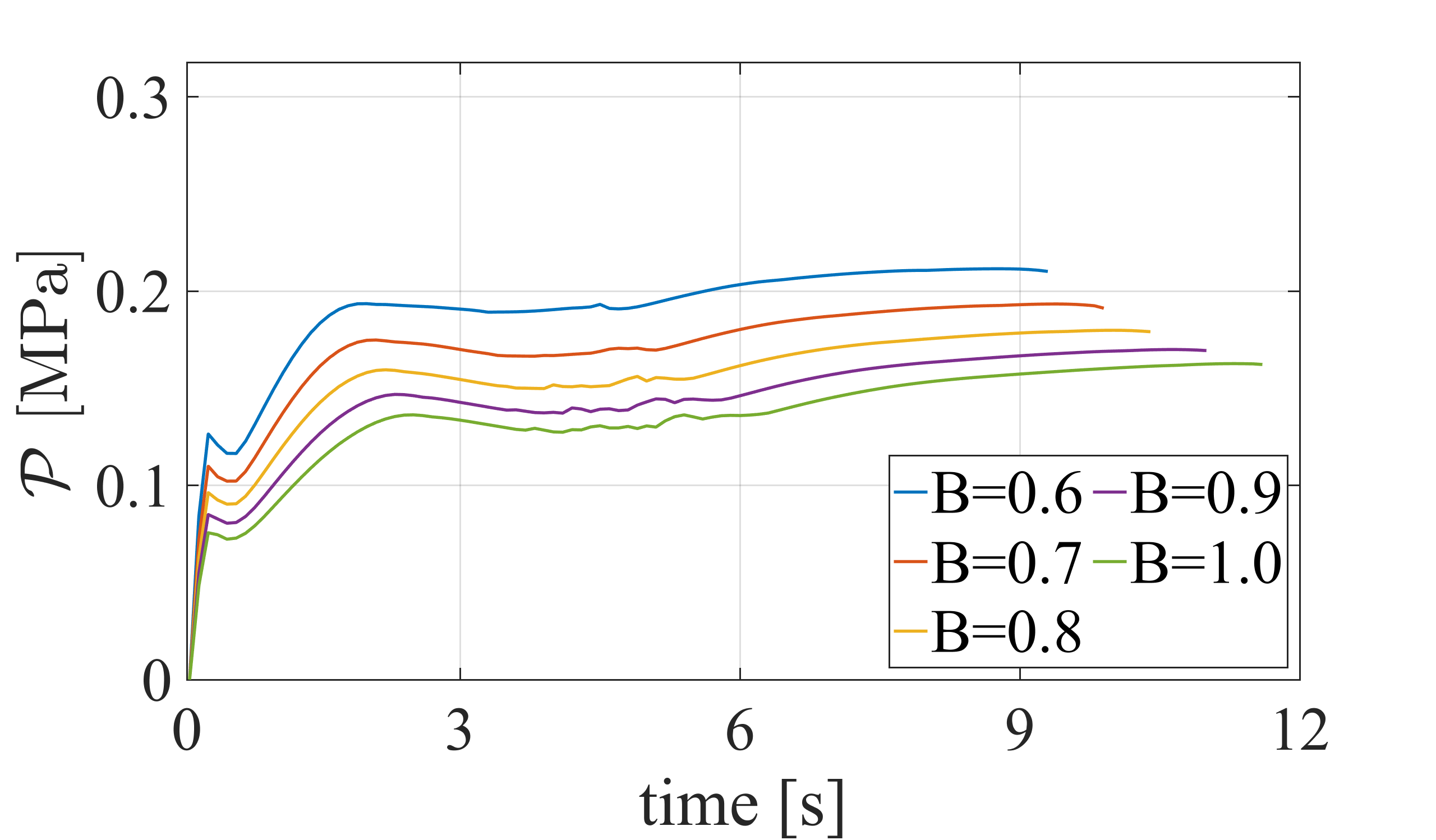}} 
	\subfloat{\includegraphics[width=0.35\textwidth]{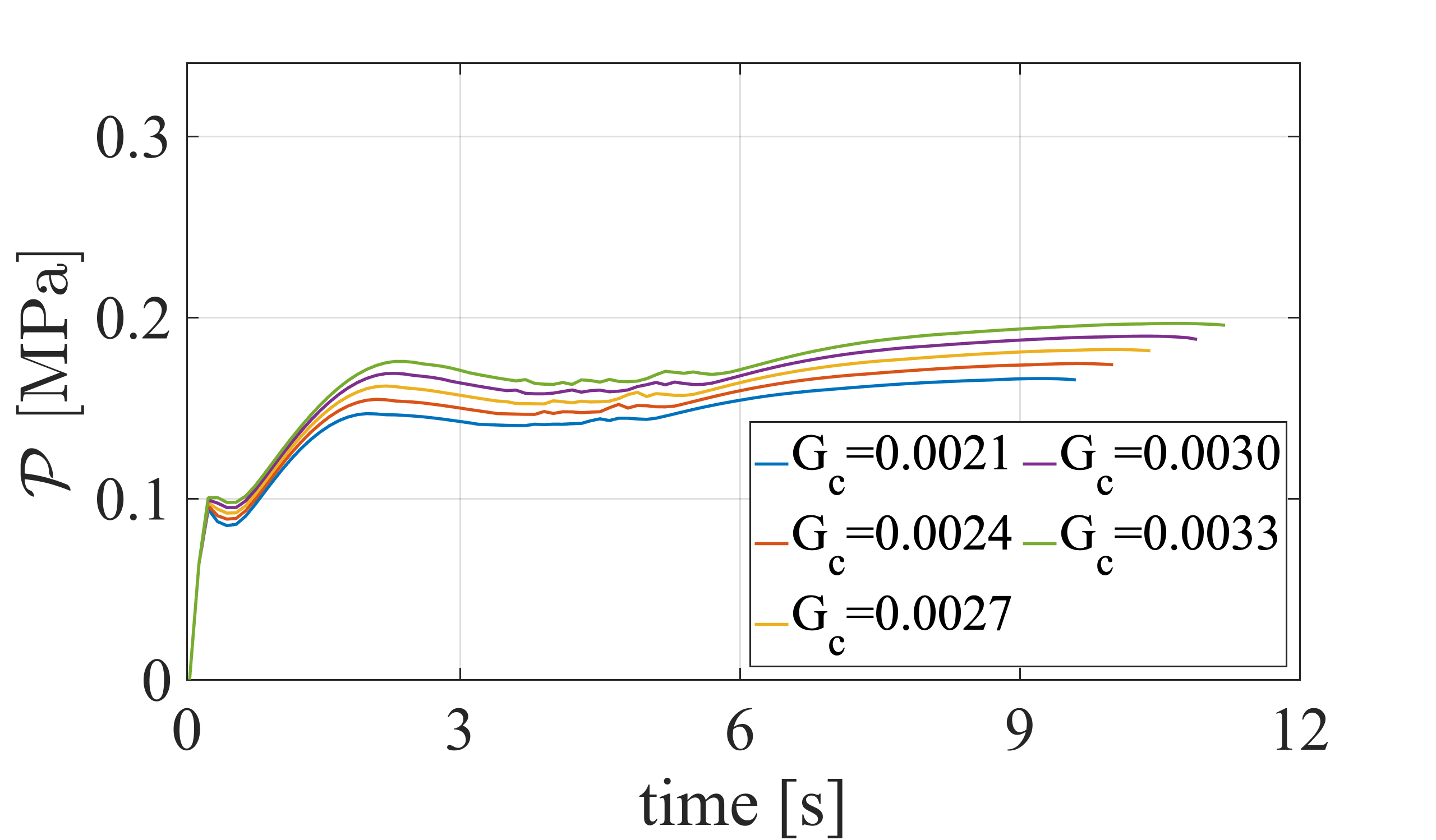}}
	\subfloat{\includegraphics[width=0.35\textwidth]{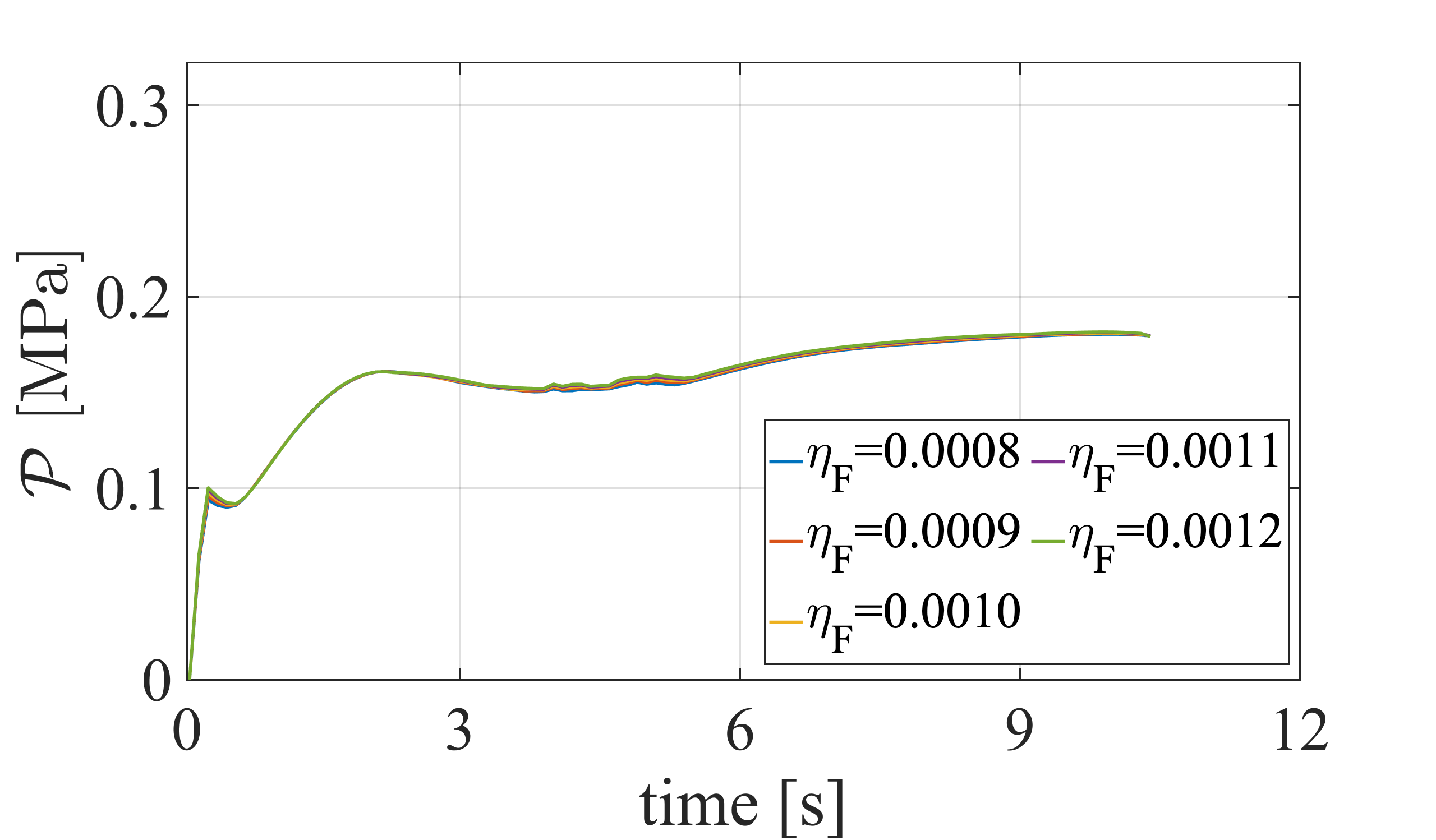}}%
	\caption{Example 4 (Case d).  The maximum pressure value $\PP$  for different values of $\mu$, $K$, $M$, $B$, $G_c$, and $\eta_F$.}
	\label{fig:exam42_curv}
\end{figure}

\begin{figure}[!]
	\centering
	{\includegraphics[clip,trim=0cm 23.3cm 0cm 17cm, width=16.3cm,height=3.6cm]{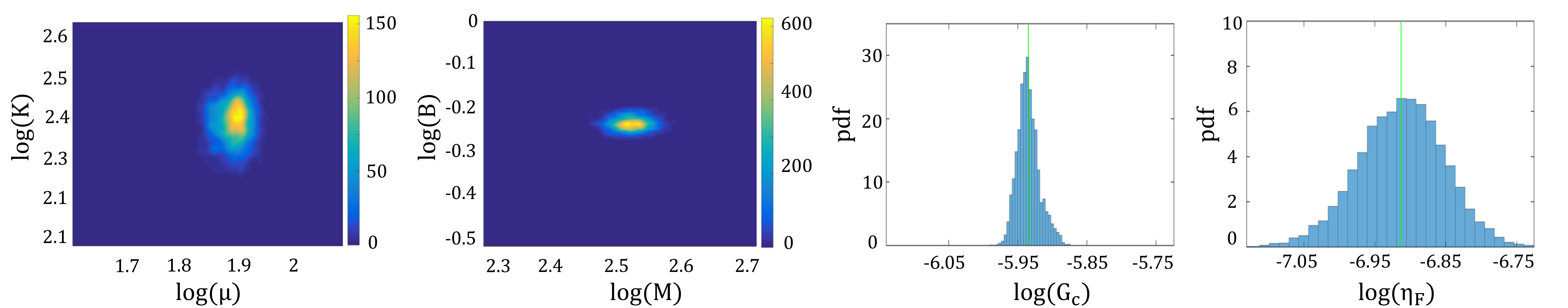}}  
	\caption{Example 4 (Case d). From left to right: the posterior density  of  the mechanical parameters, Biot's coefficients/modulus, $G_c$, and $\eta_F$. The green lines are the mean values.}
	\label{fig:exam42_hist}
\end{figure}  
\cleardoublepage
\section{Conclusions}
\label{sec_conc}
 In this paper, we presented Bayesian inversion for parameter estimation in hydraulic phase-field modeling including transversely isotropic and orthotropy anisotropic fracture. Here, three specific model equations in the sense of pressure, displacement, and crack phase-field have been coupled. Direction-dependent responses due to the preferred fiber orientation in the poroelastic material 
are enforced via an additional anisotropic energy density function for both mechanical and phase-field equations. Furthermore, a new consistent additive split for the bulk anisotropic energy density function is introduced. More precisely, the crack driving state function for the 
poroelastic material was modeled such that the compression mode of the 
anisotropic energy is avoided to be degraded. 
Furthermore, we explained a fully monolithic solution for pressure and displacement, then a staggered approach has been employed for the crack phase-field equation. 

Based on this forward model, we presented a probabilistic setting for hydraulic phase-field fracture. We adjusted the DRAM algorithm to determine various effective parameters in crack propagation. 
To this end, we used pressure during the fluid injection as the reference observation and strive to estimate six variables, including Lam{\'e} constants, Biot's coefficient, Biot's modulus, dynamic fluid viscosity, and Griffith's energy release rate. The approach compared to usual MCMC techniques, e.g., Metropolis-Hastings showed more efficiency due to the proposal adaptation and the delayed rejection; therefore, a more reliable posterior density was obtained.

In 
total, we investigated four different examples including 
various test cases.
In the last two examples, we applied our approach to transversely isotropic and orthotropic anisotropic poroelastic materials. In these cases, the uncertainty arising for the penalty-like parameters in addition to other unknowns is considered.
Our findings showed that pressure evolves in the direction of the preferred fiber orientation. Additionally, the fracture profile is aligned with the highest strength direction of the poroelastic material.  Penalty-like parameters for the anisotropic response, as well as other material properties, are well estimated through the proposed Bayesian inversion.

\section*{Appendix A. Finite Element Discretization} 
\setcounter{equation}{0}
\renewcommand{\theequation}{A.\arabic{equation}}

In the following, we deal with a multi-field problem to be solved with three-field unknowns represented by $(\Bu, p, d)$ to be solved from \req{weakForm}. Here, we aim to provide a detailed consistent linearization procedure within the finite element discretization setting. We use a Galerkin finite element method to discretize the equations with employing $H^1$-conforming bilinear (2D) elements, i.e., the
ansatz and test space uses $Q_1^c$--finite elements. We refer interested readers to \cite{Cia87} for more details. Hence, the discrete spaces have the property $\BV_{\Bu,h} \subset \BV_{\Bu}$, $W_h \subset W$ and $V_{p,h}\subset V$, see \req{space1}. In the finite element setting, the continuous primal fields are described based on piecewise polynomial discrete functions so-called nodal shape function $N^i(\bm{\xi})$ connected with the node $i$. 

Let a continuous domain $\calB$ is approximated to $\calB_h$ such that $\calB\approx\calB_h$. Approximated domain $\calB_h$ is decomposed with non-overlapping finite numbers of bilinear quadrilateral element $\calB_e\subset\calB_h$ such that 

\[ \calB\approx\calB_h = \bigcup_{e}^{n_{e}}\calB_e. \]
The finite element discretized solutions are approximated by

\begin{equation}\label{SDfem4}
{\Bu}^{h}=\sum_{i} \BN_{\Bu}^{i}\;\hat{{\Bu}}^{\;i}, \quad
{p}^{h}=\sum_{i} N_{p}^{i}\;\hat{{p}}^{\;i}, \quad
d^{h}=\sum_{i} N_{d}^{i}\;\hat{\bm d}^{\;i},
\end{equation}
with following basis functions
\begin{equation}\label{SDfem6}
\BN^{\bullet,i}=\begin{bmatrix}
N^{\bullet,i} & 0 \\[5pt]
0 & N^{\bullet,i}
\end{bmatrix},\quad
N_{d}^{i}=\begin{bmatrix}
N_{d}^{i}
\end{bmatrix}, \quad
N_{p}^{i}=\begin{bmatrix}
N_{p}^{i}
\end{bmatrix}.
\end{equation}
Accordingly, its constitutive state variables represented by $(\Bve^{h}, \BP^{h}, \nabla_{\Bx} d^{h})$
\begin{equation}\label{SDfem7}
\begin{aligned}
&\Bve^{h}(\Bu)
=\nabla^{sym}_{\Bu} {\Bu}^{h}
=\sum_{i}\BB_{\Bu}^{i}\hat{{\Bu}}^{\;i}_{G}, \\
&\BP^{h}(d_L)=\nabla_{\Bx} p^{h}=\sum_{i}\BB_{p}^{i}\hat{ p}^{\;i}_{L},\\
&\BG^{h}(d_L)=\nabla_{\Bx} d^{h}=\sum_{i}\BB_{d}^{i}\hat{ d}^{\;i}_{L}\;,
\end{aligned}
\end{equation}
where $\BB_{\Bu}^{i}$, $\BB_{p}^{i}$ and $\BB_{d}^{i}$ are the matrix representation the $i^{th}$ nodal shape function's derivative, corresponds to the deformation, pressure and crack phase-field, respectively. To do so, the matrix  $\BB$ in two-dimensional setting takes the following explicit form
\begin{equation}\label{SDfem9}
\BB_{\Bu}^{i}=\begin{bmatrix}
N_{\Bu,1}^{i} &\quad0 \\\\
0 &\quad N_{\Bu,2}^{i} \\\\
N_{\Bu,2}^{i} &\quad N_{\Bu,1}^{i}
\end{bmatrix},\quad
\BB_{p}^{i}=\begin{bmatrix}
N_{p,1}^{i} \\[5pt]
N_{p,2}^{i}
\end{bmatrix},\quad
\BB_{d}^{i}=\begin{bmatrix}
N_{d,1}^{i} \\[5pt]
N_{d,2}^{i}
\end{bmatrix}.
\end{equation}

The set of the discretized equilibrium equations based on \textit{residual} force vector denoted by $\BR^{\bullet}$ for all primary fields, i.e., $(\Bu,p,d)$, has to be determined. Thus,  we have


\begin{equation}\label{fem_R}
\begin{array}{ll}
\begin{aligned}
{\widehat{\bf{R}}^{\Bu}}
= \bigcup_{e=1} \sum_{i} \int_{\calB_e} \Big({{\bm B_\Bu^{i}}^{T} {{\bm\sigma}_h(\bm u)} \ dV} 
-\displaystyle\int_{\calB} {(\bm N_u^i)^{T}}{\bm {\bar b}}\; dV\Big)
-\displaystyle\int_{\Gamma_{N}} {(\bm N_u^i)^{T}}{\bm {\bar \tau}}dA
{=}{\bm 0}\;, 
\end{aligned}\\ [3mm]\\
\begin{aligned}
{\widehat{\bf{R}}^{p}}
= \bigcup_{e=1} \sum_{i} & \displaystyle\int_\calB \big({N_p^{i}}\big)^{T}\Big[\Big(\frac{1}{M}(p-p_n) 
+ B \big(tr(\Bve)-tr(\Bve_n)\big) 
-\Delta t \;\bar{r}_F	\Big) \;dV \\ 
&+\displaystyle\int_\calB  \big(\Delta t {\bm B_p^{i}}\big)^{T} \;\big(\BK\big) \;\nabla p \cdot  dV \\ 
& + \displaystyle\int_{\partial_N\calB} \big({N_p^{i}}\big)^{T}\bar{f} \; dA = {\bm 0} \ , \\
\end{aligned}\\ [3mm]\\
\begin{aligned}
{\widehat{\bf{R}}^d}= 
\bigcup_{e=1} \sum_{i} \Big(
&\int_{\calB_e} \Delta t\big({{N_d^{i}}\big)^{T}\Big[ {g^{\prime}(d^{h}_+)\calH}+(d^{h}-1))\Big] \ \; dV} \\
-&\int_{\calB_e} \big({{N_d^{i}}\big)^{T}\eta(d^{h}-d_n^{h}) \ \; dV} \\
+&\int_{\calB_e} \Delta tl^2\big({{\bm B_d^{i}}\big)^{T}\big(1+\beta_a\cdot\BM+\beta_g\cdot\BG\big) {\nabla d^{h}}\; dV}
{=}{\bm 0}\;, 
\end{aligned}
\end{array}
\end{equation}
with
\begin{equation*}\label{SDfemND10}
	g^{\prime}(d^{h}_+)=2(1-\kappa)d^{h}_{+}.
\end{equation*}

In order to solve a set of nonlinear algebraic equations that arise in \req{fem_R}, we use an iterative Newton-Raphson method. To that end, the linearization of 
variational formulations concerning the three PDEs for the coupled anisotropic poroelastic given in \req{weakForm} yields
\begin{equation}
\begin{array}{ll}
\Delta G_\Bve(\BfrakU, \delta \Bu) 
&= \displaystyle\int_\calB \Big( \Delta\Bsigma:\delta \Bve \Big) dV , \\ [6mm]
\Delta G_p(\BfrakU, \delta p) &= \displaystyle\int_\calB \Big[\Big(\frac{1}{M}\Delta p 
+ B \Delta tr(\Bve)\Big)\delta p 
- \Delta t \;(\Delta\BcalF) \cdot \nabla \delta p \Big] dV, \\ [6mm]
\Delta G_d(\BfrakU, \delta d) &= (1-\kappa)\Delta t\displaystyle\int_{\calB}\Big[ 2\calH\Delta d . \delta dV\Big]+
\displaystyle\int_{\calB}\Delta t\Big[   \Delta d.\delta d+l^2 \nabla\Delta d.\nabla(\delta d)\Big]\\ [4mm]
&-\eta\Delta d.\delta d\;dV
+\displaystyle\int_{\calB}\Big[ \beta_a l^2 \nabla \Delta d.\BM.\nabla(\delta d)+
\beta_g l^2 \nabla \Delta d.\BG.\nabla(\delta d)\Big]\,
dV.\\
\end{array}
\label{weakFormB1}
\end{equation}
In the linearized form given in \req{weakFormB1}, we need to determine linearized quantities for $\Delta \Bsigma, \Delta \BcalF$ and $\Delta tr(\Bve)$. First, the linearized quantity for the trace operator reads 
\begin{equation}
\Delta tr(\Bve)=\partial_{\Bve}tr(\Bve) :\Delta(\Bve)=\BI:\Delta(\Bve)=tr(\Delta \Bve).
\end{equation}
Additionally, following \req{eq21}, the linearized Cauchy stress tensor given by
\begin{equation}\label{eqA1}
\Delta \Bsigma=\Delta \Bsigma_{eff}-B\BI\Delta p=\Delta \Bsigma_{iso}+\Delta\Bsigma_{aniso}-B\BI\Delta p=\mathbb{C}^{iso}:\Delta\Bve+\mathbb{C}^{aniso}:\Delta\Bve-B\BI\Delta p.
\end{equation}
The corresponding counterparts of the fourth-order elasticity tensor $\mathbb{C}^{iso}$ for the {isotropic} poroelastic, reads

\begin{equation}
\mathbb{C}^{iso}:=\frac{\partial {\bm \sigma}^{eff}( \bm{\varepsilon})}{\partial {\bm \varepsilon}}
=g(d_+)\frac{\partial {\bm {\widetilde{\sigma}}^{iso,+}}( \bm{\varepsilon})}{\partial {\bm \varepsilon}}
+\frac{\partial {\bm {\widetilde{\sigma}}^{iso,-}}( \bm{\varepsilon})}{\partial {\bm \varepsilon}}
=:g(d_+)\widetilde{\mathbb{C}}^{iso,+}+\widetilde{\mathbb{C}}^{iso,-},
\label{eqA2}
\end{equation}
where
\begin{equation}
\widetilde{\mathbb{C}}^{iso,\pm}(\bm\varepsilon):=
\frac{\partial\bm\sigma^\pm}{\partial\bm\varepsilon}
= K H^\pm (I^{\pm}_1(\bm\varepsilon)) \mathbb{J} 
-\mu\Big(\frac{2}{\delta} H^\pm (I^{\pm}_1(\bm\varepsilon)) \mathbb{J}
-2 \mathbb{P}^\pm(\bm\varepsilon)
\Big).
\label{eqA3}
\end{equation}
Here, $H^+$ is the standard Heaviside function, $H^-:=1-H^+$, and $\mathbb{J}:= {\textbf{I}}\otimes {\textbf{I}}$ indicates the fourth-order symmetric identity tensor
with the tension/compression fourth-order projection tensor defined as $\mathbb{P}^\pm_{\bm {\varepsilon}}:=\partial_{\Bve} \bm {\varepsilon}^\pm$, see \cite{miehe2010phase}. Accordingly, the fourth-order elasticity tensor for the anisotropic term $\mathbb{C}^{aniso}$ take the following form
\begin{equation}
\mathbb{C}^{aniso}:=\frac{\partial {\bm \sigma}^{aniso}}{\partial {\bm \varepsilon}}
=g(d_+)\widetilde{\mathbb{C}}^{aniso,+}+\widetilde{\mathbb{C}}^{aniso,-},
\label{eq241}
\end{equation}
where
\begin{equation}
\widetilde{\mathbb{C}}^{aniso,\pm}=H^\pm (I_4)(\chi_a\BM\otimes\BM)+H^\pm (I_6)(\chi_a\BG\otimes\BG).
\end{equation}\\
Finally, the linearized fluid volume flux vector $\Delta \BcalF$ takes the following form
\begin{equation}\label{eqA4}
\Delta \BcalF := {\mathbb{C}}^{\BK}:\Delta \Bve 
-\BK\;\nabla\Delta p
\WITH 
\Delta \BcalF_{i} := 
{\mathbb{C}}^{\BK}_{ijk}\;\Delta \Bve_{jk}
-\BK_{ij}\;\nabla\Delta p_{j},
\end{equation}
with
\begin{equation}\label{eqA5}
{\mathbb{C}}_{ijk}^{\BK}
=\frac{\partial \BcalF_{i}}{\partial \Bve_{jk}}
=-\frac{\partial\BK_{il}\nabla p_{l}}{\partial \Bve_{jk}}
=-\frac{\partial\big((1-d)^{\zeta}\BK_{frac}\big)_{il}\nabla p_{l}}{\partial \Bve_{jk}}
=-(1-d)^{\zeta}\widetilde{\mathbb{C}}_{iljk}^{\BK}\nabla p_{l},
\end{equation}
and
\begin{equation}\label{eqA6}
\widetilde{\mathbb{C}}^{\BK}
=\frac{\partial \BK_{frac}}{\partial \Bve}
= \frac{\omega_d h_e}{{6\eta_F}}\big(
\BI - \Bn \otimes  \Bn 
\big)
\otimes \big(\Bn\otimes\Bn\big).
\end{equation}
Thus, ${\mathbb{C}}^{\BK}$ in \req{eqA4}, using \req{eqA5}-\req{eqA6} takes the following form
\begin{equation}
{\mathbb{C}}^{\BK}=-(1-d)^{\zeta}\widetilde{\mathbb{C}}^{\BK} \ocircle \nabla p \WITH {\mathbb{C}}^{\BK}_{ijk}=-(1-d)^{\zeta}\widetilde{\mathbb{C}}_{iljk}^{\BK} \ocircle \nabla p_{l}.
\end{equation}
Here, we defined a new multiplication operator $\ocircle$ such that $(\BA\ocircle\BB)_{ijk}:=\BA_{iljk}\BB_{l}$.

Now, we are able to determine the \textit{tangent stiffness matrix}  ${\widehat{\bf{K}}} $ for the coupled multi-field problem given in \req{weakForm}. Here, we are solving weak formulation arise from $(\bm u, p)$ in the monolithic manner. Then we use a staggered approach, i.e., alternately fixing $(\bm u, p)$ by solving weak formulation corresponds to the $d$ (see Algorithm 1 for a summary). For this, we need to determine ${\widehat{\bf{K}}^{\Bu\Bu}},\; {\widehat{\bf{K}}^{\Bu p}},\; {\widehat{\bf{K}}^{p\Bu}},\;{\widehat{\bf{K}}^{pp}}$ and also ${\widehat{\bf{K}}^{dd}}$ by
\begin{equation}\label{eq8}
\begin{array}{ll}
\begin{aligned}
{\widehat{\bf{K}}^{\Bu\Bu}} 
=\frac{\partial{\widehat{\bf{R}}^\Bu}}{\partial{\widehat{\Bu}}}
=\bigcup_{e=1} \sum_{i} \int_{\calB_e} \Big({{\bm B_\Bu^{i}}^{T} \big({{\mathbb{C}^{iso}+{\mathbb{C}^{aniso}}}\big){\bm B_\Bu^{i}}} \ dV}\Big), 
\end{aligned}\\ [3mm]\\
\begin{aligned}
{\widehat{\bf{K}}^{\Bu p}}
=\frac{\partial{\widehat{\bf{R}}^{\Bu}}}{\partial\widehat{p}}
= \bigcup_{e=1} \sum_{i} \int_{\calB_e} \Big({{\bm B_\Bu^{i}}^{T} \big({-B\BI}\big){N_p^{i}} \ dV} \Big),
\end{aligned}\\ [6mm]
\begin{aligned}
{\widehat{\bf{K}}^{pp}} 
=\frac{\partial{\widehat{\bf{R}}^p}}{\partial \widehat{p}}
= \bigcup_{e=1} \sum_{i}  \displaystyle\int_\calB \Big( \big({N_p^{i}}\big)^{T}\big(\frac{1}{M}\big){N_p^{i}}  \;dV 
+\displaystyle\int_\calB   \big({\bm B_p^{i}}\big)^{T} \;\big(\Delta t\BK\big) \;{\bm B_p^{i}}   dV 
\Big),
\end{aligned}\\ [6mm]
\begin{aligned}
{\widehat{\bf{K}}^{p\Bu}} 
=\frac{\partial{\widehat{\bf{R}}^p}}{\partial\widehat{\Bu}}
= \bigcup_{e=1} \sum_{i} & \displaystyle\int_\calB \big({N_p^{i}}\big)^{T}  \big(B\BI\big) {\bm B_\Bu^{i}}\;dV \\ 
&-\displaystyle\int_\calB   \big({\bm B_p^{i}}\big)^{T} \;
\Big(\Delta t\;{\mathbb{C}}^{\BK} \;\nabla p\Big){\bm B_\Bu^{i}}   dV.
\end{aligned}
\end{array}
\end{equation}
The tangent stiffness matrix for the anisotropic crack phase-field is given by

\begin{equation}\label{eq9}
\begin{aligned}
{\widehat{\bf{K}}^{dd}}  
=\frac{\partial{\widehat{\bf{R}}^d}}{\partial \widehat{d}}
=\bigcup_{e=1} \sum_{i} \Big(
&\int_{\calB_e} \Delta t\big({{N_d^{i}}\big)^{T}\big({2(1-\kappa)\calH}+1\big){N_d^{i}} \ \; dV} \\
-&\int_{\calB_e} \big({{N_d^{i}}\big)^{T}\big(\eta\big){N_d^{i}} \ \; dV} \\
+&\int_{\calB_e} \Delta tl^2\big({{\bm B_d^{i}}\big)^{T}\big(1+\beta_a\cdot\BM+\cdot\beta_g\cdot\BG\big) {\bm B_d^{i}}\; dV}
\Big). 
\end{aligned}
\end{equation}

Residual force vector in \req{fem_R} along with tangent stiffness matrix in \req{eq8}, results to  update the solution field ${\widehat{\Bvarphi}}_{k+1}$ through

\begin{equation}\label{eq10}
{\widehat{\Bvarphi}}_{k+1}
={{\widehat{\Bvarphi}}_{k}}
-{\widehat{\bf{K}}}^{-1}{\widehat{\bf{R}}}({\widehat{\Bvarphi}_k}),
\end{equation}
where
\begin{equation}\label{eq11}
{\widehat{\bf{K}}}=\begin{bmatrix}
{\widehat{\bf{K}}^{\Bu \Bu}} & {\widehat{\bf{K}}^{\Bu p}} \\[5pt]
{\widehat{\bf{K}}^{p \Bu}} & {\widehat{\bf{K}}^{p p}}
\end{bmatrix},\quad\quad
{\widehat{\bf{R}}}=\begin{bmatrix}
{\widehat{\bf{R}}^{\Bu}}  \\[5pt]
{\widehat{\bf{R}}^{p}} 
\end{bmatrix} \AND
{\widehat{\Bvarphi}}=\begin{bmatrix}
{\Bu}  \\[5pt]
{p}
\end{bmatrix},
\end{equation}
and accordingly for the crack phase-field reads
\begin{equation}
{\widehat{d}}_{k+1}
={{\widehat{d}}_{k}}
-\big({\widehat{\bf{K}}^{dd}\big)}^{-1}\;{\widehat{\bf{R}}}({\widehat{d}}_k).
\end{equation}

\section*{References}
\bibliographystyle{elsarticle-num}
\bibliography{./ref}

\end{document}